\UseRawInputEncoding
\documentclass[reprint,
amsmath,
amssymb,
prb,
superscriptaddress,
nobibnotes
twocolumn]{revtex4-2}
\usepackage{graphicx}
\usepackage[dvipsnames]{xcolor}
\usepackage[unicode=true,colorlinks=true,citecolor=blue,urlcolor=blue]{hyperref}
\usepackage{tabstackengine}
\usepackage{MnSymbol}

\usepackage{amsthm}

\usepackage{float}
\usepackage{physics}
\usepackage{cellspace} 
\usepackage{makecell}
\usepackage{chngpage}
\usepackage{multirow}
\usepackage{ctable} 
\usepackage{bigstrut}
\usepackage[export]{adjustbox}
\usepackage{threeparttable}

\theoremstyle{theorem}
\newtheorem*{theorem}{Theorem}

\newcommand*{\red}[1][]{ \textcolor{red} #1 }
\newcommand*{\blue}[1][]{\textcolor{RoyalBlue} #1 }
\newcommand*{\green}[1][]{\textcolor{Green} #1 }

\newcommand*{\mdagger}[1][]{${M}_{#1}\mbox{-}\dagger$}
\newcommand*{\gdagger}{${G}\mbox{-}\dagger$}

\begin{document}

\title{Symmetry-protected topological exceptional chains in non-Hermitian crystals}

\affiliation{%
Department of Physics, The Hong Kong University of Science and Technology, Hong Kong, China
}
\affiliation{%
School of Physics \& State Key Laboratory of Optoelectronic Materials and Technologies, Sun Yat-sen University, Guangzhou 510275, China
}
\affiliation{These authors contributed equally to this work.}

\author{Ruo-Yang Zhang\thanks{These authors contributed equally to this work.}}
\email{ruoyangzhang@ust.hk}
\affiliation{%
Department of Physics, The Hong Kong University of Science and Technology, Hong Kong, China
}%
\affiliation{These authors contributed equally to this work.}

\author{Xiaohan Cui}
\affiliation{%
Department of Physics, The Hong Kong University of Science and Technology, Hong Kong, China
}%
\affiliation{These authors contributed equally to this work.}
\author{Wen-Jie Chen}
\affiliation{%
School of Physics \& State Key Laboratory of Optoelectronic Materials and Technologies, Sun Yat-sen University, Guangzhou 510275, China
}%
\author{Zhao-Qing Zhang}%
\affiliation{%
Department of Physics, The Hong Kong University of Science and Technology, Hong Kong, China
}%
\author{C. T. Chan}%
\email{phchan@ust.hk}
\affiliation{%
Department of Physics, The Hong Kong University of Science and Technology, Hong Kong, China
}%


\begin{abstract}

In non-Hermitian systems, the defective band degeneracies, so-called exceptional points (EPs), can form robust exceptional lines (ELs) in 3D momentum space in the absence of any symmetries. 
Here, we show that a natural orientation can be assigned to every EL according to the eigenenergy braiding around it, and prove the source-free principle of ELs as a corollary of the generalized Fermion doubling theorem for EPs on an arbitrary closed oriented surface, which indicates that if several ELs flow into a junction, the same number of outflow ELs from the junction must exist. 
Based on this principle, we discover three different mechanisms that can stabilize the junction of ELs and therefore guarantee the formation of various types of exceptional chains (ECs) under the protection of  mirror, mirror-adjoint, or ${C}_2\mathcal{T}$ symmetries.
Furthermore, we analyze the thresholdless perturbations to a Hermitian nodal line and map out all possible EC configurations that can be evolved.
By strategically designing the structure and  materials,
we further exhibit that these exotic ECs can be readily observed in non-Hermitian photonic crystals. 
Our results directly manifest the combined effect of spatial symmetry and topology on the non-Hermitian singularities and pave the way for manipulating the morphology of ELs in non-Hermitian crystalline systems.

\end{abstract}

\maketitle


\section{Introduction}

In topological physics, an important direction is to study topologically protected degeneracies with different dimensions in the band structure~\cite{armitage2018Weyl, hasan2021Weyl, fang2016Topological, park2022Nodala}, which can not only be regarded as hypothetical quasi-particles assisting in fundamental physics but also induce curious transport effects enlightening practical applications. 
As impressive progress has been made in the Hermitian gapless topology in the past decades, non-Hermitian degeneracies are increasingly attracting interest from diverse disciplines of physics, especially photonics~\cite{midya2018NonHermitian,miri2019Exceptional, ozdemir2019Parity,zhen2015Spawning,cui2019Realization} and condensed matter physics~\cite{shen2018Topological,gong2018Topological,kawabata2019Symmetry,kawabata2019Classification, zhou2019Periodic,li2019Geometric}. 
It is markedly distinguished from the Hermitian band crossings that non-Hermitian bands can be degenerate at exceptional points (EPs) where both the eigenvalues and eigenvectors of different bands coalesce~\cite{heiss2004Exceptional,ding2016Emergence,cui2019Exceptional}. Owing to their fascinating physical properties, EPs have exhibited versatile functionalities, such as ultra-sensitive sensing~\cite{wiersig2014Enhancing, hodaei2017Enhanced, chen2017Exceptional, xiao2019Enhanced}, chiral non-adiabatic transport~\cite{gao2015Observation, doppler2016Dynamically, hassan2017Dynamically, zhang2018Dynamically}, and unidirectional lasing~\cite{peng2016Chiral, miao2016Orbital, longhi2017Unidirectional}. From the perspective of topology, EPs are essentially the topological obstructions to sorting the complex-valued energy bands.  When traveling around EPs, the eigenvalues braid about each other, swapping order, and may eventually fail to return to their initial state. As uncovered by recent studies, this eigenvalue braiding along 1D loops faithfully characterizes the topological classification of the non-Hermitian gapless phases associated with EPs~\cite{wojcik2020Homotopy,li2021Homotopical,hu2021Knots,yang2021fermion,tang2021Direct,wang2021Topological,hu2021Knot,wojcik2021Eigenvalue}.

In three-dimensional (3D) non-Hermitian systems, the order-2 EPs can generally trace out robust curves, known as exceptional lines (ELs)~\cite{xu2017Weyl,cerjan2019Experimental,wang2019NonHermitian,xiao2020Exceptional}, in the momentum space even without any symmetries. Specifically, it has shown that ELs can be knotted or linked together in nontrivial ways~\cite{carlstrom2019Knotted,zhang2021Tidal,carlstrom2018Exceptional,yang2019NonHermitiana, zhang2020Bulkboundary, he2020Double, wang2021Simulating},  analogous to Hermitian nodal knots and nodal links~\cite{yan2017Nodallink, chang2017Weyllink, sun2017Double, bi2017Nodalknot,yan2017Nodallink,chang2017Weyllink,sun2017Double}. However, as the counterpart of Hermitian nodal chains~\cite{bzdusek2016Nodalchain,yu2017Nodal,chang2017Topological,gong2018Symmorphic,nodal-chain_Lu_2018_NaturePhys, xiong2020Hidden},  exceptional chains (ECs), formed by several connected ELs, have a fundamental difference from other EL configurations, \textit{i.e.}, the existence and stability of ECs demand symmetry protection. Although the connection or intersection of ELs was accidentally observed in a few very recent works~\cite{cerjan2018Effects,tang2020Exceptional,yan2021Unconventional}, the underlying mechanisms of symmetry and topology  had rarely been discussed. As a result, the mystery of  EC formation remains unraveled until now.  

In this work, we reveal that the complex eigenvalue braiding around an EL can assign a positive orientation to the EL, inspired by the recent breakthrough of the Hermitian nodal chain and link theory~\cite{wu2019NonAbelian,tiwari2020NonAbelian,bouhon2020NonAbelian,yang2020Observation,guo2021Experimental,wang2021Intrinsic,yang2020Jones}.
Via generalizing the Fermion doubling theorem of EPs~\cite{yang2021fermion} to arbitrarily oriented and closed surfaces, we prove that the directed ELs are always source-free in the 3D momentum space. As an immediate application, the source-free principle together with certain non-Hermitian spatiotemporal symmetries can enforce several directed ELs to be robustly chained with each other in order to keep the balance between inward and outward Els at the chain point. 
We uncover that by incorporating the Hermitian-adjoint into account, the non-Hermitian crystalline systems are generically described by double-antisymmetry (DAS) space groups, and we propose three DAS point group symmetry-based mechanisms that can stabilize different types of ECs with distinct local morphologies and topological features, such as symmetry-protected eigenenergy Hopf link and diverse kinds of quantized Berry phases. 
Starting with a Hermitian nodal ring, we also study the evolution roadmap towards various types of ECs by introducing thresholdless non-Hermitian perturbations. 
In addition, we further design the non-Hermitian photonic crystals (PCs) to illustrate our ideas for realizing symmetry-protected ECs. Through numerical simulations, three typical ECs \textit{i.e.} a pair of linked orthogonal EC networks protected by three mirror-adjoint symmetries, a planar EC with four non-defected chain points, and a double-earring EC protected by mirror-adjoint and $C_2\mathcal{T}$ symmetries, are observed in the PCs, hence confirming the universality of our theory for full-wave systems.

\section{Orientation of exceptional lines}
\subsection{Discriminant number, eigenvalue braid, and orientation of ELs}
\begin{figure}[b!]
\includegraphics[width=0.49\textwidth]{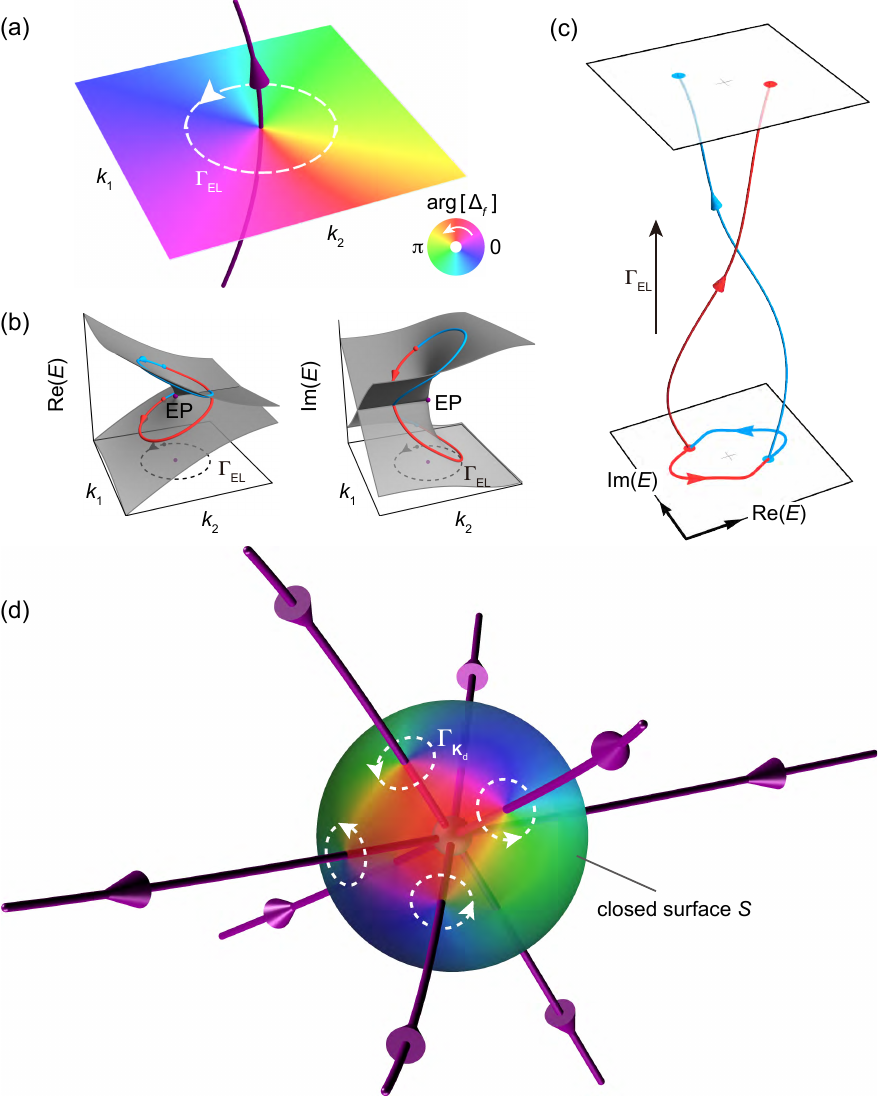}
\caption{\label{Fig-orientation} Orientation of ELs and source-free principle. (a) An order-2 EL manifests as the phase singularity of the discriminant $\Delta_f(\mathbf{k})$, where the sign of phase vortex assigns a positive orientation (arrow on the EL) to the EL in compliance with the right-hand rule of the loop $\Gamma_{\rm{EL}}$ (white dashed).
(b) Real and imaginary parts of two intersecting bands on the transverse plane in (a), and the red and light blue trajectories denote the two modes along the path $\Gamma_\mathrm{EL}$.  (c) Braiding and mode switching of two eigenvalues along the loop $\Gamma_\mathrm{EL}$ in (a). (d) Schematic of the generalized doubling theorem for EPs. Several directed ELs meet at a junction that is enclosed by a surface $S$. Colormap on the surface: $\arg[\Delta_f(\mathbf{k})]$. }
\end{figure}

Unlike nodal lines in Hermitian systems whose existence requires symmetry protection~\cite{fang2016Topological,park2022Nodala}, a line of order-2 EPs  in 3D momentum space can be topologically stable in the absence of any symmetries~\cite{kawabata2019Classification}, providing that the two crossing energy bands, $E_i$, $E_{i+1}$, braid about each other and swap their order  along a loop $\Gamma_\mathrm{EL}$ encircling the EL (see Fig.~\ref{Fig-orientation}(a-c)), 
characterized by the half-quantized interband energy vorticity~\cite{shen2018Topological,kawabata2019Classification}
   $ \nu_{i,i+1}(\Gamma_\mathrm{EL})=\frac{1}{2\pi}\oint_{\Gamma_\mathrm{EL}} d\mathbf{k}\cdot\nabla_\mathbf{k}\,\mathrm{arg}\left[E_i(\mathbf{k})-E_{i+1}(\mathbf{k})\right]=\pm\frac{1}{2}$.
The sign of the energy vorticity of the two intertwining bands endows such an elementary EL with a positive orientation, as indicated by the arrows on the ELs in Fig.~\ref{Fig-orientation}:
\begin{equation}
    \mathbf{t}_\mathrm{EL}=\mathrm{sign}[\nu_{i,i+1}(\Gamma_\mathrm{EL})]\mathbf{t}_\Gamma,
\end{equation}
where $\mathbf{t}_\Gamma$ is a tangent vector of the EL in compliance with the right-hand rule of the directed loop $\Gamma_\mathrm{EL}$.

For a generic multi-band Bloch Hamiltonian $\mathcal{H}(\mathbf{k})$, a $\mathbb{Z}$ topological invariant, dubbed as discriminant number (DN)~\cite{yang2021fermion}, was recently introduced to demarcate non-Hermitian topological phases on a 1D closed sub-manifold (a closed path $\Gamma$), 
which is defined as the sum of all interband vorticities $\nu_{ij}(\Gamma)$ along the path, and is equivalent to the phase winding number of the discriminant $\Delta_f(\mathbf{k})=\prod_{i<j}\left[E_i(\mathbf{k})-E_j(\mathbf{k})\right]^2$ of the Hamiltonian's characteristic polynomial $f(E,\mathbf{k})=\det[E-\mathcal{H}(\mathbf{k})]$~\cite{yang2021fermion,wojcik2020Homotopy}, 
\begin{equation}
    \mathcal{D}(\Gamma)=\sum_{i\neq j}\nu_{ij}(\Gamma)=2\sum_{i< j}\nu_{ij}(\Gamma)=\frac{-i}{2\pi}\oint_{\Gamma} d\mathbf{k}\cdot\nabla_\mathbf{k}\mathrm{ln}\Delta_f(\mathbf{k}).
\end{equation}
Once $\mathcal{D}(\Gamma)\neq0$, there exist lines of degenerate points $\{\mathbf{k}_d|\Delta_f(\mathbf{k}_d)=0\}$ encircled by the path $\Gamma$, prohibiting the contraction of $\Gamma$ to a point without gap closing. 

If the loop $\Gamma$ only encloses a single EL, the DN is absolutely contributed by the energy vorticity of the two bands $E_i,E_{i+1}$ forming the EL: $\mathcal{D}(\Gamma)=2\nu_{i,i+1}(\Gamma)=\pm1$, then the positive tangent vector of the EL is alternatively expressed as $\mathbf{t}_\mathrm{EL}=\mathcal{D}(\Gamma)\mathbf{t}_\Gamma$.
And for an arbitrary loop $\Gamma$ in the 3D space, the DN, $\mathcal{D}(\Gamma)\in\mathbb{Z}$, characterizes the net number of the directed ELs, as counted according to their positive directions, enclosed by the loop.

Recent studies unveiled that the braiding of eigenvalues along 1D loops (\textit{e.g.}, Fig.~\ref{Fig-orientation}(c)) can faithfully determine the 1D non-Hermitian topology with separable bands of a generic $N$-band system, and the complete classification of such topological phases is given by the braid group of $N$ strands, $B_N$, generated by the braid generators $b_{i}$ ($i\in\{1,2,\cdots,N-1\}$), denoting the braiding between $i^\mathrm{th}$ and $(i+1)^\mathrm{th}$ eigenvalues~\cite{wojcik2020Homotopy,li2021Homotopical,hu2021Knots}. For an arbitrary 1D loop $\Gamma$, the braid element, $b(\Gamma)=b_{i_1}^{\ n_1}b_{i_2}^{\ n_2}b_{i_3}^{\ n_3}\cdots\in B_N$, and the DN, $\mathcal{D}(\Gamma)\in\mathbb{Z}$, carried by the loop satisfy the relation (see proof in Supplementary Information 1)
\begin{equation}\label{relation between DN and braid}
    \mathcal{D}(\Gamma)=n_1+n_2+n_3+\cdots.
\end{equation}
Namely, the DN is equal to the sum of the exponents on the braid generators (also known as the algebraic length of the braid $b(\Gamma)$), 
representing the net number of times the mode braiding takes place. 
On the other hand, the total Berry phase of all bands along $\Gamma$, known as the global Berry phase $\Theta(\Gamma)$, also assigns a $\mathbb{Z}_2$ topological invariant to the loop~\cite{liang2013topological,hu2021Knots}.  Whenever the sequences of the initial and finial eigenstates, $\Psi_0$ and $\Psi_f$, after a cycle along $\Gamma$ are different up to a permutation $\hat{p}(\Gamma)$, $\Psi_f=\hat{p}(\Gamma)\Psi_0$, the global Berry phase is determined by the parity of the permutation~\cite{hu2021Knots}, $\det[\hat{p}(\Gamma)] = \pm 1$, and therefore also by the parity of the DN (see proof in Supplementary Information 1):
\begin{equation}\label{Eq-globalBP}
\begin{split}\exp\left[i\Theta(\Gamma)\right]&=\exp\left[\oint_\Gamma  d\mathbf{k}\cdot\mathrm{Tr}\left(\Psi^{-1}\nabla_\mathbf{k}\Psi\right)\right]\\
        &= {\det[\hat{p}(\Gamma)]} = (-1)^{\mathcal{D}(\Gamma)},
\end{split}    
\end{equation}
where $\Psi=(|\psi^R_1(\mathbf{k})\rangle,|\psi^R_2(\mathbf{k})\rangle,\cdots,|\psi^R_N(\mathbf{k})\rangle)$ and $\Psi^{-1}=(|\psi^L_1(\mathbf{k})\rangle,\cdots,|\psi^L_N(\mathbf{k})\rangle)^\dagger$ denote the matrices composed of right and left eigenstates at $\mathbf{k}$, respectively.
Thus, the $\mathbb{Z}_2$ global Berry phase actually describes whether there are even or odd number of ELs passing through the loop.

\subsection{Source-free principle of ELs}
Regarding an orientable closed surface, $S$, in the 3D  Brillouin zone (BZ), we are interested in the net number of ELs penetrating $S$, as shown by the schematic in Fig.~\ref{Fig-orientation}(d). Because the discriminant $\Delta_f(\mathbf{k})$ is a continuous single-valued function in the whole BZ, mathematically, it serves as a global section on a trivial complex line bundle $\pi:L\cong S\times\mathbb{C}\rightarrow S$  possessing zero Chern number $\mathrm{Ch}(L)=0$. Therefore, we infer from the Poincar\'e--Hopf theorem for complex line bundles~\cite{frankel2011thegeometry,Knoppel2020Riemann} that the total DN carried by all isolated degenerate points $\{\mathbf{k}_d\}$ on the surface  must vanish (see Supplementary Information 2),
\begin{equation} \label{doubling theorem}
    \sum_{\mathbf{k}_d\in S}\mathcal{D}(\Gamma_{\mathbf{k}_d})=\mathrm{Ch}(L)=0,
\end{equation}
where $\Gamma_{\mathbf{k}_d}$ stands for a small directed loop encircling the singularity $\mathbf{k}_d$ whose direction for the integral of DN is consistent with the outward normal of the surface.

If all the degenerate points on $S$ are elementary EPs with $\mathcal{D}(\Gamma_{\mathbf{k}_d})=\pm1$, Eq.~\eqref{doubling theorem} generalizes the Fermion doubling theorem for EPs~\cite{yang2021fermion} to arbitrary closed oriented surfaces, \textit{i.e.}, EPs always appear in pairs with opposite DNs on a closed oriented surface. The doubling theorem indicates that the ELs are source-free and have to form closed loops in the 3D BZ, which serves as a conservation rule regulating the morphology and evolution of ELs in the 3D space. In particular, if several oriented ELs meet at a junction under some constraints, the doubling theorem on a sufficiently small sphere enclosing the junction informs us:  the numbers of inflow and outflow ELs must be equal (Fig.~\ref{Fig-orientation}(d)), even though the ELs are formed by different pairs of bands in multi-band systems (see examples in Supplementary Information 9). Next, we will show that the intersections of ELs with various local morphologies can be guaranteed by certain spatiotemporal symmetries.

\section{symmetry-protected exceptional chains}
In previous works on non-Hermitian crystals, the magnetic space groups, including the time reversal operator $\mathcal{T}$, are usually adopted to characterize the crystalline symmetries and band degeneracies~\cite{mock2016Characterization,mock2017Comprehensive}. However, it was revealed recently that Hermitian-adjoint ``$\dagger$'' could appear as a new dimension that arouses intriguingly new physical effects and enriches the classification of non-Hermitian topological phases~\cite{kawabata2019Symmetry,shiozaki2021Symmetry}. Here, we uncover that, akin to the time-reversal, the Hermitian-adjoint transformation could be regarded  as an antisymmetry~\cite{padmanabhan2020Antisymmetry} with respective to the ordinary space groups, hence by involving both Hermitian-adjoint and time reversal as two antisymmetries, the non-Hermitian crystals can be universally described by DAS space groups~\cite{gopalan2011Rotationreversal,padmanabhan2020Antisymmetry,vanleeuwen2014Double} (see Methods for details). In what follows, we introduce three mechanisms to stabilize ECs by DAS point-group symmetries.

\begin{figure}[t!]
\includegraphics[width=0.49\textwidth]{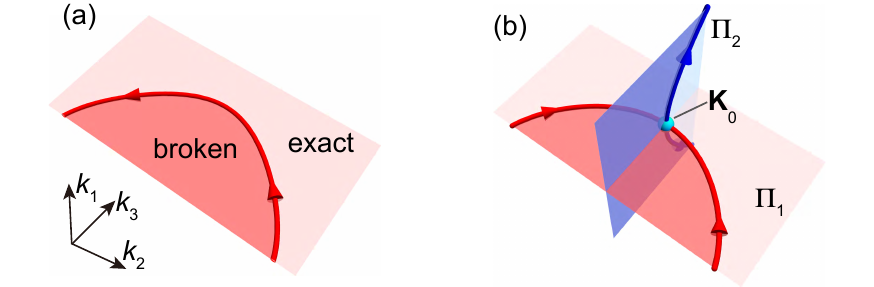}
\caption{\label{Fig-orthogonalEC}(a) Exceptional line confined in a symmetry-invariant plane by $M\hbox{-}\dagger$ or $C_2\mathcal{T}$ symmetry, separating the exact (light red) and broken (red) phases. (b) Orthogonal exceptional chain protected by two symmetries belonging to $M\hbox{-}\dagger$ or $C_2\mathcal{T}$, where ELs, confined in two perpendicular planes $\Pi_{1}$ and $\Pi_{2}$ connect at the chain point $\mathbf{K}_0$. In each plane, the lighter (darker) region denotes the exact (broken) phase.}
\end{figure}

\subsection{Orthogonal ECs protected by Mirror-adjoint (${M}\mbox{-}\dagger$) or $C_2\mathcal{T}$ symmetries}

In Hermitian systems, as a prerequisite for chaining different nodal lines together, the line nodes should be confined by symmetries (usually mirror symmetry) in the high-symmetry planes~\cite{wu2019NonAbelian,tiwari2020NonAbelian,yang2020Observation}, which inspires us that seeking suitable symmetries to fix ELs into planes would be the first step toward an EC. It is known that in a pseudo-Hermitian\cite{mostafazadeh_pseudo-hermiticity_2002,mostafazadeh2010pseudo} or $\mathcal{PT}$ symmetric\cite{bender1998} system, the parameter space can be divided into exact and broken phases wherein the eigenvalues are purely real and in complex conjugate pairs, respectively. And EPs always occur at the boundaries between the two phases. Borrowing this mechanism to a 2D subspace, we find that two DAS point-group symmetries, \textit{i.e.}  mirror-adjoint  (denoted ${M}\mbox{-}\dagger$) and $C_2\mathcal{T}$, fit the bill.

The ${M}\mbox{-}\dagger$ symmetry is a non-Hermitian generalization of  mirror symmetry, which is defined as the combination of mirror reflection and Hermitian adjoint. For the Bloch Hamiltonian $ \mathcal H(\mathbf{k})$ in the momentum space, it can be expressed as
\begin{equation}\label{mirror-dagger}
    \hat{M}\mathcal H\left(\hat{m} \mathbf{k} \right)^\dag{\hat M^{-1}} =\mathcal{H}\left( \mathbf{k} \right),
\end{equation}
where $\hat{m}$ represents the mirror reflection on spatial coordinates and vectors, and $\hat{M}$ is a Hermitian unitary reflection operator on Bloch states. ${M}\mbox{-}\dagger$ is an intrinsic non-Hermitian spatial symmetry distinct from the mirror symmetry in non-Hermitian systems, while they reduce to an identical one in Hermitian cases. 
On a ${M}$-invariant plane $\Pi_{M}=\left\{\mathbf k_m \in \mathrm{BZ} \mid \hat m \mathbf k_m=\mathbf k_m\right\}$, Eq.~\eqref{mirror-dagger} implies that $ \mathcal{H}(\mathbf{k}_m)$ is pseudo-Hermitian. Therefore, as depicted in Fig.~\ref{Fig-orthogonalEC}(a), all eigenstates on that plane are classified as being either exact or broken, and the transition boundary of the two phases forms an EL lying in 
 $\Pi_M$ (see Supplementary Information 4).

Similarly, the bosonic $C_2\mathcal{T}$ symmetry of a Bloch Hamiltonian requires
\begin{equation}
    \hat{C}_2\hat{T}\mathcal{H}(-\hat{c}_2\mathbf{k})^*( \hat{C}_2\hat{T} )^{-1} = \mathcal{H}(\mathbf{k}),
\end{equation}
where  $\hat{C}_2\hat{T}$ is the unitary part saitisfying $(\hat{C}_2\hat{T})(\hat{C}_2\hat{T})^*=1$ as the consequence of $(C_2\mathcal{T})^2=1$. 
On a $C_2\mathcal{T}$-invariant plane $\Pi_{C_2\mathcal{T}}=\{\mathbf{k}_{c}\in\text{BZ}|\,\mathbf{k}_c=-\hat{c}_2\mathbf{k}_c\}$, the Hamiltonian $\mathcal{H}(\mathbf{k}_c)$ can be considered to be 2D ``$\mathcal{PT}$-symmetric'', and hence ELs can also be confined in this plane, separating $C_2\mathcal{T}$ exact and broken phases.

In Fig.~\ref{Fig-orthogonalEC}(b), we consider a system with two such symmetries $R_1,R_2\in\{M\mbox{-}\dagger,C_2\mathcal{T}\}$, whose invariant planes $\Pi_1$ (red), $\Pi_2$ (blue) are perpendicular. 
Imagining a single EL  (red) is fixed by $R_1$ in the plane $\Pi_1$ and cuts through $\Pi_2$  at the  midpoint $\mathbf{K}_0$, $R_2$ symmetry guarantees this oriented EL to be symmetric about $\Pi_2$. In particular, its orientation must be reversed at $\mathbf{K}_0$, say,  the two red half-ELs at different sides of $\Pi_2$ are both directed outward from $\mathbf{K}_0$ (Fig.~\ref{Fig-orthogonalEC}(b)),
since the DN obeys the relation $\mathcal{D}(\Gamma)=-\mathcal{D}(\hat{R}_2\Gamma)$ 
(see Supplementary Information 3 for a more general discussion of the EL orientations under different types of symmetry protection). 
By the doubling theorem of ELs, the source-free requirement at $\mathbf{K}_0$ implies that there should exist at least another EL (blue) on $\Pi_{2}$ with two inflow half-lines connecting the red EL at $\mathbf{K}_0$, 
thereby forming an orthogonal exceptional chain. Accordingly, regions of the exact (broken) phase on the two planes are consistently joined along their intersection. Thus, we have demonstrated that any pair of symmetries belonging to $\{M\mbox{-}\dagger,C_2\mathcal{T}\}$ can protect the existence of orthogonal ECs. 

However, if either of the two symmetries is $M\mbox{-}\dagger$, the formation of the orthogonal EC requires the bands to satisfy additional conditions. To see this, we inspect the relation, imposed by the $M\hbox{-}\dagger$ symmetry, to a pair of right and left eigenstates, $|\psi^R(\mathbf{k}_m)\rangle$, $|\psi^L(\mathbf{k}_m)\rangle$ in the exact phase on a mirror plane $\Pi_M$:
\begin{equation}
    \hat{M}|\psi^R(\mathbf{k}_m)\rangle=\rho(\mathbf{k}_m)|\psi^L(\mathbf{k}_m)\rangle.
\end{equation}
On the condition of binormalization $\langle\psi^L(\mathbf{k}_m)|\psi^R(\mathbf{k}_m)\rangle=1$, the coefficient $\rho(\mathbf{k}_m)\neq0$ is ensured to be real and therefore invests each eigenstate $|\psi(\mathbf{k}_m)\rangle$ in the exact phase with a certain $M\hbox{-}\dagger$-parity, 
\begin{equation}
    \tilde{p}(\mathbf{k}_m)=\mathrm{sign}[\rho(\mathbf{k}_m)]=\pm1,
\end{equation}
generalizing the concept of mirror-parity for the eigenstates in  mirror-symmetric systems.
Intriguingly, it can be proved that (see Supplementary Information 4)
\begin{theorem}
A $M\mbox{-}\dagger$-symmetry protected order-2 exceptional line on $\Pi_M$ can only be formed by two bands with \textbf{opposite $M\mbox{-}\dagger$-parities} in the nearby exact phase.
\end{theorem}
Consequently, we know from the theorem that an orthogonal EC in Fig.~\ref{Fig-orthogonalEC}(b) 
protected by either two $M\mbox{-}\dagger$ or $(M\mbox{-}\dagger,C_2 \mathcal{T})$ symmetries
cannot be formed, unless the $M\hbox{-}\dagger$-parities of the two crossing bands take opposite signs in the exact phases on the corresponding mirror planes.

\subsection{Planar ECs protected by two ${M}\mbox{-}\dagger$ symmetries}
\begin{figure}[b!]
\includegraphics[width=0.49\textwidth]{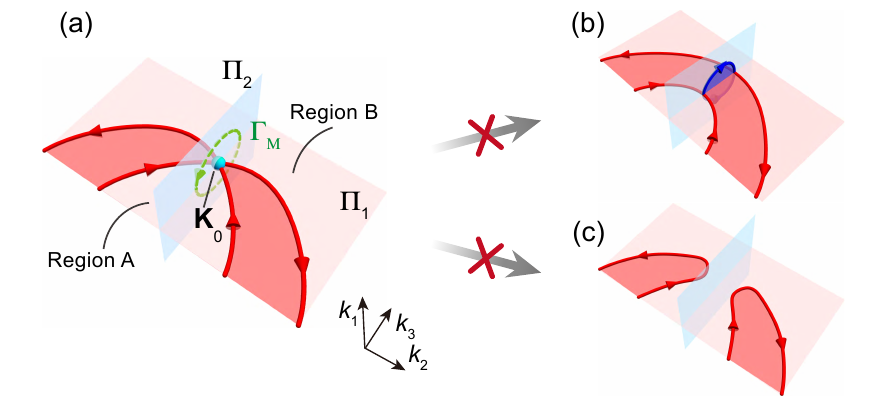}
\caption{\label{Fig-planarEC}(a) Planer exceptional chain protected by two $M_{1,2}\hbox{-}\dagger$ symmetries. $\mathbf{K}_0$ is the non-defective chain point confined on the intersection of two mirror planes $\Pi_1$ and $\Pi_2$. A and B are two regions of exact phase in $\Pi_1$ with opposite $M_1\mbox{-}\dagger$ parities $\tilde{p}^+_A \tilde{p}^+_B <0$.
(b,c) Two plausible evolutions compatible with the source-free principle of ELs, which are, however, forbidden by (b) the same $M_2\mbox{-}\dagger$-parity of two bands and by (c) the quantized Berry phase along $\Gamma_M$, respectively.}
\end{figure}

In the presence of two orthogonal mirror-adjoint symmetries, denoted $
{M}_1\hbox{-}\dagger$ and $
{M}_2\hbox{-}\dagger$, they can stabilize another type of planar EC, as shown in Fig.~\ref{Fig-planarEC}(a), with two crossing ELs confined in the same mirror plane $\Pi_1$ and the chain point is fixed along the intersection of the two mirror planes (the little group along the intersection line is isomorphic to the DAS point group $\blue{m^\dagger m^\dagger}2$), where the two ELs are formed by the coalescence of two bands whose $M\mbox{-}\dagger$ parities are opposite in the nearby exact phase in $\Pi_1$ but are identical in $\Pi_2$.

At first glance, the source-free principle of ELs does not forbid the two possible evolutions shown in Figs.~\ref{Fig-planarEC}(b) and (c). However, the contrapositive of the above theorem indicates that any degeneracy between two bands with the same $M_2\hbox{-}\dagger$-parity in the exact phase should be non-defective on $\Pi_2$, 
hence the crossing point $\mathbf{K}_0$ of the two $M_2\hbox{-}\dagger$-partner ELs on $\Pi_1$ is a non-defective diabolic point banned from expanding to an EP ring in $\Pi_2$ (i.e., the case of Fig.~\ref{Fig-planarEC}(b)).

In addition, 
we consider a vertical $M_1$-symmetric loop $\Gamma_M=\hat{m}_1{\Gamma_M}^{-1}$ enclosing $\mathbf{K}_0$ (the dashed green circle in Fig.~\ref{Fig-planarEC}(a)), where the negative power indicates the direction of the loop is reversed by reflection. $\Gamma_M$ is demanded to cross $\Pi_1$ in two different regions, $A$ and $B$, of exact phases with opposite $M_1\mbox{-}\dagger$-parities 
$\tilde{p}^+_A \tilde{p}^+_B <0$, where the superscript ``$+$'' denotes the band with larger real parts of eigenvalues in each region. 
Despite having a null DN, the loop carries a nontrivial Berry phase quantized by $M_1\hbox{-}\dagger$ symmetry,
(see Mehtods):
\begin{equation}
    \begin{split}
    \hspace{-10pt}&\exp\left[i\,\theta(\Gamma_M )\right]\\
    =&\,\exp\left[i\,\mathrm{Re}[\theta^{LR}(\Gamma_M )]\right] =\exp\left[i\,\mathrm{Re}[\theta^{RL}(\Gamma_M )]\right]\\
    =&\,\exp\left[i\,\frac{1}{2}\left(\theta^{LL}(\Gamma_M )+\theta^{RR}\right)\right]
    =\tilde{p}_A^+ \,\tilde{p}_B^+ =-1,
    \end{split}
\end{equation}
i.e., $\theta(\Gamma_M )=\pi\bmod 2\pi$, where $\theta^{\alpha\beta}(\Gamma_M)=\oint_{\Gamma_M}A^{\alpha\beta}(\mathbf{k})\cdot d\mathbf{k}$ and $A^{\alpha\beta}(\mathbf{k})=-i\langle \psi^\alpha(\mathbf{k})|\nabla_\mathbf{k}|\psi^\beta(\mathbf{k})\rangle$ ($\alpha,\beta\in\{L,R\}$) denote the four different types of Berry phases and Berry connections. And in computing $A^{LL}$ and $A^{RR}$, the left and right eigenstates are required to obey the gauge constraint $\langle\psi^L(\mathbf{k})|\psi^R(\mathbf{k})\rangle\in\mathbb{R}$. Thereupon, the intersection of the ELs cannot be gapped out in the way that Fig.~\ref{Fig-planarEC}(c) shows, justifying the stability of the planar EC.

\subsection{Mirror-symmetric exceptional chains}
The realization of $M\mbox{-}\dagger$ or $C_2\mathcal{T}$ protected ECs relies on the confinement of ELs into symmetry-invariant planes, which is somewhat similar to the method for Hermitian nodal chains~\cite{bzdusek2016Nodalchain,yu2017Nodal,chang2017Topological,gong2018Symmorphic,nodal-chain_Lu_2018_NaturePhys, xiong2020Hidden}. Next, we show that a single mirror symmetry can give rise to EC formation, whose ingenious mechanism  has no Hermitian counterpart. 

\begin{figure}[t!]
\includegraphics[width=0.49\textwidth]{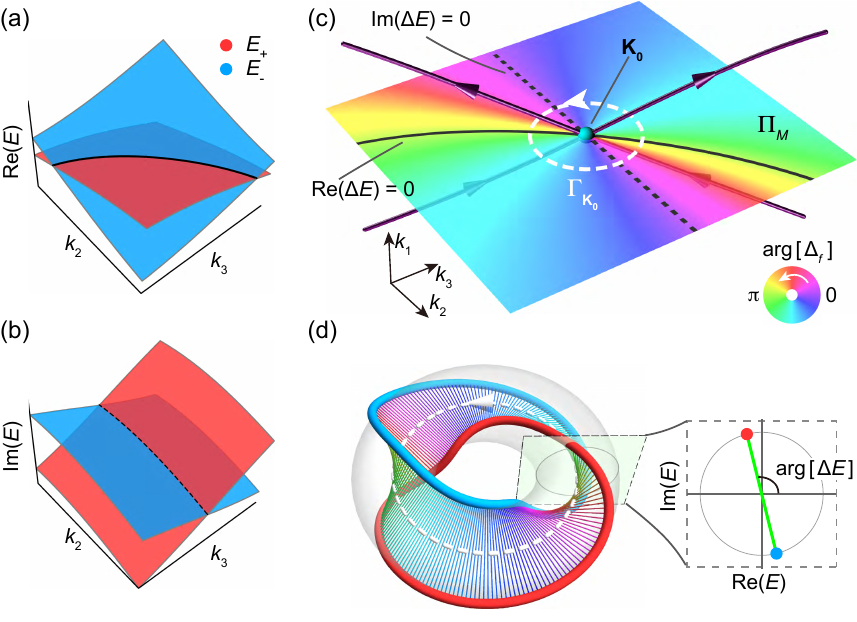}
\caption{\label{Fig-mirrorECs} Exceptional chain protected by mirror symmetry. (a,b) Real and imaginary parts of two bands $E_+$, $E_-$ with opposite mirror parities on the mirror plane $\Pi_M$. (c) Mirror-symmetric EC  intersecting at $\mathbf{K}_0$ on $\Pi_M$. Colormap on $\Pi_M$: the phase vortex of $\Delta_f(\mathbf{k}_m)$ around $\mathbf{K}_0$. (d) Eigenvalue braiding of $E_+$ (red), $E_-$ (light blue) along the loop $\Gamma_{\mathbf{K}_0}$ encircling $\mathbf{K}_0$ forms a Hopf link with two rings of opposite mirror parities. Inset: section at a point $\mathbf{k}_\theta\in\Gamma_{\mathbf{K}_0}$ gives $E_+(\mathbf{k}_\theta),E_-(\mathbf{k}_\theta)$  on the complex plane. Color of the bar depicts $\arg[\Delta_f]$, which is twice as big as the bar's twist angle $\arg[\Delta E]$.}
\end{figure}

For a general mirror-symmetric non-Hermitian system satisfying $\hat{M}\mathcal H\left(\hat{m} \mathbf{k} \right){\hat M^{-1}} =\mathcal{H}\left( \mathbf{k} \right)$, every eigenstate on the mirror plane $\Pi_M$ has a definite (even or odd) $M$-parity.
Considering two bands with opposite $M$-parities on $\Pi_M$, their eigenvalues $E_+(\mathbf{k}_m)$, $E_-(\mathbf{k}_m)$, are well-ordered by the parities (see Fig.~\ref{Fig-mirrorECs}(a,b)), 
hence $\Delta E=E_+-E_-$ is single-valued in the regions where no other band crosses.
The intersection of the two bands on the mirror plane requires both the real and imaginary parts of the two eigenvalues to be equal, 
$\mathrm{Re}[\Delta E(\mathbf{k}_m)]=0$ and $\mathrm{Im}[\Delta E(\mathbf{k}_m)]$=0, which corresponds to two curves on $\Pi_M$ with their crossing point (e.g., the cyan dot $\mathbf{K}_0$) determining the degeneracy of the two bands, as illustrated in Fig.~\ref{Fig-mirrorECs}(c). Consequently, in contrast to the best-known fact that Hermitian band crossings on the mirror plane form nodal lines~\cite{fang2016Topological}, two non-Hermitian bands with opposite $M$-parities always stably intersect at isolated points (cyan dot) inside the mirror plane.
Moreover, since the eigenstates' coalescence is forbidden by their opposite $M$-parities, these isolated nodal points are non-defective, whereas an EP inside the mirror plane can only be formed by two bands with the same $M$-parity.

\begin{figure*}[t!]
\includegraphics[width=0.8\textwidth]{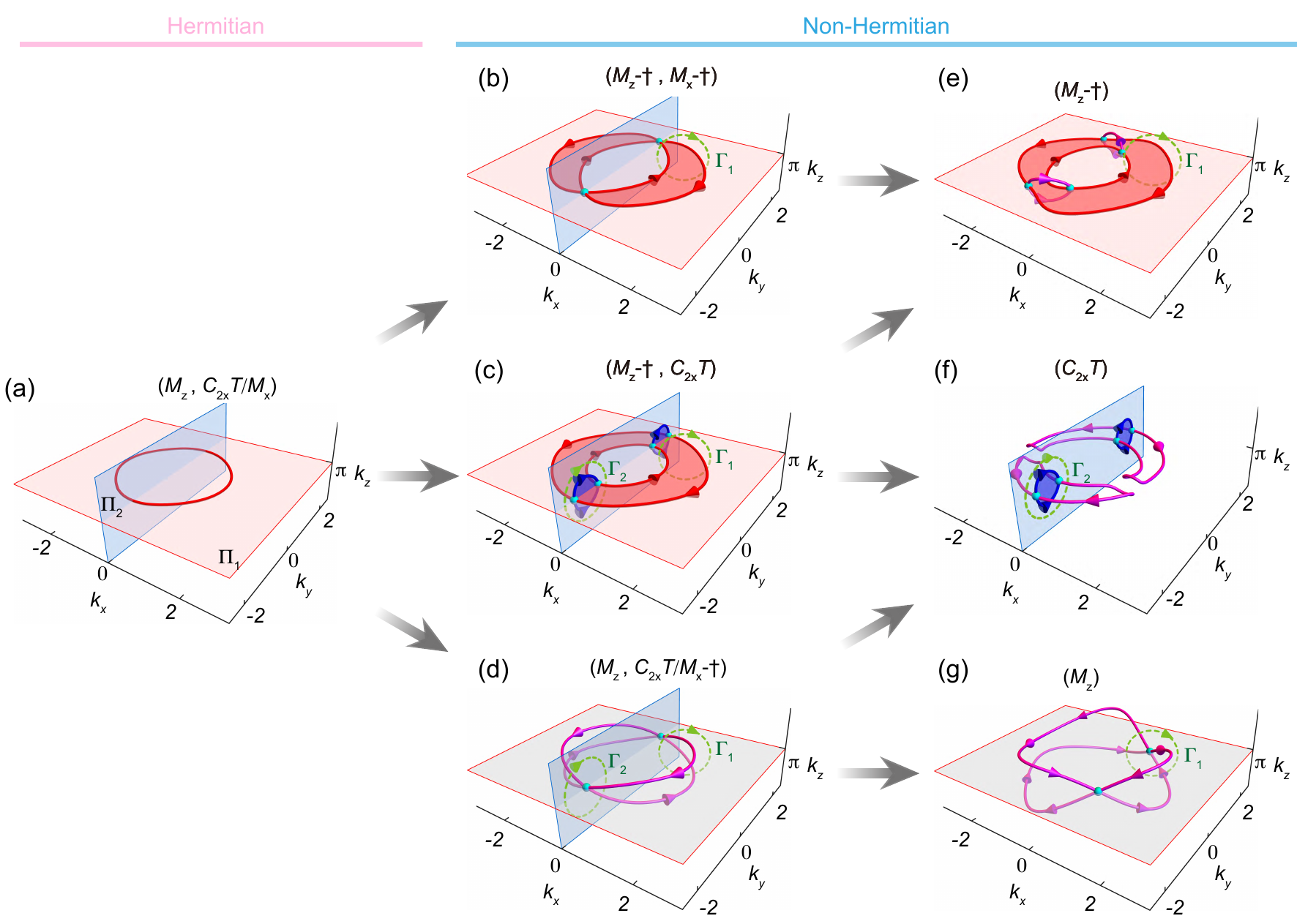}
\caption{\label{local configurations} Different types of ECs evolved from a nodal ring. (a) A Hermitian nodal ring protected by $M_z$ symmetry lays on the mirror plane $\Pi_z$. A second symmetry of $C_{2x} \mathcal{T}$ or $M_x$ presents with the corresponding high-symmetry plane $\Pi_x$. (b-d) ECs induced by non-Hermitian perturbation respecting two different symmetries, where the chain points (cyan dots) are fixed on the intersection line between $\Pi_x$ and $\Pi_z$. (e-g) ECs protect by a single symmetry.
The parenthesis above each configuration gives the required symmetries of that case.
}
\end{figure*}

Indeed, the mirror-symmetry protected nodal point is a phase singularity of $\Delta E(\mathbf{k}_m)$ with $\nu_{+-}(\Gamma_{\mathbf{K}_0})=\frac{1}{2\pi}\oint_\Gamma d\arg[\Delta E]=\pm1$ 
along any loop $\Gamma_{\mathbf{K}_0}$ (white dashed line) on $\Pi_M$ solely encircling the degenerate point $\mathbf{K}_0$. Thus, the DN of $\Gamma_{\mathbf{K}_0}$, which is entirely attributed to $\nu_{+-}(\Gamma_{\mathbf{K}_0})$, is quantized to an even number (see the phase vortex of $\Delta_f(\mathbf{k}_m)$ on $\Pi_M$ in Fig.~\ref{Fig-mirrorECs}(c))
\begin{equation}
    \mathcal{D}(\Gamma_{\mathbf{K}_0})= 2\nu_{+-}(\Gamma_{\mathbf{K}_0})=\pm2,
\end{equation}
indicating the non-defective node $\mathbf{K}_0$ is an intersection of two ELs that pierce the mirror plane from the same side and form a mirror symmetry protected EC. 
Different from the $M$-$\dag$ or $C_2 \mathcal{T}$ cases, the two crossing ELs are distributed antisymmetrically about $\Pi_M$, \textit{i.e.}, they have $M$-symmetric shapes but opposite orientations, as the consequence of $\mathcal{D}(\Gamma)=\mathcal{D}(\hat{m}\Gamma)$ for two mirror-partner loops imposed by the mirror symmetry (see Supplementary Information 3). 

According to the relation Eq.~\eqref{relation between DN and braid}, the DN $\mathcal{D}(\Gamma_{\mathbf{K}_0})=\pm2$ indicates that the two eigenmodes of opposite $M$-parities braid twice along the loop $\Gamma_{\mathbf{K}_0}$. Therefore, when returning to the initial states after a round, their eigenvalue trajectories are interwoven into a mirror-symmetry-protected Hopf link~\cite{hu2021Knots,wang2021Topological}. As shown in Fig.~\ref{Fig-mirrorECs}(d), the colored bars connecting the two trajectories represent the vector $\overrightarrow{\Delta E}(\mathbf{k}_\theta)=(\mathrm{Re}[\Delta E],\mathrm{Im}[\Delta E])$ ($\mathbf{k}_\theta\in\Gamma_{\mathbf{K}_0}$) on the complex energy plane, whose altitude angle gives $\arg[\Delta E]$. Therefore, the eigenvalue Hopf link delineates the twist of the vector $\overrightarrow{\Delta E}$ around the loop. 

\section{Exceptional chains arising from a nodal ring}
In the previous sections, we have introduced three DAS symmetries, $M$-$\dagger$, ${C}_2\mathcal{T}$, and $M$, that can protect ECs. Here, starting from an $M_z$-protected Hermitian nodal ring (Fig.~\ref{local configurations}(a)) in the mirror plane $\Pi_z$ (light red), we will enumerate all possible EC structures that can be deterministically generated by thresholdless non-Hermitian perturbations preserving $M$, $M$-$\dagger$ or ${C}_2\mathcal{T}$ symmetries. 

We first introduce $M_z$-$\dagger$-invariant non-Hermitian perturbations. As shown in Fig.~\ref{local configurations}(b,c), since the system becomes pseudo-Hermitian on $\Pi_z$, 
the nodal line immediately splits into a pair of  ELs confined in $\Pi_z$. Considering an $M_z$-$\dagger$-symmetric loop $\Gamma_1$ around the EL pair, it carries a zero DN, $\mathcal{D}(\Gamma_1)=0$, and also a $\pi$ quantized biorthogonal Berry phase, $\theta(\Gamma_1)=\pi$, both inherited from the Hermitian system, which oblige the two ELs to be oppositely directed 
but exempt from pair annihilation.

If the system under non-Hermitian perturbations also preserves a second symmetry $R_2\in\{C_{2x}\mathcal{T},M_x\text{-}\dagger\}$ with an $R_2$-invariant plane, $\Pi_x$, normally crossing $\Pi_z$, 
the EL pair in $\Pi_z$ should reverse their directions when traversing $\Pi_x$, then two different types of ECs can be deterministically formed.
When $R_2=M_x$-$\dagger$, evolving from the $M_x$ symmetry of the Hermitian case, the absence of nodal lines in $\Pi_x$ in Fig.~\ref{local configurations}(a) indicates the two original Hermitian bands share the same $M_x$-parity.  So an identical $M_x$-$\dagger$-parity is inherited by the non-Hermitian bands, which gives rise to a planar EC with two non-defective chain points (cyan dot) confined on the intersection line of $\Pi_x$ and $\Pi_z$, as shown in Fig.~\ref{local configurations}(b).
If the $M_x$-$\dagger$ symmetry is broken, the ban on EPs appearing in $\Pi_x$ is lifted; then at once, each non-defective chain point spawns a vertical EP ring lying out of high-symmetry planes, as shown in Fig.~\ref{local configurations}(e). Amazingly, the vertical EP rings remain connecting with the planar ELs and the EC survives. This remarkable robustness of EC is because the directions of the $M_z$-$\dagger$-protected ELs cannot change abruptly and the sourceless requirement (Eq.~\eqref{doubling theorem}) at the direction-reversal points of the in-plane ELs compels the emergence of the out-of-plane ELs from those points, uncovering that the $M_z$-$\dagger$ symmetry itself can protect ECs. 

When $R_2=C_{2x}\mathcal{T}$, the non-Hermitian perturbation expands the intersection points between the Hermitian nodal line and the plane $\Pi_x$ into two exceptional rings (blue) fixed in $\Pi_x$ bridging the gaps between the pair of horizontal ELs in $\Pi_z$ (see Fig.~\ref{local configurations}(c)), akin to the formation of an exceptional ring from a Dirac point in 2D $\mathcal{PT}$ symmetric systems~\cite{szameit2011PTsymmetry,zhen2015Spawning,okugawa2019Topologicala}. 
Thus, a double-earring EC can be realized.
In addition, if the $M_z\hbox{-}\dagger$ symmetry is broken while $C_{2x}\mathcal{T}$ persists as shown in Fig.~\ref{local configurations}(f), a $C_{2x}\mathcal{T}$-protected $\pi$-quantized biorthogonal Berry phase, $\theta^{LR}=\oint_{\Gamma_2}A^{LR}(\mathbf{k})\cdot d\mathbf{k}=\pi$, along a loop $\Gamma_2$ in $\Pi_x$ protects the earring exceptional ring against shrinking to disappear (see Methods), and the EC stays alive in the presence of only the $C_{2x}\mathcal{T}$ symmetry. Moreover, since the annihilation between the two horizontal ELs is now permissible in that the biorthogonal Berry phase $\theta(\Gamma_1)$ is no longer quantized, the EC can split into two separate ones shown by Fig.~\ref{local configurations}(f). 

Next, we consider the case in which $M_z$-symmetric non-Hermitian perturbations are introduced. As exhibited in Fig.~\ref{local configurations}(d,g), the nodal line splits into two ELs antisymmetrically distributed at the two sides of $\Pi_z$. 
Provided that the two ELs meet inside the mirror plane, $M_z$ symmetry can protect their stable intersections at non-defective points on $\Pi_z$ and hence guarantees the formation of EC, as illustrated in the above section. By imposing another symmetry $R_2\in\{C_{2x}\mathcal{T},M_x\text{-}\dagger\}$, the chain point can be further fixed on the line intersection of the two planes $\Pi_x$ and $\Pi_z$ by the DAS point group $m_z\blue{m_x^\dagger 2_y^\dagger}$ or $\red{2'_xm'_y}m_z$ (see Supplementary Information 6). 
Breaking $R_2$ while preserving $M_z$, the chain points can move freely in the $M_z$-invariant plane $\Pi_z$, as shown in Fig.~\ref{local configurations}(g), which cannot disappear unless annihilating in pairs of opposite DNs $\mathcal{D}=\pm2$. Conversely, if we break $M_z$ while maintaining $R_2=C_{2x}\mathcal{T}$, the $M_z$-symmetric EC in Fig.~\ref{local configurations}(d) can evolve to the configuration in Fig.~\ref{local configurations}(f).

The concrete Hamiltonians gnerating the ECs in Fig.~\ref{local configurations} are given in Supplementary Information 6, where we also offer a rigorous analysis of the determinate evolution paths towards local ECs from a DAS point group perspective.
Furthermore, if we break the symmetries protecting the chain point in certain manners, the ECs may evolve into more fascinating exceptional links~\cite{carlstrom2018Exceptional,yang2019NonHermitiana,yang2020Jones} (see examples in Supplementary Information 7).

\begin{figure*}[t!]
\includegraphics[width=0.9\textwidth]{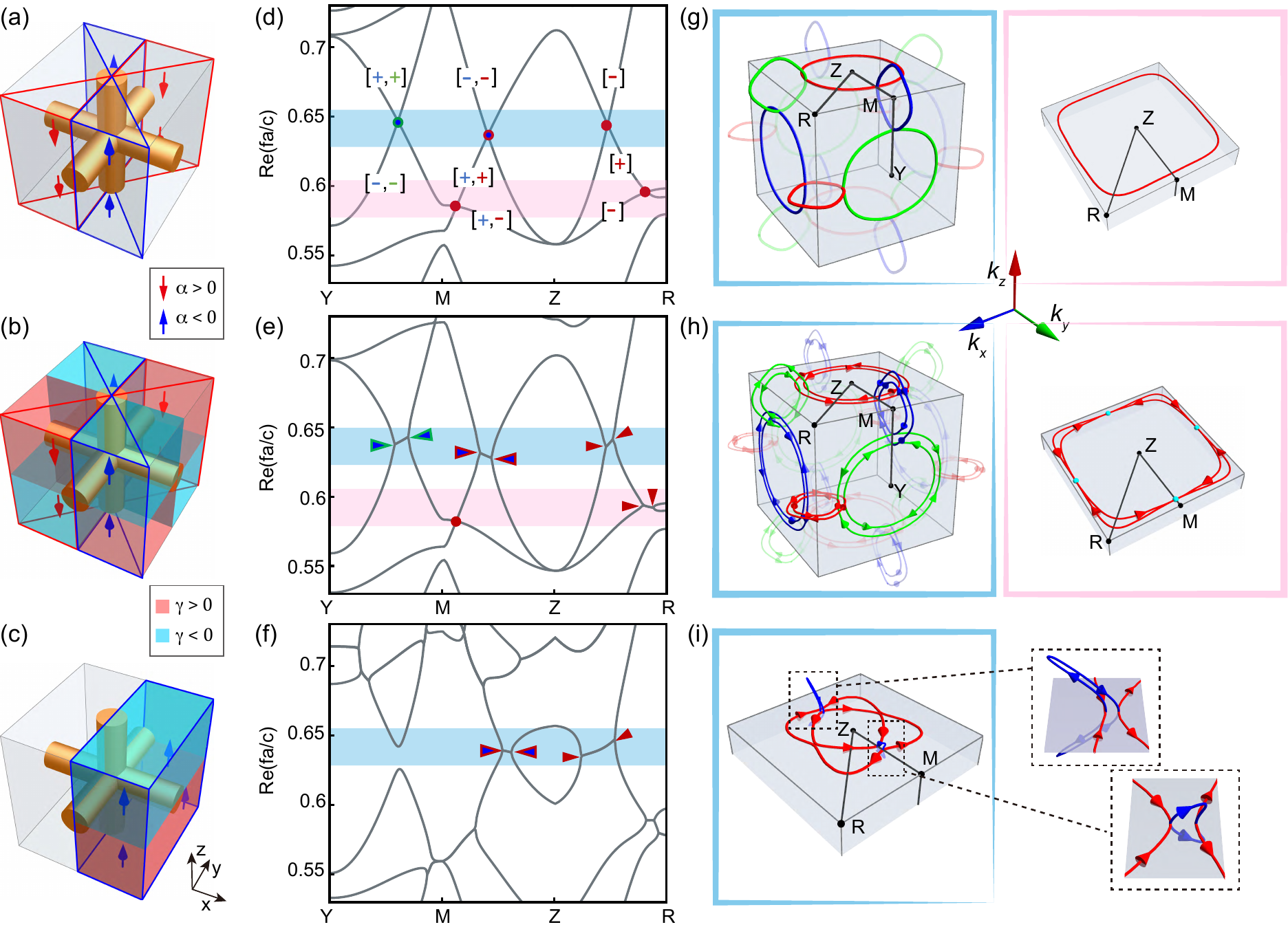}
\caption{\label{PC-realization}
Different types of ECs in non-Hermitian photonic crystals. (a-c) Schematics of three cubic PCs' unit cells with lattice constant $a$. Orange cylinders:  metallic meshes. Regions in gray, red, and cyan colors: conservative ($\gamma=0$), lossy ($\gamma>0$), and gain ($\gamma<0$) materials, respectively.  
Regions with down-red (up-blue) arrows: gyrotropic media with $\alpha<0$ ($\alpha>0$). 
(d-f) Band structures (real part) along high symmetry lines $YMZR$ corresponding to the PCs in (a-c).
(g-i) The nodal structures in the BZ of the PCs in (a-c). 
(a,d,g) The Hermitian PC in (a) respecting $M_{x,y,z}$ symmetries with $\alpha=\pm 0.5$ and $\gamma=0$. Labels $[\pm]$ in blue, green and red in (d): the $M_{x,y,z}$-parities of the bands, respectively. 
Left panel of (g): nodal chain formed around f=0.65c/a (light blue shaded region in (d)); right panel: nodal ring formed around $f=0.6c/a$ (pink shaded region in (d)). 
(b,e,h) The non-Hermitian PC respecting $M_{x,y,z}\hbox{-}\dagger$ symmetries with $\alpha=\pm 0.5$, $\gamma=\pm 0.5$. 
Left panel of (h): a pair
of linked orthogonal EC network formed around f=0.65c/a; right panel (arrows: directions of the ELs): a planar EC with four non-defective chain point (cyan dots) formed around $f=0.6c/a$ (pink shaded region in (d)).
Triangles in blue, green, and red in (e): EPs on the $M_{x,y,z}$-invariant planes, respectively; triangles in two colors: EC points. 
(c,f,i) The non-Hermitian PC respecting $M_z\hbox{-}\dagger$ and $C_{2x}\mathcal{T}$ symmetries with $\alpha= -0.3$, $\gamma=\pm 0.5$. (i) A double-earring EC formed around $f=0.65c/a$ (light blue shaded region in (f)).
}
\end{figure*}


\section{Photonic crystal realization of exceptional chains}
In this section, we will show that various types of ECs protected by different DAS crystalline symmetries can be realized in non-Hermitian photonic crystals. 
It has been experimentally proved that a Hermitian metallic-mesh PC supports nodal chains protected by both $\mathcal{PT}$ symmetry and three mirror symmetries $M_{x,y,z}$~\cite{nodal-chain_Lu_2018_NaturePhys}, and the nodal rings on different mirror planes can be expanded into exceptional torus if the $\mathcal{PT}$-symmetric non-Hermiticity is introduced~\cite{zhou2019Exceptional}.
Here, we first modify the metallic mesh PC so as to design a non-Hermitian PC with three mirror-adjoint symmetries. 
One thing to note is that a reciprocal PC with the three $M_{x,y,z}\hbox{-}\dagger$ symmetries must be $\mathcal{PT}$ symmetric, inasmuch as  ``reciprocity'' is equivalent to the time-reversal-adjoint symmetry, $\mathcal{T}\hbox{-}\dagger$,~\cite{kawabata2019Symmetry} and $(M_x\hbox{-}\dagger)(M_y\hbox{-}\dagger)(M_z\hbox{-}\dagger) (\mathcal{T}\hbox{-}\dagger)=\mathcal{PT}$. Therefore, to realize ELs on the mirror planes in all three directions instead of EP surfaces, we need to break the $\mathcal{PT}$ symmetry and accordingly the reciprocity of the PC, which can be achieved by properly arranging the non-Hermitian gyrotropic materials with the relative permittivity and permeability surrounding the metallic mesh:
\begin{equation}
    {\varepsilon _r}\left( \mathbf{r} \right) = \left( {\begin{array}{*{20}{c}}
{1 + i\gamma(\mathbf{r}) }&{i\alpha(\mathbf{r}) }&0\\
{ - i\alpha(\mathbf{r}) }&{1 + i\gamma(\mathbf{r}) }&0\\
0&0&{1 + i\gamma(\mathbf{r}) }
\end{array}} \right),
\ \ {\mu _r} = 1.
\end{equation}

Figure~\ref{PC-realization}(a) displays the non-reciprocal Hermitian metallic mesh PC with $\gamma=0$, where the prism regions are magnetized with the red (blue) arrows denoting the gyration vectors corresponding to $\alpha=0.5\,(-0.5)$. The Hermitian space group of the PC is $Pmmm\blue{1^\dagger}$ (No. 1698~\cite{vanleeuwen2014Double})
including three mirror symmetries, $M_{x,yz}$,  but no $\mathcal{PT}$ symmetry. Figure~\ref{PC-realization}(d) shows the bulk band structure of the Hermitian PC along high symmetry lines (see Fig.~\ref{PC-realization}(g)), which is numerically calculated by COMSOL. 
Since 
the two bands near the frequency of $f=0.65c/a$ have opposite mirror parities (labels on the band structure) along every intersection line of two orthogonal mirror planes, as shown in Fig.~\ref{PC-realization}(g), the nodal rings in different mirror planes should connect together~\cite{chang2017Topological,gong2018Symmorphic,wu2019NonAbelian} and form a globally connected nodal chain network, where the nodal rings are retrieved from the 2D band structure on the mirror planes (see Supplementary Information 10). For the lower two bands at about $f=0.6c/a$ with opposite $M_z$ parities but identical $M_{x,y}$ parities, the nodal ring only appears on the $k_z=\pi/a$ plane, as shown by the right panel of Fig.~\ref{PC-realization}(g).   

Introducing gain (loss) with $\gamma=0.5\,(-0.5)$ into the background materials in red (cyan) color in Fig.~\ref{PC-realization}(b), the prototype Hermitian PC becomes non-Hermitian and the three mirror symmetries $M_{x,y,z}$ convert to the corresponding mirror-adjoint symmetries, $M_{x,y,z}$-$\dagger$, so the DAS space group of PC becomes $P\blue{m^\dagger m^\dagger m^\dagger}$ (No. 3558~\cite{vanleeuwen2014Double}). The non-Hermitian perturbed band structure is plotted in Fig.~\ref{PC-realization}(e). As such, the nodal rings in different mirror planes are split into paired exceptional rings with opposite orientations.  
For the upper pair of bands near $f=0.65c/a$, the two EP rings on each plane can be marked as ``inner'' and ``outer'' according to their relative positions. In particular, every inner (outer) EP ring on a mirror plane is chained with another two outer (inner) rings in the two perpendicular planes, resulting in a pair of triply orthogonal EC networks linked with each other (see the toy tight-binding model generating the same EC structure in Supplementary Information 8), as depicted by the left panel in Fig.~\ref{PC-realization}(h). Moreover, the numerical tests also verify that the numbers of inflow and outflow ELs at each chain point are balanced (see Supplementary Information 10), corroborating the source-free principle of ELs.
The multi-band nature of the PC also demonstrates that our theory of EC formation is universal for generic full-wave crystalline systems. 

For the lower pair of bands around $f=0.6c/a$, thanks to their identical $M_{x}\hbox{-}\dagger$ ($M_{y}\hbox{-}\dagger$) parity on the mirror plane of $k_x=0$ ($k_y=0$) descended from the Hermitian mirror parity, the Hermitian degenerate points of the two bands along $k_x=0$ (M-Z direction in Fig.~\ref{PC-realization}(d)) stay intact after introducing the non-Hermiticity (Fig.\ref{PC-realization}(e)), and so do the Hermitian degenerate points along $k_y=0$. Consequently, as seen from the right panel of Fig.~\ref{PC-realization}(h), the pair of exceptional rings splitting from the Hermitian nodal ring intersect at four $M_{x,y}\hbox{-}\dagger$-protected non-defective points and therefore form a planar EC in the $k_z=\pi/a$ plane.   

In the last row of Fig.~\ref{PC-realization}, we designed another metallic mesh PC with a different distribution of the gyrotropic and nonconservative materials (see the unit cell in Fig.~\ref{PC-realization}(c)), such that its  space group is $P\blue{m^\dagger}\green{m'^\dagger}\red{2'}$ (No. 8967) possessing $M_z\hbox{-}\dagger$ and $C_{2x}\mathcal{T}$ symmetries. Because of the lack of $M_{x,y}$ symmetries in the prototype Hermitian PC ($\gamma =0$), only the $M_z$-invariant planes can support Hermitian nodal rings in this case. After bringing gain and loss into the PC, the nodal ring in the $k_z=\pi/a$ plane divides into two planar exceptional rings joining together by two vertical EP loops in the $k_x=0$ plane protected by the $C_{2x}\mathcal{T}$ symmetry, as can be seen from Fig.~\ref{PC-realization}(i). As a result, a double-earring EC arises in the non-Hermitian PC, reproducing the configuration in Fig.~\ref{local configurations}(c). 


\section{Discussion}
In conclusion, we showed that the source-free principle of directed ELs and the symmetry constraints are two crucial conditions for the formation of ECs.
Based on this idea, we developed a general theory for constructing ECs of various local morphologies, including orthogonal EC, planar EC, and mirror-symmetric EC, protected by $C_2\mathcal{T}$, mirror-adjoint,  or mirror symmetries. We also investigated all possible EC configurations evolving from a Hermitian nodal ring with thresholdless non-Hermitian perturbations. Furthermore, via designing non-Hermitian PCs respecting certain symmetries, we realized three representative and exotic ECs in the PCs, \textit{i.e.}, a pair of linked orthogonal EC networks, a planar EC with non-defective chain points, and a double-earring EC. It is worth emphasizing that the source-free principle of ELs and symmetry constraints not only account for the formation of two-band ECs but are also valid to the multiband cases, such as the  exceptional nexus of the ELs formed by the intersections of triple bands~\cite{tang2020Exceptional} (see Supplementary Information 9).

Our theory integrates the non-Hermitian topology with the spatiotemporal crystalline symmetries and opens the avenue for symmetry-protected non-Hermitian topological phases and topological degeneracies, such as symmetry-protected higher-order EPs~\cite{delplace2021SymmetryProtected,mandal2021Symmetry,sayyad2022Realizinga}, in non-Hermitian crystals. In light of the analogy between non-Hermitian physics and other systems or theories~\cite{lee2019Topological,fruchart2021Nonreciprocal,peng2022Topological}, our framework may also be transplanted to these systems, giving birth to new physical effects. For example, it was recently discovered that the circular polarization singularity (C point) of 3D optical polarization fields can map to a non-Hermitian EP~\cite{peng2022Topological}. Therefore, the present method would also be applied to realize symmetry-protected chains of C points for 3D optical fields in the real space.

\begin{table*}
\centering
\setlength\cellspacetoplimit{2pt}
\setlength\cellspacebottomlimit{2pt}
\addtolength\tabcolsep{0pt}
\caption{Classification of double-antisymmetry space (point) groups for non-Hermitian crystals. The serial number (\#) of each category listed in the table follows Ref.~\cite{vanleeuwen2014Double}. \label{DAS group table}}
\begin{tabular}{Sc|Sc|Sc|Sc}\hline\hline
  & \textbf{Gray space groups} &  \textbf{Single-antisymmetry groups} & \textbf{Multicolor groups} \\
 \hline
 \multirow{4}{*}[-18pt]{\makecell{Colorless\\ crystallographic\\ space groups\\ $\mathbf{Q}$ \\[5pt] {\scriptsize Category (1)} }}  
   & \makecell{\red{TRS} gray groups $\mathbf{Q}\red{1'}$\\{\scriptsize Category (2)} }  & \makecell{Magnetic groups $\red{\mathbf{Q}'}$\\{\scriptsize Category (3)} }   &  \makecell{\red{TRS} adjoint groups $\blue{\mathbf{Q}^\dagger}\red{1'}=\green{\mathbf{Q}'^\dagger}\red{1'}$\\{\scriptsize Category (9)} } \\\cline{2-4}
    &  \makecell{\blue{Hermitian} gray groups $\mathbf{Q}\blue{1^\dagger}$\\{\scriptsize Category (4)} }  & \makecell{Adjoint groups $\blue{\mathbf{Q}^\dagger}$\\{\scriptsize Category (7)} }   & \makecell{\blue{Hermitian} magnetic groups $\red{\mathbf{Q}'}\blue{1^\dagger}=\green{\mathbf{Q}'^\dagger}\blue{1^\dagger}$\\{\scriptsize Category (6)} }  \\\cline{2-4}
    &  \makecell{\green{Reciprocal} gray groups $\mathbf{Q}\green{1'^\dagger}$\\{\scriptsize Category (8)} } &  \makecell{Magnetic-adjoint groups $\green{\mathbf{Q}'^\dagger}$\\{\scriptsize Category (11)} }  & \makecell{\green{Reciprocal} magnetic groups $\red{\mathbf{Q}'}\green{1'^\dagger}=\blue{\mathbf{Q}^\dagger}\green{1'^\dagger}$\\{\scriptsize Category (10)} }  \\\cline{2-4}
    &  \makecell{\blue{Hermitian}---\red{TRS} groups $\mathbf{Q}\red{1'}\blue{1^\dagger}$\\{\scriptsize Category (5)} }   &    & \makecell{Indecomposable DAS groups $\mathbf{Q}^{\red{(\prime)\blue{(\dagger)}}}$ \\{\scriptsize Category (12)} }  \\
\hline\hline
\end{tabular}
\end{table*}

\section*{Methods}
\subsection{Non-Hermitian crystalline symmetries and DAS space groups.}
An antisymmetry $A$ with respect to a group $\mathcal{G}$ is defined as an operator satisfying three conditions~\cite{padmanabhan2020Antisymmetry}:
\begin{enumerate}
\setlength{\itemindent}{-1.5em}
\setlength{\itemsep}{0pt}
\vspace{-1mm}
    \item $A$ itself is not an element of $\mathcal{G}$: $A\not\in\mathcal{G}$;
    \item Involutivity (self-inverseness): $A^2=I$;
    \item $A$ commutes with all elements of $\mathcal{G}$: $[A,G]=0$, $\forall G\in\mathcal{G}$.
\end{enumerate}\vspace{-1mm}
For example, the time reversal operator $\mathcal{T}$ ($\red{1'}$ in Hermann--Mauguin (HM) notation) is an antisymmetry for spinless space (point) groups, hence involving $\mathcal{T}$ into the crystallographic space (point) groups gives rise to single-antisymmetry space (point) groups, i.e., the widely acknowledged magnetic space (point) groups.

For non-Hermitian systems, the Hermitian-adjoint operation has been shown to be a new symmetry dimension that can greatly enrich the non-Hermitian topological phases~\cite{kawabata2019Symmetry}. In contrast to the spatiotemporal transformations which are linear (antilinear) operators acting on the states in Hilbert space, the Hermitian-adjoint ($\blue{1^\dagger}$ in HM notation) is an antiautomorphic map on the set of bounded linear operators (mathematically, a noncommutative ring) on the Hilbert space. Nevertheless, the spatiotemperal transformations, say $G$, may also be regarded as maps between linear operators via unitary transformation: $G(\mathcal{H})=\hat{G}\mathcal{H}\hat{G}^{-1}=\hat{G}\mathcal{H}\hat{G}^\dagger$. In this sense, \textbf{the Hermitian-adjoint map indeed manifests as an antisymmetry to any group $\mathcal{G}$ of unitary transformations} since it is self-inverse $\blue{1^\dagger 1^\dagger}(\mathcal{H})=(\mathcal{H}^\dagger)^\dagger=\mathcal{H}$ and commutative with any unitary transformation $\blue{1^\dagger}G(\mathcal{H})=(\hat{G}\mathcal{H}\hat{G}^\dagger)^\dagger=\hat{G}\mathcal{H}^\dagger\hat{G}^\dagger=G\blue{1^\dagger}(\mathcal{H})$, $\forall G\in\mathcal{G}$.

Consequently, by taking both time-reversal $\red{1'}$ and Hermitian-adjoint $\blue{1^\dagger}$ into account, we find that the expanded space (point) group for a non-Hermitian crystal is generally isomorphic to a double-antisymmetry space (point) group~\cite{vanleeuwen2014Double}. 
In particular, the product of time-reversal and Hermition-adjoint operations (denoted $\green{1'^\dagger}$ in HM notation) is also an antisymmetry to space groups. Physically, it just represents the reciprocity transformation of the system  $\mathcal{T}\mbox{-}{\dagger}$~\cite{kawabata2019Symmetry}. These three antisymmetry operations together with identity transformation constitute a group isomorphic to the dihedral group $D_2$ with the multiplication table:
\begin{equation}
\addtolength\tabcolsep{5pt}
\begin{tabular}{|Sc|ScScScSc|}
\hline
 & $1$ &  $\red{1'}$ &  $\blue{1^\dagger}$ &  $\green{1'^\dagger}$ \\\hline
 1 & 1 & $\red{1'}$ & $\blue{1^\dagger}$ & $\green{1'^\dagger}$ \\
  $\red{1'}$ & $\red{1'}$ & 1  & $\green{1'^\dagger}$ & $\blue{1^\dagger}$ \\
  $\blue{1^\dagger}$ & $\blue{1^\dagger}$ & $\green{1'^\dagger}$ & 1 & $\red{1'}$ \\
  $\green{1'^\dagger}$ & $\green{1'^\dagger}$ & $\blue{1^\dagger}$ & $\red{1'}$ & 1 \\
\hline
\end{tabular}
\end{equation}

In the DAS space (point) groups for non-Hermitian crystals, the symmetries are sorted into four types: 
\begin{enumerate}
\setlength{\itemindent}{-0.5em}
\setlength{\itemsep}{0pt}
\vspace{-1mm}
    \item[(i)]  The pure spatial symmetries, deonoted $G$ or $g$:\\[3pt] 
\hspace*{40pt} $\hat{G}\mathcal{H}(\hat{g}^{-1}\mathbf{x})\hat{G}^{-1}=\mathcal{H}(\mathbf{x})$;\hspace*{\fill}
    \item[(ii)]  The spatial-adjoint symmetries, denoted $G\mbox{-}\dagger$ or $\blue{g^\dagger}$:\\[3pt] \hspace*{40pt}$\hat{G}\mathcal{H}(\hat{g}^{-1}\mathbf{x})^\dagger\hat{G}^{-1}=\mathcal{H}(\mathbf{x})$;
     \item[(iii)]  The spatiotemporal symmetries,  denoted $G\mathcal{T}$ or $\red{g'}$:\\[3pt] \hspace*{40pt}$\hat{G}\hat{T}\mathcal{H}(\hat{g}^{-1}\mathbf{x})^*(\hat{G}\hat{T})^{-1}=\mathcal{H}(\mathbf{x})$;
     \item[(iv)]  The spatiotemporal-adjoint  symmetries, denoted $G\mathcal{T}\mbox{-}\dagger$ or $\green{g'^\dagger}$:\\[3pt]
     \hspace*{40pt}$\hat{G}\hat{T}\mathcal{H}(\hat{g}^{-1}\mathbf{x})^\intercal(\hat{G}\hat{T})^{-1}=\mathcal{H}(\mathbf{x})$.
\end{enumerate}
Here, $\hat{g}=\{R\,|\,\mathbf{w}\}$ denotes the spatial transformation, with a rotation part $R$ and a translation part $\mathbf{w}$, acting on the position vector $\mathbf{x}$, and $\hat{G}$, $\hat{T}$ denote the unitary parts associated with the $G$, $\mathcal{T}$ operators, respectively, acting on the internal degrees of freedom. The last three types of symmetries can be viewed as coloring the original space-group symmetries into red, blue, and green, respectively. And the DAS space (point) groups are obtained by coloring the elements of the original space (point) groups. There are 624 DAS point groups and 17803 DAS space groups in total~\cite{gopalan2011Rotationreversal,padmanabhan2020Antisymmetry,vanleeuwen2014Double}.

Based on these four types of symmetries, the DAS space (point) groups can be classified into 12 categories~\cite{vanleeuwen2014Double}. In Table~\ref{DAS group table}, we list the 12 categories with vesting each categories  with a physics meaningful designation. The first category (1) includes all the original colorblind space groups $\mathbf{Q}$ without any ``colored'' symmetries. The second column of the table shows four categories of gray space groups (namely, all elements in a group are uniformly colored), each of which possesses at least one antisymmetry, describing the time-reversal symmetric (TRS) systems, Hermitian systems, reciprocal systems, and Hermitian TRS systems, respectively. 
The third column shows the three categories of single-antisymmetry groups. For example, the groups in category (3) are just the black-white magnetic groups which only contain symmetries of types (i) and (iii) but  do not have TRS. Similarly, in category (4), an adjoint group describes the non-Hermitian systems only possessing symmetries of types (i) and (ii). The space group of the PC shown in Fig.~\ref{PC-realization}(b), $P\blue{m^\dagger m^\dagger m^\dagger}$, just belongs to this category. The categories in the fourth column describe the groups containing at least three types of symmetries. In categories (6), (9), and (10), each group also contains an antisymmetry, contrary to the indecomposable groups in category (12) that violate all the three antisymmeries. For example, the space group of the PC shown in Fig.~\ref{PC-realization}(c), $P\blue{m^\dagger}\green{ m'^\dagger} \red{2'}$, belongs to the category (12).

\subsection{Quantized Berry phases protected by DAS point-group symmetries}

\begin{table*}
\begin{adjustwidth}{-.5in}{-.5in}  
\centering
\setlength\cellspacetoplimit{3pt}
\setlength\cellspacebottomlimit{3pt}
\addtolength\tabcolsep{2pt}
\caption{Quantized Berry phases of a \textbf{self-closed} band protected by different types of twofold DAS point-group symmetries $G$. The corresponding $G$-invariant subspace is $\Pi_G=\qty{\mathbf{k}_G\,|\,\hat{G}\mathbf{k}_G=\mathbf{k}_G}$. $p(\phi)\in\qty{\pm1}$ and $\tilde{p}(\phi)\in\qty{\pm1}$ denote the $G$-parities and \gdagger-parities, respectively, of the eigenstates at the $G$-invariant points $\mathbf{k}(\phi)$  ($\phi\in\qty{0,\pi}$). 
\label{berry phase table}}
\begin{tabular}{Sc|Sc|Sc|Sc|Sc}\hline\hline
 \textbf{Symmetry type} & \textbf{Loop} & \textbf{Other conditions} & \textbf{Quantized Berry phases }($0\ \text{or}\ \pi\ \bmod 2\pi$) & \textbf{Examples}\\
 \hline
 $G$  & $\Gamma=\hat{g}\Gamma^{-1}$  & none & $\theta^{LR}=\theta^{RL}=\theta^{LL}=\theta^{RR}=\arg\qty(p(0)p(\pi))$ & $\mathcal{P}$, $M$ \\\hline
 $G\mbox{-}\dagger$  & $\Gamma= \hat{g}\Gamma^{-1}$  & \makecell{two intersections of $\Gamma$ and \\ $\Pi_G$ are in the exact phase} &  $\mathrm{Re}\qty[\theta^{LR}]=\mathrm{Re}\qty[\theta^{RL}]=\frac{1}{2}\qty(\theta^{LL}+\theta^{RR})=\arg\qty(\tilde{p}(0)\tilde{p}(\pi))$ & \mdagger \\\hline
  $G\mathcal{T}$  & $\Gamma=-\hat{g}\Gamma$  & $\Gamma$ is in the exact phase & $\mathrm{Re}\qty[\theta^{LR}]=\mathrm{Re}\qty[\theta^{RL}]=\theta^{LL}=\theta^{RR}$ & $\mathcal{PT}$, $C_2\mathcal{T}$ \\\hline
  $G\mathcal{T}\mbox{-}\dagger$  & $\Gamma=-\hat{g}\Gamma$  & none & $\theta^{LR}=\theta^{RL}=\frac{1}{2}\qty(\theta^{LL}+\theta^{RR})$  & $\mathcal{PT}\mbox{-}\dagger$ \\
\hline\hline
\end{tabular}
\end{adjustwidth}
\end{table*}

The quantization conditions for Berry phases protected by purely spatial and spatiotemporal point-group symmetries, such as $\mathcal{PT}$ and mirror symmetries, have been widely investigated in Hermitian systems. However, when extending our perspective to non-Hermitian crystalline systems, we urgently need to test whether the results established in the Hermitian case still hold and whether the new types of symmetries that are intrinsically non-Hermitian can also preserve the quantized Berry phases.

Moreover, since the left and right eigenvectors of the same eigenstate become different in non-Hermitian systems, more variants of Berry phases can be introduced, making the situation more complicated. Apart from the global biorthogonal Berry phase $\Theta$ in Eq.~\eqref{Eq-globalBP}, we can define  \textbf{four different types of Berry phases, $\theta^{LL}(\Gamma)$, $\theta^{RR}(\Gamma)$, $\theta^{LR}(\Gamma)$, $\theta^{RL}(\Gamma)$, of a single continuous band along a loop $\Gamma=\qty{\mathbf{k}(\phi)|\,\mathbf{k}(\pi)=\mathbf{k}(-\pi),-\pi\leq\phi\leq\pi}$} through integrating the corresponding Berry connections , $A^{LL}$, $A^{RR}$, $A^{LR}$, $A^{RL}$~\cite{shen2018Topological}:
\begin{gather}
    \theta^{\alpha\beta}(\Gamma)=\oint_\Gamma d\mathbf{k}\cdot A^{\alpha\beta}(\mathbf{k}),\\A^{\alpha\beta}(\mathbf{k})={-i\bra{ {\psi}_n^\alpha(\mathbf{k})}\nabla_\mathbf{k}\ket{{\psi}_n^\beta(\mathbf{k})}},\quad (\alpha,\beta\in\qty{L,R}),\label{berry connection}
\end{gather}
where the eigenvectors $\ket{{\psi}_n^{\alpha/\beta}(\mathbf{k})}$ 
should be normalized by  
$\braket{{\psi}^\alpha_n(\mathbf{k})}{{\psi}^\beta_n(\mathbf{k})}=1$,
which is dependent on the different $\alpha,\beta$. For the biorthogonal Berry connections ($\alpha\neq\beta$), $A^{LR}$, $A^{RL}$, $\ket{\psi_n^{\alpha/\beta}(\mathbf{k})}$ respects the binormalization condition $\bra{\psi_n^L(\mathbf{k})}\ket{\psi_n^R(\mathbf{k})}=1$. Whereas for left or right Berry connections ($\alpha=\beta\in\qty{L,R}$),  $A^{LL}$, $A^{RR}$, $\ket{\psi_n^{\alpha}(\mathbf{k})}$ is just self-normalized: $\bra{\psi_n^\alpha(\mathbf{k})}\ket{\psi_n^\alpha(\mathbf{k})}=1$.

The eigenstates along the loop $\Gamma$, abbreviated as $|\psi^{R/L}_n(\phi)\rangle :=|\psi^{R/L}_n(\mathbf{k}(\phi))\rangle$, form a continuous band from $\phi=-\pi$ to $\pi$. To guarantee the Berry phases are well-defined, namely, the results are identical in the sense of modulo $2\pi$ for any continuous gauge of the eigenvectors along $\Gamma$, we require that the continuous band concerned is \textbf{self-closed}, namely the eigenstate returns to the initial one, i.e., $\ket{\psi_n^\alpha(\pi)}=\ket{\psi_n^\alpha(-\pi)}$, after travelling on the concerned band along the loop one turn ($\phi$ evolves from $-\pi$ to $\pi$), implying the DN along the loop to be an even number: $\mathcal{D}(\Gamma)\ \in\  2\mathbb{Z}$.

In addition, according to the relation ${A^{\alpha\beta}}^*={A^{\alpha\beta}}^\dagger=i\bra{\nabla_\mathbf{k}\psi^\beta_n}\ket{\psi^\alpha_n}=-i\bra{\psi^\beta_n}\ket{\nabla_\mathbf{k}\psi^\alpha_n}=A^{\beta\alpha}$, 
the two biorthogonal Berry connections, as well as the two biorthogonal Berry phases, are always complex-conjugate
\begin{equation}
    A^{LR}(\mathbf{k})=A^{RL}(\mathbf{k})^*,\qquad \theta^{LR}(\Gamma)=\theta^{RL}(\Gamma)^*,
\end{equation}  
while the ordinary Berry connections and Berry phases for left (right) eigenvectors always take real values
\begin{equation}
    A^{LL}(\mathbf{k}),\  A^{RR}(\mathbf{k})\in\mathbb{R}^3,\qquad \theta^{LL}(\Gamma),\ \theta^{RR}(\Gamma)\in \mathbb{R}.
\end{equation}

In Supplementary Information 5, we proved that the four types of two-fold DAS point-group symmetries could protect the quantization of different kinds of Berry phases along symmetry-invariant loops, as summarized in Table~\ref{berry phase table}. Here, ``two-fold'' symmetry means that $G^2=I$. In computing $\frac{1}{2}(\theta^{LL}+\theta^{RR})$, the left and right eigenstates are required to satisfy the gauge constraint. In the Supplementary Information, we also showed that the self-closeness of the band can be guaranteed by some rather lenient Properties of the bands in each case.

\begin{acknowledgements}
We thank Profs. Guancong Ma, Biao Yang, Kun Ding, and Shubo Wang for the fruitful discussions.
This work is supported by the Research Grants Council of Hong Kong (Grant No. 16303119 and 16307420) and by the Croucher Foundation (Grant No. CAS20SC01). Wen-Jie Chen acknowledges the supports from National Natural Science Foundation of China (Grant No. 11874435), Guangzhou Science, Technology and Innovation Commission (Grant No. 201904010223).
\end{acknowledgements}

\bibliography{reference.bib}

\end{document}



\title{Supplementary information:\\[5pt] {Topological exceptional chains}}



\maketitle

\tableofcontents
\newpage
\section{Relations between eigenvalue braiding, discriminant number and global Berry phase}

\subsection{Relation between eigenvalue braiding and discriminant number}
Here, we prove the relation between the Abelian topological invariant, discriminant number $\mathcal{D}(\Gamma)\in\mathbb{Z}$~\cite{yang2021fermion}, and the non-Abelian topological invariant along the loop with separable bands, the braid of the complex eigenvalues of $N$ bands: $b(\Gamma)=b_{i_1}^{\,n_1}b_{i_2}^{\,n_2}b_{i_3}^{\,n_3}\,\cdots\, b_{i_L}^{\,n_L}\in B_N$  ($L$ can be an arbitrary positive integer,  $b_{i_m}\in\qty{b_1,b_2,\cdots,b_{N-1}}$ ($m\in\qty{1,\cdots,L}$), and $n_m\in\mathbb{Z}$)~\cite{wojcik2020Homotopy,li2021Homotopical}:

\begin{theorem}\label{theorem-algebraic length}
For a $N$-level Hamiltonian $\mathcal{H}(\mathbf{k})$ with separable bands along a loop $\Gamma$, the discriminant number, $\mathcal{D}(\Gamma)\in\mathbb{Z}$ gives the algebraic length (i.e., the sum of the exponents on the braid generators) of the eigenvalue braid $b(\Gamma)=b_{i_1}^{\,n_1}b_{i_2}^{\,n_2}b_{i_3}^{\,n_3}\,\cdots\, b_{i_L}^{\,n_L}\in B_N$:
  \begin{equation}\label{algebraic length}
      \mathcal{D}(\Gamma)=\sum_{l=1}^{L}n_l.
  \end{equation}
Therefore, the homomorphism $F:B_N\ni b(\Gamma)\rightarrow \mathcal{D}(\Gamma)\in\mathbb{Z}$ gives an Abelianization of the braid group.
\end{theorem}
Physically, the theorem manifests that \textbf{the discriminant number $\mathcal{D}(\Gamma)$ along a loop $\Gamma$ counts the net number of directed ELs encircled by $\mathcal{D}(\Gamma)$ and also equals the net number of times that eigenmode braiding occurs (taking count of the sign of the braiding) along the loop}.
\vspace{3pt}

\begin{figure*}[b!]
\includegraphics[width=0.7\textwidth]{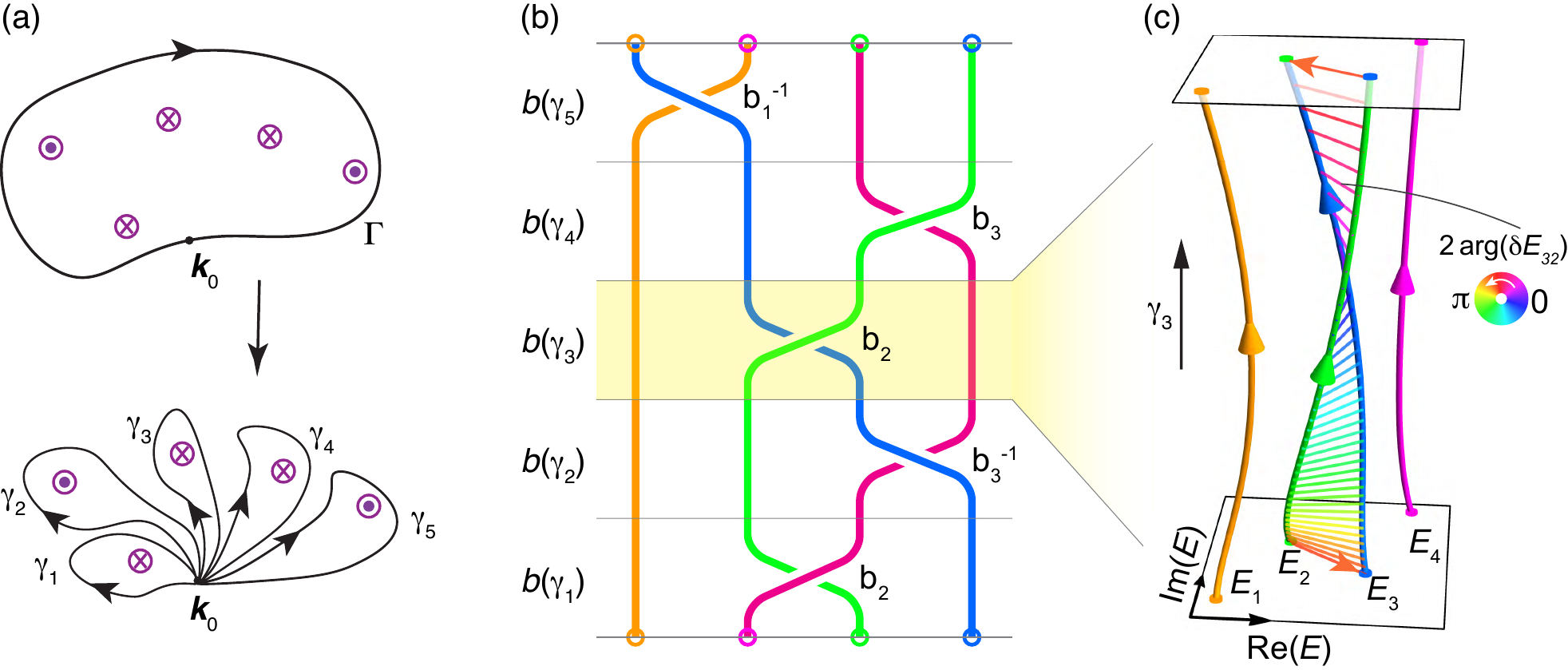}
\caption{\label{fig-relation-proofmehtod1} 
Schematic for the proof \textbf{Method 1} of the \textbf{Theorem}~\ref{theorem-algebraic length}. (a) A Hamiltonian with $N=4$ bands has $M=5$ ELs flowing in or out of the screen in a loop $\Gamma$ that can be continuously transformed into sub-loops: $\gamma_1\circ\gamma_2\circ\cdots\circ\gamma_5$. (b) The braid diagrams for sub-loops encircling a single elementary EL. (c) The eigenenergy braiding in the complex energy plane for the braid $b(\gamma_3)=b_2$ along the sub-loop $\gamma_3$, and the colors of the connecting bars between the strand $E_2$ and $E_3$ denote the phase of the relative energy.     }
\end{figure*} 

\begin{proof}
\textbf{Method 1.}  We first give an intuitive proof by virtue of their relation to the ELs in 3D parameter space. As shown in Fig.~\ref{fig-relation-proofmehtod1}(a), imagine that the loop $\Gamma$ encircles a total of $M$ ELs. Without loss of generality, we can assume all the ELs are of order-2 and elementary, i.e., only one pair of eigenvalues braid around each EL, since all nonelementary degeneracies can split into elementary ones under small perturbations without changing the topology along $\Gamma$. Through a continuous deformation that maintains the basepoint $\mathbf{k}_0$ (i.e., the starting and ending point of $\Gamma$) and bypasses the ELs, we can transform the loop $\Gamma$ as the concatenation of a series of sub-loops,  $\Gamma=\gamma_1\circ\gamma_2\circ\cdots\circ\gamma_M$, such that every sub-loop $\gamma_m$ is based at $\mathbf{k}_0$ and encloses a single elementary EL. Then, both the discriminant number and the braid element can be broken down into the contributions of all sub-loops:
\begin{equation} \label{Eq-relation-nvD}
    b(\Gamma)=\prod_{m=1}^M b(\gamma_m)=\prod_{m=1}^M {b}_{i'_m}^{\, n'_{m}},\qquad \mathcal{D}(\Gamma)=\sum_{m=1}^M\mathcal{D}(\gamma_m),
\end{equation}
Since each sub-loop only encircles one elementary EL, we have $b(\gamma_m)={b}_{i'_m}^{\, n'_{m}}\in\qty{b_1,b_1^{\,-1},b_2,b_2^{\,-1},\cdots,b_{N-1},b_{N-1}^{\,-1}}$ and $\mathcal{D}(\gamma_m)\in\qty{\pm1}$, as shown in Fig.~\ref{fig-relation-proofmehtod1}(b). 
The decomposition of $\mathcal{D}(\Gamma)$ also exhibits that \textbf{the discriminant number is equal to the net number of directed elementary ELs passing through the loop.}

As shown in Fig.~\ref{fig-relation-proofmehtod1}(c), the braiding direction of the two swapping eigenvalues (i.e, the sign of the exponent) is indeed consistent with the winding direction of the relative eigenenergy $\delta E_{i'_m+1,i'_m}=E_{i'_m+1}-E_{i'_m}$ on the complex plane\footnote{Obtaining the braid word depends on the convention in which direction we project the 3D eigenvalue paths onto the 2D plane. However, the algebraic length of a braid is independent to the convention. Here, we project the eigenvalues onto the real axis so that braiding occurs when $\delta E_{i'_m+1,i'm}$ traverses the imaginary axis.}. Hence, the exponent of the braid is just given by the energy vorticity $\nu_{i'_m+1,i'}(\gamma_m)=\oint_{\gamma_m}d\arg(E_{i'_m+1}-E_{i'})$ and hence by the discriminant number along the sub-loop:
\begin{equation*}
    n'_m=2\,\nu_{i'_m+1,i'_m}(\gamma_m)=\mathcal{D}(\gamma_m)\in\qty{\pm1}.
\end{equation*}
Since the algebraic length (exponent sum) is a braid invariant, the different factorizations, e.g., $\Pi_{l=1}^L b_{i_l}^{\,n_l}$ and $\Pi_{m=1}^{\,M}b(\gamma_m)=\Pi_{m=1}^{\,M}b_{i'_m}^{n'_m}$, of the same braid $b(\Gamma)$ do not change the algebraic length, leading to the relation in Eq.~\eqref{algebraic length}:
\begin{equation*}
\sum_{l=1}^{L}n_l=\sum_{m=1}^M n'_m=\sum_{m=1}^M \mathcal{D}(\gamma_m)=\mathcal{D}(\Gamma).
\end{equation*}

\begin{figure*}[b!]
\includegraphics[width=0.99\textwidth]{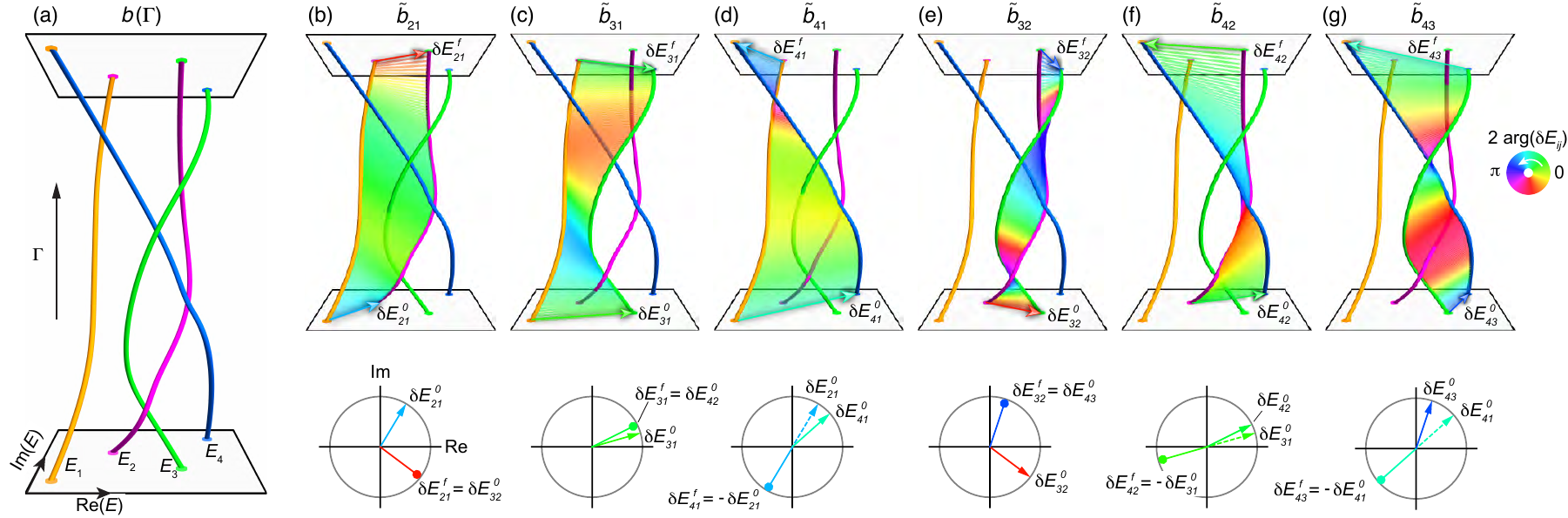}
\caption{\label{fig-relation-proofmehtod2} 
Schematic for the proof \textbf{Method 2} of the \textbf{Theorem}~\ref{theorem-algebraic length}.
(a) The eigenenergy braiding in the complex eigenenergy plane along the loop $\Gamma$. The braid $b(\Gamma)$ of $N=4$ strands can be decomposed into $6$ sub-braids $\tilde{b}_{ij}$ which are shown in (b-g), where the lower panels show the initial (final) normalized relative eigenenergy $\delta E_{ij}^0$ ($\delta E_{ij}^f$) in the complex energy plane.  }
\end{figure*} 

\textbf{Method 2.} In this method, instead of embedding the loop into 3D parameter space in the method 1, we directly use the 1D information along the loop to prove the relation. As illustrated in Fig.~\ref{fig-relation-proofmehtod2}, we first decompose the braid into the sub-braids $\tilde{b}_{ij}$ between pairwise strands $(i,j)$ (the strands are numbered by their order of the real parts at the bottom). Denoting $n_{ij}$ as the number of times that braiding occurs in $\tilde{b}_{ij}$, the algebraic length of $b(\Gamma)$ can be alternatively evaluated by
\begin{equation*}
    \sum_{l=1}^L n_l=\sum_{j<i} n_{ij}.
\end{equation*}
It is easy to see that the relative phase winding between each pair of eigenvalue strands $(i,j)$ ($j<i$) satisfies
\begin{equation}\label{relative phase change}
    \Delta\theta_{ij}=\oint_{\Gamma}d\arg(E_i-E_j)=2\pi\floor*{\frac{n_{ij}}{2}}+\qty(\theta^f_{ij}-\theta^0_{ij})=
    \left\{\begin{aligned}
    &\pi n_{ij}+ \qty(\theta^f_{ij}-\theta^0_{ij})\ &\text{if}\ n_{ij}\in\text{even}\\
    &\pi (n_{ij}-1)+ \qty(\theta^f_{ij}-\theta^0_{ij})\ &\text{if}\ n_{ij}\in\text{odd}
    \end{aligned}\right.,
\end{equation}
where $\floor*{*}$ represents the floor function, $\theta^0_{ij}=\arg(\delta E^0_{ij})\in(-\pi/2,\pi/2)$ and $\theta^f_{ij}=\arg(\delta E^f_{ij})\in(-\pi/2,3\pi/2)$ gives the values of $\arg(E_i-E_j)$ at the bottom and top ends of the strands (see Fig.). The first term $2\pi\floor*{\frac{n_{ij}}{2}}$ in Eq.~\eqref{relative phase change} records the contribution of intact winding periods in the sub-braid $\tilde{b}_{ij}$, and the second term $(\theta^f_{ij}-\theta^0_{ij})$ gives the residual winding angle less than one period. If we regard $\delta E_{ij}$ as a vector on the complex plane, the initial vector $\delta E^0_{ij}$ always appears in the right-hand side of the imaginary axis due to the requirement of $j<i$. Hence, we let $\theta^0_{ij}$ take values in the single-valued interval $(-\pi/2,\pi/2)$, and accordingly $\theta^f_{ij}$ should be in $(-\pi/2,3\pi/2)$.

If the pair of strands $(i,j)$ goes to the endpoints ($f_i,f_j$) on the top (numbered by the order of the strands at the top as shown Fig.~\ref{fig-relation-proofmehtod2}(b-g)), owing to the periodicity of the braid, the final relative phase $\theta^f_{ij}$ of the pair $(i,j)$ are related to the initial relative phase of the pair $(f_i,f_j)$:
\begin{equation}\label{final relative phase}
    \theta^f_{ij}=\theta^0_{f_i f_j}=\arg(E^0_{f_i}-E^0_{f_j})=\theta_{\mathrm{sort}\{f_i,f_j\}}^0+\qty(n_{ij}-2\floor*{\frac{n_{ij}}{2}})\pi=
    \left\{\begin{aligned}
    &\theta_{f_if_j}^0\ &\text{if}\ n_{ij}\in\text{even}\\
    &\theta_{f_jf_i}^0+\pi\ &\text{if}\ n_{ij}\in\text{odd}
    \end{aligned}\right.,
\end{equation}
where the function $\mathrm{sort}\{f_i,f_j\}$ sorts the pair of indices in descending order. 
This result can be understood as follows. When the two strands braid an even number of times ($n_{ij}\in\text{even}$), the final order of the two strands, i.e., the order of $(f_i,f_j)$, remains the same as the initial order of $(i,j)$, so $j<i$ ensures $f_j<f_i$. Consequently, the final vector $\delta E^f_{ij}$ of the pair $(i,j)$ is coincident with the initial vector $\delta E^0_{f_if_j}$, as depicted in Figs.~\ref{fig-relation-proofmehtod2}(b,c,e). In contrast, a braid with an odd number of times ($n_{ij}\in\text{odd}$) reverses the sequence of the two strands at the bottom and top ends, so $j<i$ leads to $f_j>f_i$, making the final vector $\delta E^f_{ij}$ lie at the antipodal point of the initial vector $\delta E^0_{f_jf_i}$ on the left side of the imaginary axis (see the lower panels in Figs.~\ref{fig-relation-proofmehtod2}(d,f,g)).

Substituting Eq.~\eqref{final relative phase} into Eq.~\eqref{relative phase change}, we obtain
\begin{equation*}
    \Delta\theta_{ij}=\pi\,n_{ij}+\qty(\theta_{\mathrm{sort}\{f_i,f_j\}}^0-\theta^0_{ij}),
\end{equation*}
and also the discriminant number
\begin{equation}\label{dis with relative phase}
    \mathcal{D}(\Gamma)=\sum_{j<i} \frac{1}{\pi}\oint_\Gamma d\arg(E_i-E_j)=\sum_{j<i}\frac{\Delta\theta_{ij}}{\pi}=\sum_{j<i}n_{ij}+\frac{1}{\pi}\sum_{j<i}\qty(\theta_{\mathrm{sort}\{f_i,f_j\}}^0-\theta^0_{ij}).
\end{equation}
In the following, we show that the last term has no contribution to the DN. 
The unordered pair $\{f_i,f_j\}$ of the two strands at the top end of the subbraid $\tilde{b}_{ij}$ can be obtained by the braid permutation $\hat{\mathcal{P}}$ : $\{f_i,f_j\}=\hat{\mathcal{P}}\{i,j\}$. And since the permutation $\hat{\mathcal{P}}$ is an automorphism of the set $[N]=\{1,\cdots, N\}$, the set of all unordered pairs in $[N]$ is invariant against permutation:
\begin{equation*}
    S=\Big\{\{i,j\}\Big|i,j\in [N]\ \text{and}\ i\neq j\Big\}=\Big\{\hat{\mathcal{P}}\{i,j\}\Big|i,j\in [N]\ \text{and}\ i\neq j\Big\}=S_\mathcal{P}.
\end{equation*}
Therefore, 
\begin{equation*}
    \sum_{j<i}\theta_{\mathrm{sort}\{f_i,f_j\}}^0=
    \sum_{\qty{i,j}\in S}\theta_{\mathrm{sort}\qty[\hat{\mathcal{P}}\qty{i,j}]}^0=\sum_{\qty{i,j}\in S_\mathcal{P}}\theta_{\mathrm{sort}\qty{i,j}}^0=\sum_{\qty{i,j}\in S}\theta_{\mathrm{sort}\qty{i,j}}^0=
    \sum_{j<i}\theta_{ij}^0,
\end{equation*}
which ensures the cancellation of the last term in Eq.~\eqref{dis with relative phase}; accordingly, Eq.~\eqref{algebraic length} is proved. 
\end{proof}

\subsection{Relation between global Berry phase and discriminant number}
We consider a $N$-band non-Hermitian Hamiltonian $\mathcal{H}(\mathbf{k})$, and the right and left eigenstates  $\left|\psi^{R/L}_n(\mathbf{k})\right\rangle$ corresponding to the $n$th eigenenergy $E_{n}(\mathbf{k})$ satisfy
\begin{equation}\label{Eq-eigenequation}
\begin{aligned}
\mathcal H(\mathbf{k})\left|\psi^R_{n}(\mathbf{k})\right\rangle &=E_{n}(\mathbf{k})\left|\psi^R_{n}(\mathbf{k})\right\rangle, \\
\mathcal H^{\dagger}(\mathbf{k})\left|\psi^L_{n}(\mathbf{k})\right\rangle &=E_{n}^{*}(\mathbf{k})|\psi^L_{n}(\mathbf{k})\rangle.
\end{aligned}
\end{equation}
The total biorthogonal Berry phase of all bands along the loop $\Gamma$,  termed global biorthogonal Berry phase, is given by~\cite{hu2021Knots}
\begin{equation}\label{global berry phase}
    \Theta(\Gamma)=-i\sum_n^N \oint_\Gamma d\mathbf{k}\cdot\langle \psi_n^L|\nabla_\mathbf{k}|\psi^R_n\rangle= -i\oint_\Gamma d\mathbf{k}\cdot\mathrm{Tr}\left(\Psi^{-1}\nabla_\mathbf{k}\Psi\right),
\end{equation}
where the eigenvectors $|\psi^{R/L}_n(\mathbf{k})\rangle$ should satisfy the biorthonormal relation $\left\langle \psi^L_{n'}(\mathbf{k}) \mid \psi^R_{n}(\mathbf{k})\right\rangle=\delta_{nn'}$ and are continuous through the loop (including at the concatenate points of two bands with different indices), and $\Psi=(\left|\psi^R_{1}(\mathbf{k})\right\rangle,\left|\psi^R_{2}(\mathbf{k})\right\rangle,\cdots,\left|\psi^R_{N}(\mathbf{k})\right\rangle)$ and $\Psi^{-1}=(\left|\psi^L_{1}(\mathbf{k})\right\rangle,\left|\psi^L_{2}(\mathbf{k})\right\rangle,\cdots,\left|\psi^L_{N}(\mathbf{k})\right\rangle)^\dagger$. In the Hermitian case, the global Berry phase always trivially equals zero, while in the non-Hermitian case, the global Berry phase $\Theta$ can take nontrivial quantized value stemming from the nontrivial eigenvalue permutation~\cite{hu2021Knots}. Therefore, the global Berry phase offers a $\mathbb{Z}_2$ topological invariant for a non-Hermitian Hamiltonian along 1D loops. Here, we can proceed to show that the $\mathbb{Z}_2$ global Berry phase can also be determined by the discriminant number.

\begin{theorem}
For a Hamiltonian with separable bands along a loop $\Gamma$, the $\mathbb{Z}_2$ global biorthogonal Berry phase $\Theta(\Gamma)$, the parity of the braid permutation $\hat{\mathcal{P}}(\Gamma)$, and the parity of the discriminant number $\mathcal{D}(\Gamma)$ satisfy the relation:  
\begin{equation}
   \exp\qty[i\Theta(\Gamma)]=\det\qty[\hat{\mathcal{P}}(\Gamma)]=(-1)^{\mathcal{D}(\Gamma)}\in\qty{\pm1}.
\end{equation}
\end{theorem}
The theorem sheds light on the physical meaning of the global Berry phase, i.e., \textbf{the global Berry phase $\Theta(\Gamma)$ (trivial or nontrivial) along a loop $\Gamma$ corresponds to the parity (even or odd) of the  number of mode swapping along $\Gamma$ and also to the parity of the number of ELs encircled by $\Gamma$.}
\vspace{3pt}

\begin{proof}
Following Ref.~\cite{hu2021Knots}, the global Berry phase factor is derived as 
\begin{equation*}
\exp\qty[i\Theta(\Gamma)]=\exp\left[\oint_\Gamma d\mathbf{k}\cdot\mathrm{Tr}\left(\Psi^{-1}\nabla_\mathbf{k}\Psi\right)\right]=\exp\left[\oint_\Gamma d\ln\big(\det\Psi\big)\right]=\exp\left[\ln\frac{\det\Psi^f}{\det\Psi^0}\right]
=\frac{\det\Psi^f}{\det\Psi^0},
\end{equation*}
where $\Psi^0$ and $\Psi^f$ are the matrices of the ordered eigenvectors at the starting and end points of the loop $\Gamma$. For generic braiding non-Hermitian bands, the initial and final eigenstates $\Psi^0$ and $\Psi^f$ are identical up to a permutation $\hat{\mathcal{P}}(\Gamma)$:
$
\Psi^f=\hat{\mathcal{P}}(\Gamma)\Psi^0
$ due to the periodicity. 
Thus, the global Berry phase is just determined by the \textbf{parity of the permutation}: 
$
\exp\qty[i\Theta(\Gamma)]=\det\qty[\hat{\mathcal{P}}(\Gamma)] = \pm1
$. 

In addition, the parity of a permutation equals $\det\qty[\hat{\mathcal{P}}(\Gamma)]=(-1)^{N(\hat{\mathcal{P}}(\Gamma))}$, where $N(\hat{\mathcal{P}}(\Gamma))=\sum_{l=1}^{L}|n_l|$ denotes the total number of eigenvalue swaps in the permutation (without regard to the sign of the swapping), which can be directly related to the algebraic length of the braid $b(\Gamma)=\sum_{l=1}^L b_{i_l}^{n_l}$: 
\begin{equation*}
    \det\qty[\hat{\mathcal{P}}(\Gamma)]=(-1)^{N(\hat{\mathcal{P}}(\Gamma))}=(-1)^{\sum_{l=1}^{L} |n_l|}=(-1)^{\sum_{l=1}^{L} n_l}=(-1)^{\mathcal{D}(\Gamma)},
\end{equation*}
where we have used the result in \textbf{Theorem}~\ref{theorem-algebraic length}, i.e., the  algebraic length of the braid equals the discriminant number along the loop. Thereupon, the proof of the relation is accomplished.
\end{proof}

\begin{remark}
Under a generic continuous gauge transformation to the eigenstates along a loop, $\Psi'(k)=U(k)\Psi(k)$ with $U(k)=\mathrm{diag}(\alpha_1(k),\cdots,\alpha_N(k))$ and $\alpha_i(0)=\alpha_i(2\pi)+2n_i\pi$ ($n_i\in\mathbb{Z}$), the global Berry phase is changed accordingly $\Theta'(\Gamma)=\Theta(\Gamma)+2\pi\sum_in_i$. Therefore, the global Berry phase should be recognized as a $\mathbb{Z}_2$ invariant with an unfixed $2n\pi$ phase freedom if we do not specify the gauge of the eigenstates along the loop. In Refs.~\cite{mailybaev2005Geometric,li2019Geometric}, the global Berry phase along a loop $\Gamma_\mathrm{EP}$ around a single order-2 EP is proved to be exactly proportional to the DN, $\Theta(\Gamma)=\pi\mathcal{D}(\Gamma_\mathrm{EP})$ if the eigenstates satisfy a specific continuous gauge inside the whole enclosed area of $\Gamma_\mathrm{EP}$.  However, when the loop encircles multiple EPs with arbitrary orders in a multi-band system, how to construct such a 2D continuous gauge  remains as an open question.\\[3pt]

\end{remark}

\clearpage
\section{Source-free principle of exceptional lines}

\subsection{Doubling theorem for degenerate points in 2D Brillouin zone}

In Ref.~\cite{yang2021fermion}, it is revealed that  the Nielsen-Ninomiya fermion doubling theorem can be generalized to characterize the EPs in the Brillouin zone (BZ) for 2D non-Hermitian periodic systems:

\begin{theorem}[\bf Doubling theorem of EPs in 2D BZ~\cite{yang2021fermion}]
For a 2D periodic non-Hermitian systems, if the Hamiltonian in the momentum space $\mathcal{H}(\mathbf{k})$ only has isolated degeneracies at the points $\qty{\mathbf{k}_d}$ in the 2D BZ, the total indices of all degenerate points vanish.
\begin{equation}
    \sum_{\mathbf{k}_d\in\mathrm{BZ}}\mathcal D(\Gamma_{\mathbf{k}_d})=0\label{DTonT2},
\end{equation}
where the index of a degenerate point $\mathbf{k}_d$ is given by the DN along a small counterclockwise loop $\Gamma_{\mathbf{k}_d}$ solely encircling $\mathbf{k}_d$. If all degeneracies are elementary EPs that are stable under generic perturbations, the elementary EPs must appear in pairs with opposite DNs: $\mathcal D=\pm1$.
\end{theorem}
 
\begin{figure*}[b!]
\includegraphics[width=0.67\textwidth]{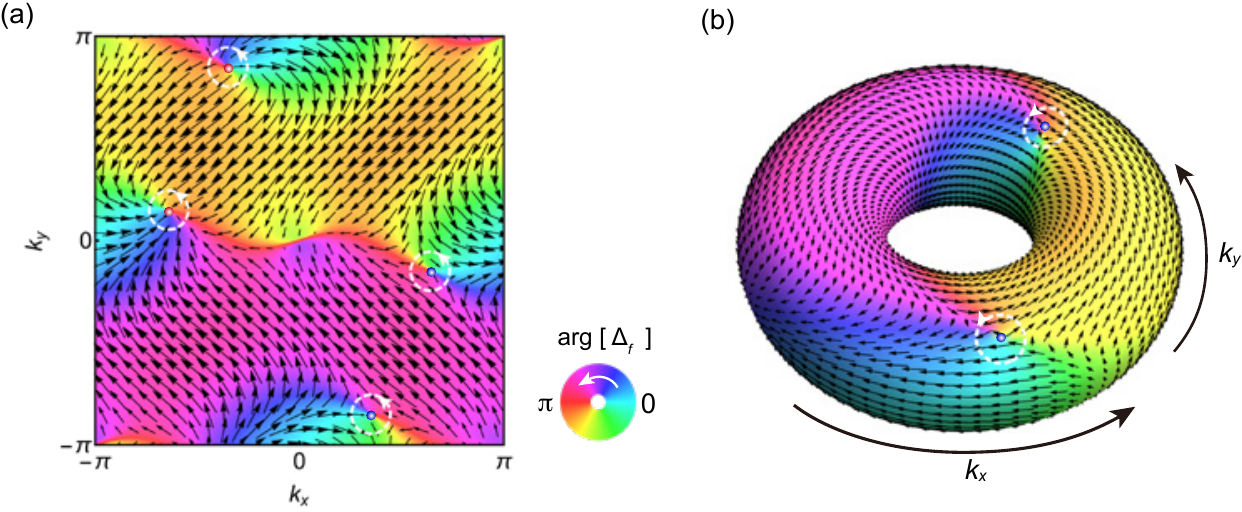}
\caption{\label{PH theorem_torus} 
Doubling theorem of EPs in 2D BZ. (a) Schematic of elementary order-2 EPs in 2D BZ. (b) Mapping the disciminant $\Delta_f(\mathbf{k})$ to a tangent vector field $\mathbf{v}_{\Delta_f}(\mathbf{k})=\mathrm{Re}\qty[\Delta_f(\mathbf{k})]\hat{\mathbf{k}}_x+\mathrm{Im}\qty[\Delta_f(\mathbf{k})]\hat{\mathbf{k}}_y$ on the torus $T^2$. The background colormap shows the phase distribution of $\Delta_f(\mathbf{k})$. The arrows display the vector field $\mathbf{v}_{\Delta_f}(\mathbf{k})$.  The red and blue dots at the phase and vector field singularities represent the EPs with $\mathcal{D}(\Gamma_{\mathbf{k}_d})=\pm 1$, respectively, and the white dashed loops denote $\Gamma_{\mathbf{k}_d}$.
}
\end{figure*}

Indeed, if we map the discriminant $\Delta_f(\mathbf{k})$ to a real tangential vector field $\mathbf{v}_{\Delta_f}(\mathbf{k})=\mathrm{Re}\qty[\Delta_f(\mathbf{k})]\hat{\mathbf{k}}_x+\mathrm{Im}\qty[\Delta_f(\mathbf{k})]\hat{\mathbf{k}}_y$ on the 2D BZ (a torus $T^2$) (see the schematic in Fig.~\ref{PH theorem_torus}), the doubling theorem of EPs can be understood as an direct application of the \textbf{Poincar\'e-Hopf theorem for tangent bundles}~\cite{needham2021visual,frankel2011thegeometry} (also known as hairy ball theorem), which states that for any continuous tangential vector field $\mathbf{v}(\mathbf{k})$ (mathematically, a global section of the tangent bundle) on a 2D oriented compact manifold $S$, the sum of the indices of all zero points $\qty{\mathbf{k}_d}$ of $\mathbf{v}$ equals the the Euler characteristic, $\chi(S)$, of the manifold $S$:
\begin{equation}
  \sum_{x_i\in M}\mathrm{index}_{\mathbf{v}}(\mathbf{k}_d)=\chi(S),
\end{equation}
where the index of a zero point $\mathbf{k}_d$, $\mathrm{index}_{\mathbf{v}}(\mathbf{k}_d)$, is given by the winding number of $\vec{v}$ around $\mathbf{k}_d$ (see the white dashed loops in Fig.~\ref{PH theorem_torus}). Therefore, the vanishing of the total DNs of all degeneracies in a 2D BZ  precisely stems from the zero Euler characteristic of a torus $\chi(T^2)=0$.

\subsection{Generalized doubling theorem for the EPs on an arbitrary orientable closed surface}
If we consider an arbitrary  2D orientable closed surface, rather than the 2D BZ, inside the 3D momentum space, the Poincare-Hopf theorem for tangential vector fields becomes generally inapplicable to decide the existence of EPs on the surface, since the discriminant $\Delta_f(\mathbf{k})$ \textbf{cannot} map to a tangential vector field on the surface with nontrivial Euler characteristic. For instance, the discriminant on a sphere in 3D BZ can be nowhere-vanishing (the Hamiltonian is non-degenerate everywhere), contradicting to the Euler characteristic of a sphere $\chi(S^2)=2$. 

Nevertheless, we will see that the doubling theorem for EPs remains valid on any closed and oriented surfaces by using the generalized index theorem for complex line bundles. In fact, as long as the Hamiltonian $\mathcal{H}(\mathbf{k})$ is continuous everywhere in the 3D momentum space, the characteristic polynomial of $\mathcal{H}(\mathbf{k})$, $f(E,\mathbf{k})=\det\qty[E-\mathcal{H}(\mathbf{k})]=E^N+a_{N-1}(\mathbf{k})E^{N-1}+\cdots+a_1(\mathbf{k})E+a_0(\mathbf{k})$, and the discriminant $\Delta_f(\mathbf{k})\in\mathbb{C}$, which is a polynomial in $a_0(\mathbf{k})\cdots a_{N_1}(\mathbf{\mathbf{k}})$ with integer coefficients, is also \textbf{continuous} in the whole space. Given a 2D closed and oriented surface $S$ in the momentum space, the discriminant $\Delta_f(\mathbf{k})$, as a \textbf{continuous and single-valued complex function}, induces a continuous map from the surface to the product manifold $L=S\times\mathbb{C}$:
\begin{equation}\label{discriminant map}
    \tilde{\Delta}_f:\ S\ \rightarrow\  L=S\times\mathbb{C}.
\end{equation}
Mathematically, the product manifold $L=S\times\mathbb{C}$ can be regarded as a \textbf{trivial\footnote{A fiber bundle is \textbf{trivial} just means that the bundle space is the direct product of the base space and the fiber and hence its fibers are ``untwisted''.} complex line bundle}: 
\begin{equation}
    \pi:\ L=S\times\mathbb{C}\ \rightarrow\ S,
\end{equation}
with the 2D surface $S$ being the base space, $L$ being the bundle space, the complex plane (1D complex vector space) $\mathbb{C}$ being the fiber, and $\pi$ the bundle projection. And the ``discriminant map'' $\tilde{\Delta}_f$ (Eq.~\eqref{discriminant map}) indeed forms a \textbf{global section} of the line bundle $L$ such that $\pi\circ\tilde{\Delta}_f=\mathrm{id}: S\ \rightarrow\ S$ gives the identity map, i.e., $\pi\circ\tilde{\Delta}_f(\mathbf{k})=\mathbf{k}$, $\forall \mathbf{k}\in S$. 

\begin{remark}
The global continuity and single-value nature of the discriminant $\Delta(\mathbf{k})$ are crucial for regarding it as a global section of a \textbf{trivial} complex line bundle. Indeed, in a nontrivial bundle, 
it is generically impossible to express a global section as a continuous single-complex-valued function in the whole base space. Instead, a global section can only be parameterized to local complex-valued functions (known as gauges in physics) in different patches. In the glued region of two patches, two different parameterized functions are related by a ``gauge transformation''. The most well-known application of this notion in physics is the nonexistence of a globally continuous  gauge for the electron wave functions surrounding a Dirac magnetic monopole~\cite{wu1976dirac}. 
\end{remark}

For complex line bundles, the number of zeros of a global section is also determined by the global topology of the bundle, which generalizes the Poincar\'e-Hopf index theorem as follows:

\begin{theorem} [\bf Poincare-Hopf theorem for complex line bundle~\cite{frankel2011thegeometry,Knoppel2020Riemann}]
For any global section $\tilde{\Delta}:\ S\ \rightarrow\ L$ of a complex line bundle $\pi:\ L\ \rightarrow\ S$ on a 2D closed and oriented surface $S$, if the global section only has isolated zeros at $\qty{\mathbf{k}_d}$ on $S$, the index sum of all zero points are identically given by the first Chern number $\mathrm{Ch}(L)$ of the bundle:
\begin{equation}
	\sum_{\mathbf{k}_d\in S}\mathcal{D}_{\tilde{\Delta}}(\Gamma_{\mathbf{k}_d})=\mathrm{Ch}(L),
\end{equation}
where the index of a zero point $\mathbf{k}_d$ is just given by phase winding number of the local fiber coordinate (i.e., a local complex-valued function $\Delta(\mathbf{k})$) of the section $\tilde{\Delta}$ along a loop $\Gamma_{\mathbf{k}_d}$ around $\mathbf{k}_d$: $\mathcal{D}_{\tilde{\Delta}}(\Gamma_{\mathbf{k}_d})=\oint_{\Gamma_{\mathbf{k}_d}} d\arg\Delta(\mathbf{k})$, and the direction of the loop $\Gamma_{\mathbf{k}_d}$ for integration should comply with the outward normal of the oriented surface.
\end{theorem}

Since the bundle for the ``discriminant section'' $\tilde{\Delta}_f$~\eqref{discriminant map} is always trivial:   $L\cong S\times\mathbb{C}$  regardless of the Hamiltonian, and the Chern number  of the trivial line bundle always vanishes $\mathrm{Ch}(L)\equiv0$, the generalization of the doubling theorem for EPs on any compact oriented surface $S$ is established: 

\begin{theorem}[\bf Doubling theorem of EPs on 2D compact oriented surface]\label{generalized doubling theorem}
For a continuous non-Hermitian Hamiltonian $\mathcal{H}(\mathbf{k})$ defined on any 2D compact oriented surface $S$, if all degeneracies of $\mathcal{H}(\mathbf{k})$ (i.e., the zeros of the discriminant $\Delta_f(\mathbf{k})$) are isolated points at $\qty{\mathbf{k}_d}$ on $S$, the total DNs of all degenerate points vanish.
\begin{equation}\label{DTonT2}
    \sum_{\mathbf{k}_d\in\mathrm{BZ}}\mathcal D(\Gamma_{\mathbf{k}_d})=\mathrm{Ch}(S\times\mathbb{C})=0.
\end{equation}
In particular, allowing for generic perturbations, the degeneracies on the surface $S$ always emerge as elementary EP pairs with opposite DNs: $\mathcal D=\pm1$.
\end{theorem}

When a 2D surface $S$ is embedded in the 3D momentum space, the EPs on $S$ are indeed the loci where ELs pass across the surface (see the schematic in Fig.1(d) of the main text), and the EPs' DNs determine the orientations of the ELs. Then we obtain from \textbf{Theorem}~\ref{generalized doubling theorem} that
\begin{corollary}[\bf Source-free principle of ELs]
  For any 2D orientable closed surface $S$ in the 3D momentum space, if all ELs intersecting with the $S$ are elementary, the numbers of the directed elementary ELs entering and exiting the surface are always identical. Therefore, ELs should either form closed loops or join together at chain points, at each of which the inflow and outflow ELs must be balanced.
\end{corollary}
\vspace{15pt}

\section{Correlated directions of symmetry-partner ELs}
\vspace{-15pt}
\begin{table}[h!]
\begin{adjustwidth}{-1.5in}{-1.5in}  
\centering
\setlength\cellspacetoplimit{5pt}
\setlength\cellspacebottomlimit{5pt}
\addtolength\tabcolsep{5pt}
\caption{Reltations between the DNs of two symmetry-partner loops and relations between the positive directions of two symmetry-partner ELs. \label{DN table}}
\begin{tabular}{Sc|Sc|Sc|Sc|Sc}\hline\hline
 \textbf{Symmetry type} & \textbf{Spectral symmetry} &  \textbf{DNs} & \textbf{Directions of partner ELs} & \textbf{Examples}\\
 \hline
 $G$  &  $E_i(\mathbf{k})=E_i(\hat{g}\mathbf{k})$    & $\mathcal{D}(\hat{g}\Gamma)=\mathcal{D}(\Gamma)$   & $\mathbf{t}_\mathrm{EL}(\hat{g}\mathbf{k}_\mathrm{EP})=\det[\hat{g}]\,\hat{g}\mathbf{t}_\mathrm{EL}(\mathbf{k}_\mathrm{EP})$ & $\mathcal{P}$, $M$ \\\hline
 $G\mbox{-}\dagger$  & $E_i(\mathbf{k})=E_i(\hat{g}\mathbf{k})^*$  & $\mathcal{D}(\hat{g}\Gamma)= -\mathcal{D}(\Gamma)$   &  $\mathbf{t}_\mathrm{EL}(\hat{g}\mathbf{k}_\mathrm{EP})=-\det[\hat{g}]\,\hat{g}\mathbf{t}_\mathrm{EL}(\mathbf{k}_\mathrm{EP})$ & \mdagger \\\hline
  $G\mathcal{T}$  &  $E_i(\mathbf{k})=E_i(-\hat{g}\mathbf{k})^*$  & $\mathcal{D}(-\hat{g}\Gamma)=-\mathcal{D}(\Gamma)$   & $\mathbf{t}_\mathrm{EL}(-\hat{g}\mathbf{k}_\mathrm{EP})=-\det[\hat{g}]\,\hat{g}\mathbf{t}_\mathrm{EL}(\mathbf{k}_\mathrm{EP})$ & $C_2\mathcal{T}$ \\\hline
  $G\mathcal{T}\mbox{-}\dagger$  &  $E_i(\mathbf{k})=E_i(-\hat{g}\mathbf{k})$ &  $\mathcal{D}(-\hat{g}\Gamma)=\mathcal{D}(\Gamma)$  & $\mathbf{t}_\mathrm{EL}(-\hat{g}\mathbf{k}_\mathrm{EP})=\det[\hat{g}]\,\hat{g}\mathbf{t}_\mathrm{EL}(\mathbf{k}_\mathrm{EP})$  & $C_2\mathcal{T}\mbox{-}\dagger$ \\
\hline\hline
\end{tabular}
\end{adjustwidth}
\end{table}

As we showed in the Methods of the main text, by introducing time reversal ($\mathcal{T}$) and Hermitian-adjoint ($\dagger$) operations, the non-Hermitian crystalline symmetries are characterized by double-antisymmetry (DAS) space (point) groups. In DAS point groups, the symmetries are sorted into four types obeying the following expressions in the momentum space:
\begin{enumerate}
    \item[(i)]  The pure spatial symmetries, deonoted $G$: $\hat{G}\mathcal{H}(\hat{g}^{-1}\mathbf{k})\hat{G}^{-1}=\mathcal{H}(\mathbf{k})$;
    \item[(ii)]  The spatial-adjoint symmetry, denoted $G\mbox{-}\dagger$: $\hat{G}\mathcal{H}(\hat{g}^{-1}\mathbf{k})^\dagger\hat{G}^{-1}=\mathcal{H}(\mathbf{k})$;
     \item[(iii)]  The spatiotemporal (magnetic-group) symmetries, denoted $G\mathcal{T}$: $\hat{G}\hat{T}\mathcal{H}(-\hat{g}^{-1}\mathbf{k})^*(\hat{G}\hat{T})^{-1}=\mathcal{H}(\mathbf{k})$;
     \item[(iv)]  The spatiotemporal-adjoint  symmetries, denoted $G\mathcal{T}\mbox{-}\dagger$: $\hat{G}\hat{T}\mathcal{H}(-\hat{g}^{-1}\mathbf{k})^\intercal(\hat{G}\hat{T})^{-1}=\mathcal{H}(\mathbf{k})$.
\end{enumerate}
Here, $\hat{g}\in O(3)$ denotes the spatial rotation acting on the vectors in the real and momentum spaces, and $\hat{G}$, $\hat{T}$ denote the unitary parts associated with the $G$, $\mathcal{T}$ operators, respectively, acting on the Bloch states.

The four types of symmetries connect the eigenenergies of the Bloch states at two symmetry-partner points in the momentum space, hence giving rise to the different types of spectral symmetries of the energy bands as listed in the second column in Table~\ref{DN table}. The spectral symmetries also correlate the discriminants $\Delta_f(\mathbf{k})=\prod_{j<i}\qty[E_i(\mathbf{k})-E_j(\mathbf{k})]^2$ of the characteristic polynomial $f(E,\mathbf{k})=\det\qty[\mathcal{H}(\mathbf{k})-E]$ at the two partner points:
\begin{alignat}{3}
    G:&\quad \Delta_f(\hat{g}\mathbf{k})
    &&=\Delta_f(\mathbf{k}),\\
    G\mbox{-}\dagger:&\quad \Delta_f(\hat{g}\mathbf{k})
    &&=\Delta_f(\mathbf{k})^*,\\
    G\mathcal{T}:&\quad \Delta_f(-\hat{g}\mathbf{k})
    &&=\Delta_f(\mathbf{k})^*,\\
    G\mathcal{T}\mbox{-}\dagger:&\quad \Delta_f(-\hat{g}\mathbf{k})
    &&=\Delta_f(\mathbf{k}).
\end{alignat}
Hence, the point-group symmetries also connect the DNs along two symmetry-partner loops, provided that all states on the loops are non-degenerate. Say, a $G\mbox{-}\dagger$ symmetry connects the DNs along two $G\mbox{-}\dagger$-partner loops $\Gamma$ and $\Gamma'=\hat{g}\Gamma$:
\begin{equation}
    \begin{split}
    \mathcal{D}(\hat{g}\Gamma) &=\oint_{\hat{g}\Gamma} d\mathbf{k}\cdot\nabla_\mathbf{k}\arg\qty[\Delta_f(\mathbf{k})]=\oint_{\Gamma} d\mathbf{k}\cdot\nabla_\mathbf{k}\arg\qty[\Delta_f(\hat{g}\mathbf{k})]\\
    &=\oint_{\Gamma} d\mathbf{k}\cdot\nabla_\mathbf{k}\arg\qty[\Delta_f(\mathbf{k})^*]
    =-\oint_{\Gamma} d\mathbf{k}\cdot\nabla_\mathbf{k}\arg\qty[\Delta_f(\mathbf{k})]
    =-\mathcal{D}(\Gamma).
\end{split}
\end{equation}
The relations between the DNs along two symmetry-partner loops arising from the other three types of symmetries can be similarly deduced, as shown in the third column in Table~\ref{DN table}.

Owing to the spectral symmetry induced by a DAS point-group symmetry $\tilde{G} \in \qty{G, G\mbox{-}\dagger, G\mathcal{T}, G\mathcal{T}\mbox{-}\dagger}$, the ELs should also be $\tilde{G}$-symmetrically distributed in the momentum space. Next, we explain how the positive directions of two $\tilde{G}$-partner ELs endowed by the DNs around them are correlated with each other by the $\tilde{G}$ symmetry. Given a directed loop, $\Gamma$, encircling a single EL and a tangent vector of the EL $\mathbf{t}_\Gamma(\mathbf{k}_\mathrm{EP})$ ($\mathbf{k}_\mathrm{EP}$ is a point on the EL) whose orientation is determined by the right-hand rule of the loop, the positive direction of the EL at $\mathbf{k}_\mathrm{EP}$ is defined by $\mathbf{t}_\mathrm{EL}(\mathbf{k}_\mathrm{EP})=\mathcal{D}(\Gamma)\mathbf{t}_\Gamma(\mathbf{k}_\mathrm{EP})$. Correspondingly, the positive direction of the $\tilde{G}$-partner EL at $\mathbf{k}'_\mathrm{EP}=\tilde{g}\mathbf{k}_\mathrm{EP}$ is $\mathbf{t}_\mathrm{EL}(\mathbf{k}'_\mathrm{EP})=\mathcal{D}(\Gamma')\mathbf{t}_{\Gamma'}(\mathbf{k}'_\mathrm{EP})$, where $\Gamma'=\tilde{g}\Gamma$  is the $\tilde{G}$-partner loop of $\Gamma$ and $\mathbf{t}_{\Gamma'}(\mathbf{k}'_\mathrm{EP})$ is the corresponding tangent vector . Importantly,  \textbf{the tangent vector $\mathbf{t}_\Gamma(\mathbf{k}_\mathrm{EP})$ associated with a loop by the right-hand rule is a pseudovector under rotations:} 
\begin{equation}\label{pseudo tangetn vector}
    \mathbf{t}_{\Gamma'}(\mathbf{k}'_\mathrm{EP})=\det[\tilde{g}]\,\tilde{g}\,\mathbf{t}_\Gamma(\mathbf{k}_\mathrm{EP}),
\end{equation}
which means a sign change $\det[\tilde{g}]=-1$ appears if the tangent vector undergoes an improper rotation $\tilde{g}$.
Here, $\tilde{g}=\hat{g}$ or $-\hat{g}$ depending on the type of the symmetry $\tilde{G}$. Using Eq.~\eqref{pseudo tangetn vector} and the relations between the DNs, we obtain the relations between the position tangent vectors of two symmetry-partner ELs with the results listed in the fourth column of Table~\ref{DN table}.

\section{Mirror-adjoint symmetry}\label{section-mirror-dagger}

We first consider a mirror operation $\hat m: z \rightarrow - z$, which flips the sign of the $z$ coordinate in real space. It is known that a $M$-symmetric momentum-space Hamiltonian $ \mathcal H(\mathbf{k})$ fulfills 
\begin{equation}\label{MzH-definition}
    \hat{M}\mathcal H\left(\hat{m} \mathbf{k} \right){\hat M^{-1}} =\mathcal{H}\left( \mathbf{k} \right),
\end{equation}
where $\hat{M}=\hat{M}^\dagger=\hat{M}^{-1}$ is a Hermitian unitary operator denoting the mirror reflection on  the internal degrees of freedom. All the eigenstates $\psi(\mathbf k)$ of a $M$-symmetric Hermitian Hamiltonian are also eigenstates of the mirror operator $\hat M$
\begin{equation}
    \hat{M}\psi(\mathbf{k})=p\psi(\hat{m}\mathbf{k})=\pm \psi(\hat{m}\mathbf{k}),
\end{equation}
where $p=\pm 1$ denotes the mirror parity. 
On the $M$-invariant plane $\Pi_M$ that is defined as
\begin{equation}
\Pi_M=\left\{\mathbf k_m \in \mathrm{BZ} \mid \hat m \mathbf k_m=\mathbf k_m\right\} \subset \mathrm{BZ},
\end{equation}
two bands with opposite mirror parity can cross and form band-structure nodal lines, which separates the plane $\Pi_M$ into two regions based on the sign of the mirror parity of the lower band~\cite{wu2019NonAbelian}. 

For any symmetry or antisymmetry of a Hermitian Hamiltonian $\mathcal{H}$, such as $\hat{A}\mathcal{H}\hat{A}^{-1}=\pm\mathcal{H}$, it can be ramified to two distinct symmetries in non-Hermitian systems, $\hat{A}\mathcal{H}\hat{A}^{-1}=\pm\mathcal{H}$ and $\hat{A}\mathcal{H}^\dagger\hat{A}^{-1}=\pm\mathcal{H}$ due to $\mathcal H^\dagger \ne \mathcal H$. For example, non-Hermiticity enables a Hermitian-adjoint counterpart of the Altland-Zirnbauer(AZ) symmetry classes, forming the AZ$^\dagger$  symmetry classes~\cite{kawabata2019Symmetry}. This principle is also applicable to spatial symmetries. In this section, we introduce the Hermitian-adjoint counterpart of the mirror symmetry, dubbed \textit{mirror-adjoint symmetry} (\mdagger): 
\begin{equation}\label{MzHdagger-definition}
    \hat{M}\mathcal H^\dagger\left(\hat{m} \mathbf{k} \right){\hat M^{-1}} =\mathcal{H}\left( \mathbf{k} \right),
\end{equation}
which is distinct from the mirror symmetry given by Eq.~\eqref{MzH-definition} provided that the Hamiltonian is non-Hermitian. 
Equation~\eqref{MzHdagger-definition} implies that the Hamiltonian is \textit{pseudo-Hermitian} on the mirror-invariant plane $\Pi_M$, where the eigenvalues on $\Pi_M$ are either purely real and complex-conjugate paired, and hence the mirror plane $\Pi_M$ can be partitioned into exact and broken phases. 


\subsection{Mirror-adjoint parity of states in exact phase}

We consider a non-Hermitian Hamiltonian $\mathcal{H}(\mathbf{k})$ with a pair of right and left eigenstates  $\left|\psi^{R/L}_n(\mathbf{k})\right\rangle$ corresponding to the eigenenergy $E_n(\mathbf{k})$ satisfying Eq.~\eqref{Eq-eigenequation}. In this section, unless otherwise stated, we will not postulate that the eigenstates satisfy specific normalization conditions so as to be compatible with the self-biorthogonal relation at EPs. Thus, the left and right eigenvectors are biorthogonal $\left\langle \psi^L_{n'}(\mathbf{k}) \mid \psi^R_{n}(\mathbf{k})\right\rangle=0$ for $n'\neq n$, but they do not need to obey the binormalization condition $\left\langle \psi^L_{n}(\mathbf{k}) \mid \psi^R_{n}(\mathbf{k})\right\rangle=1$ in general.

For a \mdagger-symmetric Hamiltonian satisfying Eq.~\eqref{MzHdagger-definition}, we have 
\begin{equation}
    \mathcal{H}^\dagger(\hat{m}\mathbf{k})\hat{M}\left|\psi^R_n(\mathbf{k})\right\rangle=\hat{M}\mathcal{H}(\mathbf{k})\left|\psi^R_n(\mathbf{k})\right\rangle=E_n(\mathbf{k})\hat{M}\left|\psi^R_n(\mathbf{k})\right\rangle,
\end{equation}
 meaning that 
$
      \hat{M}\left|\psi^R_n(\mathbf{k})\right\rangle=\rho_n(\mathbf{k})|\psi^L_{n'}(\hat{m}\mathbf{k})\rangle
$
is proportional to a left eigenstate $|\psi^L_{n'}(\hat{m}\mathbf{k})\rangle$ of $\mathcal{H} (\hat m \mathbf k)$ with eigenvalue $E_{n'}^*(\hat m \mathbf{k})=E_n(\mathbf{k})$ at the mirror point $\hat{m}\mathbf{k}$ of $\mathbf{k}$. Note that
\begin{equation}
    |\rho_n(\mathbf{k})|^2=\frac{\langle|\psi_n^R(\mathbf{k})|\hat{M}^\dagger \hat{M}| \psi_n^R(\mathbf{k})\rangle}{\langle\psi^L_{n'}(\hat{m}\mathbf{k}))|\psi^L_{n'}(\hat{m}\mathbf{k})\rangle}=\frac{\langle|\psi_n^R(\mathbf{k})|\psi_n^R(\mathbf{k})\rangle}{\langle\psi^L_{n'}(\hat{m}\mathbf{k}))|\psi^L_{n'}(\hat{m}\mathbf{k})\rangle}\neq 0\ \text{or}\ \infty,\label{norms between left and right}
\end{equation}
as long as the norms of $|\psi_n^R(\mathbf{k})\rangle$ and $|\psi^L_{n'}(\hat{m}\mathbf{k})\rangle$ are both nonzero and finite.
Therefore, the \mdagger\   symmetry ensures that the band structure is complex-conjugate symmetric about the mirror plane $\Pi_M$.

If the state $|\psi^R_n(\mathbf{k}_m)\rangle$ is situated in the exact phase of the mirror plane $\Pi_M$, we have $E_{n'}(\mathbf{k}_m)=E_{n'}(\hat{m}\mathbf{k}_m)^*=E_{n}(\mathbf{k}_m)$. Provided that $|\psi^R_n(\mathbf{k}_m)\rangle$ is non-degenerate, i.e., $n'=n$,  the mirror operation becomes a map between the right and left eigenvectors of the same state:
\begin{equation}\label{mirror at exact phase}
      \hat{M}\left|\psi^R_n(\mathbf{k}_m)\right\rangle=\rho_n(\mathbf{k}_m)|\psi^L_{n}(\mathbf{k}_m)\rangle\quad\Leftrightarrow\quad
      \hat{M}\left|\psi^L_n(\mathbf{k}_m)\right\rangle=\frac{1}{\rho_n(\mathbf{k}_m)}|\psi^R_{n}(\mathbf{k}_m)\rangle.
\end{equation} 
Therefore, we know that in the exact phase, the expectations of the mirror operator in a pair of right and left eigenvectors take \textbf{the same nonzero real value}:
\begin{equation}
    \langle \hat{M}\rangle_n=\frac{\langle \psi^R_n(\mathbf{k}_m)|\hat{M}|\psi^R_n(\mathbf{k}_m)\rangle}{\langle \psi^R_n(\mathbf{k}_m)|\psi^R_n(\mathbf{k}_m)\rangle} =\frac{\langle \psi^L_n(\mathbf{k}_m)|\hat{M}|\psi^L_n(\mathbf{k}_m)\rangle}{\langle \psi^L_n(\mathbf{k}_m)|\psi^L_n(\mathbf{k}_m)\rangle}=\rho_n(\mathbf{k}_m) \frac{\langle \psi^R_n(\mathbf{k}_m)|\psi^L_n(\mathbf{k}_m)\rangle}{\langle \psi^R_n(\mathbf{k}_m)|\psi^R_n(\mathbf{k}_m)\rangle}\in \mathbb{R}_{\neq0},
\end{equation}
considering the facts that $\rho_n(\mathbf{k}_m)\neq0$, $\langle\psi_n^R(\mathbf{k}_m)|\psi_n^L(\mathbf{k}_m)\rangle\neq0$, and $\langle\psi_n^R(\mathbf{k}_m)|\psi_n^R(\mathbf{k}_m)\rangle\neq\infty$ in the exact phase on $\Pi_M$. Since the expectation value of a state is invariant against gauge transformation $|\psi\rangle\rightarrow c|\psi\rangle$ for any $c\in\mathbb{C}$, we have
\begin{definition}[\bf \mdagger-parity]
 \textbf{The \mdagger-parity of an eigenstate in the exact phase on the mirror plane} is defined as the sign of the mirror expectation value:
\begin{equation}
    \begin{aligned}
    \tilde{p}_n(\mathbf{k}_m)&=\qquad\qquad\mathrm{sign}\left[\langle \hat{M}\rangle_n\right]\qquad\quad \in \{\pm1\}.\\
    &=\mathrm{sign}\left[\langle \psi^R_n(\mathbf{k}_m)|\hat{M}|\psi^R_n(\mathbf{k}_m)\rangle\right]\\
    &=\mathrm{sign}\left[\langle \psi^L_n(\mathbf{k}_m)|\hat{M}|\psi^L_n(\mathbf{k}_m)\rangle\right]\\
     &=\mathrm{sign}\left[\rho_n(\mathbf{k}_m)\langle \psi^R_n(\mathbf{k}_m)|\psi^L_n(\mathbf{k}_m)\rangle\right],
\end{aligned}
\end{equation}
which generalizes the notion of mirror-parity for the eigenstates of a mirror-symmetric Hamiltonian on mirror planes.
\end{definition}

If the left and right eigenvectors observe the binormalization condition, i.e, $\langle \psi^R_n(\mathbf{k}_m)|\psi^L_n(\mathbf{k}_m)\rangle=1$, we have $\rho_n(\mathbf{k}_m)\in\mathbb{R}_{\neq0}$ and $\tilde{p}_n(\mathbf{k}_m)=\mathrm{sign}[\rho_n(\mathbf{k})]$. As $\tilde{p}_n(\mathbf{k}_m)=\pm1$ is quantized, \textbf{the \mdagger-parity offers a constant index labeling the bands in a continuous region of the exact phase on the mirror plane}.

For any state $|\psi^R_n(\mathbf{k}_m)\rangle$ located in the broken phase on $\Pi_M$, its mirror-partner left eigenvector $|\psi^L_{n'}(\mathbf{k}_m)\rangle=\frac{1}{\rho_n(\mathbf{k}_m)}\hat{M}|\psi^R_n(\mathbf{k}_m)\rangle$ is on a different branch of band (i.e., $n'\neq n$) with the complex-conjugate eigenvalue  $E_{n'}(\mathbf{k}_m)=E_{n}(\mathbf{k}_m)^*\neq E_{n}(\mathbf{k}_m)$. Thus, the mirror expectation value of the states in the broken phase vanishes:
\begin{equation}
    \langle M_z\rangle_n=\frac{\langle \psi^R_n(\mathbf{k}_m)|\hat{M}|\psi^R_n(\mathbf{k}_m)\rangle}{\langle \psi^R_n(\mathbf{k}_m)|\psi^R_n(\mathbf{k}_m)\rangle} =\rho_n(\mathbf{k}_m)\frac{\langle \psi^R_n(\mathbf{k}_m)|\psi^L_{n'}(\mathbf{k}_m)\rangle}{\langle \psi^R_n(\mathbf{k}_m)|\psi^R_n(\mathbf{k}_m)\rangle}=0.
\end{equation}
The null mirror expectation value also holds for the states at 
EPs due to the self-biorthogonality of EP $\langle \psi^R_n(\mathbf{k}_m)|\psi^L_{n'}(\mathbf{k}_m)\rangle =0$. Thus, the \mdagger-parity of a state is well-defined if and only if the state is in the exact phase on the mirror plane.

Consider a \mdagger-symmetric non-Hermitian Hamiltonian $\mathcal{H}(\mathbf{k},\epsilon)=\mathcal{H}_0(\mathbf{k})+\epsilon\,\Delta\mathcal{H}(\mathbf{k})$ perturbed from a mirror-symmetric Hermitian Hamiltonian $\mathcal{H}_0(\mathbf{k})=\hat{M}\mathcal{H}_0(\hat{m}\mathbf{k})\hat{M}$, where $\epsilon>0$ gives a dimensionless parameter measuring the strength of the non-Hermitian perturbation. As $\epsilon$ decreases to 0, an exact eigenstate $|\psi^R(\mathbf{k}_m,\epsilon)\rangle$ of $\mathcal{H}(\mathbf{k}_m,\epsilon)$ on $\Pi_M$ reduces to an eigenvector $|\psi^R_0(\mathbf{k}_m,\epsilon)\rangle$ of the Hermitian Hamiltonian $\mathcal{H}_0(\mathbf{k}_m)$. Correspondingly, the \mdagger-parity, $\tilde{p}(\mathbf{k}_m,\epsilon)$ of $|\psi^R(\mathbf{k}_m,\epsilon)\rangle$ is inherited from the mirror-parity, $p(\mathbf{k}_m)=\tilde{p}(\mathbf{k}_m,0)$, of $|\psi^R_0(\mathbf{k}_m)\rangle$ in the unperturbed Hermitian case. Provided not encountering phase transition, $\tilde{p}(\mathbf{k}_m,\epsilon)$ remains a quantized constant during the evolution of $\epsilon$, therefore \textbf{the \mdagger-parity of a band in the exact phase on the mirror plane should be equal to the mirror-parity of the original band in the unperturbed Hermitian case}.

\subsection{Relation between \mdagger-parity and degeneracies on the mirror plane}
Before analyzing the interaction between \mdagger\ 
symmetry and EPs, we first review some useful properties of the eigenstates and generalized eigenstates of an EP. At an EP of order 2 (EP2) with eigenvalue $E_0$, the right and left eigenstates $|\chi^{R/L}_0\rangle$ and the  right and left generalized eigenstates $|\chi_1^{R/L}\rangle$ satisfy
\begin{equation}
    \begin{aligned}
{\qty(\mathcal{H}_\mathrm{EP}-E_0)} |\chi^R_0\rangle
&=0,\\
{\qty(\mathcal{H}_\mathrm{EP}-E_0)} |\chi^R_1\rangle
&=|\chi^R_0\rangle,
\end{aligned}\qquad
\begin{aligned}
{\qty(\mathcal{H}_\mathrm{EP}^\dagger-E_0^*)} |\chi^L_0\rangle &=0,\\
{\qty(\mathcal{H}_\mathrm{EP}^\dagger-E_0^*)} |\chi^L_1\rangle &=|\chi^L_0\rangle.
\end{aligned}
\end{equation}
where $\mathcal{H}_\mathrm{EP}$ is short for $\mathcal{H}(\mathbf{k}_\mathrm{EP})$.

\begin{property}\label{property at EP}
The eigenvectors and generalized eigenvectors at an EP2 obey the relations
\begin{align}
    \langle\chi^L_0|\chi^R_0\rangle &=0,\label{orthogonal at EP}\\ \langle\chi^L_1|\chi^R_0\rangle & =\langle\chi^L_0|\chi^R_1\rangle\neq0,\label{cross orthogonal at EP}\\
\langle\psi^L_{E'}|\chi^R_i\rangle &=\langle\psi^R_{E'}|\chi^L_i\rangle =0,\quad (i=0,1)\label{cross orthogonal2 at EP}
\end{align}
where $|\psi^{R/L}_{E'}\rangle$ denotes another pair of right and left eigenvectors at $\vb{k}_\mathrm{EP}$ with a different eigenvalue $E'\neq E_0$.
\end{property}
\begin{proof}
\begin{align*}
    \langle\chi^L_0|\chi^R_0\rangle=\langle\chi^L_0|(\mathcal{H}_\mathrm{EP}-E_0)|\chi^R_1\rangle=0\quad \Rightarrow\quad
    \text{Eq.~\eqref{orthogonal at EP}}\hspace{26pt}&\\[3pt]
    \langle\chi^L_1|\chi^R_0\rangle=\langle\chi^L_1|(\mathcal{H}_\mathrm{EP}-E_0)|\chi^R_1\rangle=\langle\chi^L_0|\chi^R_1\rangle\quad\Rightarrow\quad
    \text{the equality in Eq.~\eqref{cross orthogonal at EP}}\hspace{-45pt}&\\[3pt]
    \left.\begin{aligned}
    0=\langle\psi^L_{E'}|(\mathcal{H}_\mathrm{EP}-\mathcal{H}_\mathrm{EP})|\chi^R_0\rangle=(E'-E_0)\langle\psi^L_{E'}|\chi^R_0\rangle\quad \Rightarrow\quad \langle\psi^L_{E'}|\chi^R_0\rangle=0\\
0=\langle\psi^L_{E'}|(\mathcal{H}_\mathrm{EP}-\mathcal{H}_\mathrm{EP})|\chi^R_1\rangle=(E'-E_0)\langle\psi^L_{E'}|\chi^R_1\rangle-\langle\psi^L_{E'}|\chi^R_0\rangle\quad\Rightarrow\quad \langle\psi^L_{E'}|\chi^R_1\rangle=0
\end{aligned}\right\}\ &\Rightarrow\quad \text{Eq.~\eqref{cross orthogonal2 at EP}}
\end{align*}
Similarly, we can prove $\langle\psi^R_{E'}|\chi^L_i\rangle=0\quad (i=0,1)$. This result indicates that $|\chi^L_0\rangle,|\chi^L_1\rangle$ span the 2D orthogonal complement $V_{N-2}^\perp$ of the $(N-2)$-D subspace  $V_{N-2}=\mathrm{span}\qty[|\psi^R_{E_1}\rangle,\cdots,|\psi^R_{E_{N-2}}\rangle]$ ($E_i\neq E_0$) for a $N$-level Hamiltonian.  On the other hand, since the eigenvectors and generalized eigenvectors of $\mathcal{H}_\mathrm{EP}$ are linearly independent, we have
$|\chi^R_0\rangle\not\in V_{N-2}$, implying that the projection of $|\chi^R_0\rangle$ to $V_{N-2}^\perp$ is nonzero:
\begin{equation*}
    \vspace{-15pt}\hspace{50pt}\underbrace{|\langle\chi^L_0|\chi^R_0\rangle|}_{=0}+|\langle\chi^L_1|\chi^R_0\rangle|\neq0\quad\Rightarrow\quad\text{ inequality in Eq.~\eqref{cross orthogonal at EP}}
\end{equation*}

\end{proof}
In general, there are four free complex coefficients, $a_0,a_1,b_0,b_1$, to uniquely fix the left and right eigenvectors and  generalized eigenvectors at an EP2:
\begin{equation}\label{gauge transformation}
    \begin{aligned}
|\chi'^R_0\rangle &\rightarrow a_0|\chi^R_0\rangle\\
|\chi'^R_1\rangle &\rightarrow a_0|\chi^R_1\rangle+a_1|\chi^R_0\rangle
\end{aligned}\qquad
\begin{aligned}
|\chi'^L_0\rangle &\rightarrow b_0|\chi^L_0\rangle\\
|\chi'^L_1\rangle &\rightarrow b_0|\chi^L_1\rangle+b_1|\chi^L_0\rangle
\end{aligned}
\end{equation}
It is safe to introduce two orthonormal conditions consistent with \textbf{Property}~\ref{property at EP} to reduce two undetermined degrees of freedom in the eigenvectors：
\begin{equation}
    \langle \chi^L_1|\chi^R_1\rangle=0,\qquad  \langle\chi^L_1|\chi^R_0\rangle=\langle\chi^L_0|\chi^R_1\rangle=1.\label{normlization1}
\end{equation}

Now, we consider the \mdagger\ symmetry-protected EP2s on the mirror plane. In a \mdagger-symmetric system, the EP2s on the mirror plane can be classified into two classes: (1) real EP2s lying on the transition line of exact and broken phase of two complex-conjugate paired bands; (2) complex EP2s formed by two bands both in the broken phases (which do not form complex-conjugate pair). Here, we only consider the first class, for which the eigenvalue of the EP2 must be real $E_0=E_0^*\in\mathbb{R}$. 

\vspace{20pt}
\begin{property}\label{property}
The eigenvectors and generalized eigenvectors of a \mdagger \ symmetry protected real EP2 on the mirror plane observe the relations:
\begin{gather}
\hat{M}|\chi^R_0\rangle=\rho_0|\chi^L_0\rangle,\qquad \hat{M}|\chi^L_0\rangle=\frac{1}{\rho_0}|\chi^R_0\rangle,\\
\hat{M}|\chi^R_1\rangle=\rho_0\qty(|\chi^L_1\rangle+c|\chi^L_0\rangle),\qquad
\hat{M}|\chi^L_1\rangle=\frac{1}{\rho_0}\qty(|\chi^R_1\rangle-c|\chi^R_0\rangle).\label{M operation on generalize vector}
\end{gather}
Moreover, the normalization condition Eq.~\eqref{normlization1} guarantees that
\begin{enumerate}
    \item[(1)] $\rho_0\neq0$ and $c$ are real numbers;
    \item[(2)] $\mathrm{sign}\qty[\rho_0]$ is invariant
against the gauge transformation~\eqref{gauge transformation} of the $|\chi_i^{R/L}\rangle$ ($i=1,2$).
\end{enumerate}
\end{property} 
\begin{proof}
The proof of the first line for the eigenvectors at the real EP2 is the same as Eq.~\eqref{mirror at exact phase} for the eigenstates in the exact phase. And similar to Eq.~\eqref{norms between left and right}, we know that $\rho_0\neq0$. For the generalized eigenvectors, we have
\begin{equation*}
    (\mathcal{H}_\mathrm{EP}^\dagger-E_0)\hat{M}|\chi^R_1\rangle=\hat{M}(\mathcal{H}_\mathrm{EP}-E_0)|\chi^R_1\rangle=\hat{M}|\chi^R_0\rangle=\rho_0|\chi^L_0\rangle,
\end{equation*}
which indicates that $\hat{M}|\chi^R_1\rangle$ is a left generalized eigenvector given by
$
    \hat{M}|\chi^R_1\rangle=\rho_0\qty(|\chi^L_1\rangle+c|\chi^L_0\rangle)
$ with an undetermined coefficient $c$. Acting $\hat{M}$ on the two sides, we obtain the second equality in Eq.~\eqref{M operation on generalize vector}:
\begin{equation*}
\hat{M}|\chi^L_1\rangle=\frac{1}{\rho_0}|\chi^R_1\rangle-c\,M|\chi^L_0\rangle=\frac{1}{\rho_0}|\chi^R_1\rangle-\frac{c}{\rho_0}|\chi^R_0\rangle.
\end{equation*}
The normalization condition Eq.~\eqref{normlization1} leads to
\begin{gather*}
    1=\langle\chi^L_0|\chi^R_1\rangle=\langle\chi^L_0|\hat{M}\hat{M}|\chi^R_1\rangle=\frac{\rho_0}{\rho_0^*}\underbrace{\langle\chi^R_0|\chi^L_1\rangle}_{=1}+\frac{c\rho_0}{\rho_0^*}\underbrace{\langle\chi^R_0|\chi^L_0\rangle}_{=0}\quad\Rightarrow\quad \rho_0=\rho_0^*\in\mathbb{R},\\
    0=\langle\chi^L_1|\chi^R_1\rangle=\langle\chi^L_1|\hat{M}\hat{M}|\chi^R_1\rangle=\qty(\frac{1}{\rho_0^*}\langle\chi^R_1|-\frac{c^*}{{\rho_0}^*}\langle\chi^R_0|)\qty(\rho_0|\chi^L_1\rangle+c\rho_0|\chi^L_0\rangle)=\frac{c\rho_0}{\rho_0^*}\langle\chi^R_1|\chi^L_0\rangle-\frac{c^*\rho_0}{{\rho_0}^*}\langle\chi^R_0|\chi^L_1\rangle\nonumber\\
\Rightarrow\quad \frac{c\rho_0}{\rho_0^*}=\frac{c^*\rho_0}{\rho_0^*}\quad\Rightarrow\quad c=c^*\in\mathbb{R}.
\end{gather*}
Since $\rho_0=\bra{\chi^R_1}\hat{M}\ket{\chi_0^R}$ under the condition Eq.~\eqref{normlization1}, the gauge transformation~\eqref{gauge transformation} yields
\begin{equation*}
    \rho'_0=\bra{\chi'^R_1}\hat{M}\ket{\chi'^R_0}=\qty(\bra{\chi_1^R}a_0^*+\bra{\chi_0^R}a_1^*)\hat{M}\qty(a_0\ket{\chi_0^R})=\qty(\bra{\chi_1^R}a_0^*+\bra{\chi_0^R}a_1^*)\qty(a_0\rho_0\ket{\chi_0^L})=|a_0|^2\rho_0.
\end{equation*}
Therefore, $\mathrm{sign}[\rho'_0]=\mathrm{sign}[\rho_0]$ confirms that the sign of $\rho_0$ is gauge independent under the constraint of Eq.~\eqref{normlization1}.
\end{proof}
Starting at a real EP2 on the transition boundary between exact and broken phases on the mirror plane, the perturbation of the two branches of eigenvalues and eigenstates bifurcating from the EP2 can be explicitly expressed as the \textbf{Puiseux series} (Theorem 2.3 in Ref.~\cite{Seyranian2003Multiparameter} and also see Refs.\cite{mailybaev2005Geometric,seyranian2005Coupling}):
\begin{align}
E_\pm(\mathbf{k}_\mathrm{EP}+\delta\mathbf{k}) &=E_0\pm\sqrt{\mu(\delta\hat{\mathbf{k}})}\,\delta k^{1/2}+\mathcal{O}(\delta k),\\\label{expansion1}
|\psi^R_\pm(\mathbf{k}_\mathrm{EP}+\delta\mathbf{k})\rangle &=|\chi^R_0\rangle\pm\sqrt{\mu(\delta\hat{\mathbf{k}})}\,\delta k^{1/2}|\chi^R_1\rangle+\mathcal{O}(\delta k),\\
|\psi^L_\pm(\mathbf{k}_\mathrm{EP}+\delta\mathbf{k})\rangle &=|\chi^L_0\rangle\pm\sqrt{\mu(\delta\hat{\mathbf{k}})^*}\,\delta k^{1/2}|\chi^L_1\rangle+\mathcal{O}(\delta k),\label{expansion2}
\end{align}
with
\begin{equation}
    \mu(\delta\hat{\mathbf{k}})=\langle\chi_0^L|\nabla_{\mathbf{k}}\mathcal{H}(\mathbf{k}_\mathrm{EP})|\chi_0^R\rangle\cdot\delta\hat{\mathbf{k}}=\sum_{i}\langle\chi_0^L|\partial_{k_i}\mathcal{H}(\mathbf{k}_\mathrm{EP})|\chi_0^R\rangle \delta k_i/\delta k,
\end{equation}
where $\delta\hat{\mathbf{k}}$ denotes the unit vector in the direction of $\delta\mathbf{k}$. $\mu(\delta\hat{\mathbf{k}})\neq0$ is usually satisfied providing that $\delta\mathbf{k}$ is not along the tangent direction of the EP line.  Note that in the derivation of the eigenvectors' expansions, the normalization conditions $\langle\chi_1^L|\psi^R_\pm(\mathbf{k}_\mathrm{EP}+\delta\mathbf{k})\rangle\equiv1$ and $\langle\chi_1^R|\psi^L_\pm(\mathbf{k}_\mathrm{EP}+\delta\mathbf{k})\rangle\equiv1$, which are consistent with Eq.~$\eqref{normlization1}$ at the EPs, have been imposed.

\begin{figure*}[t!]
\includegraphics[width=0.75\textwidth]{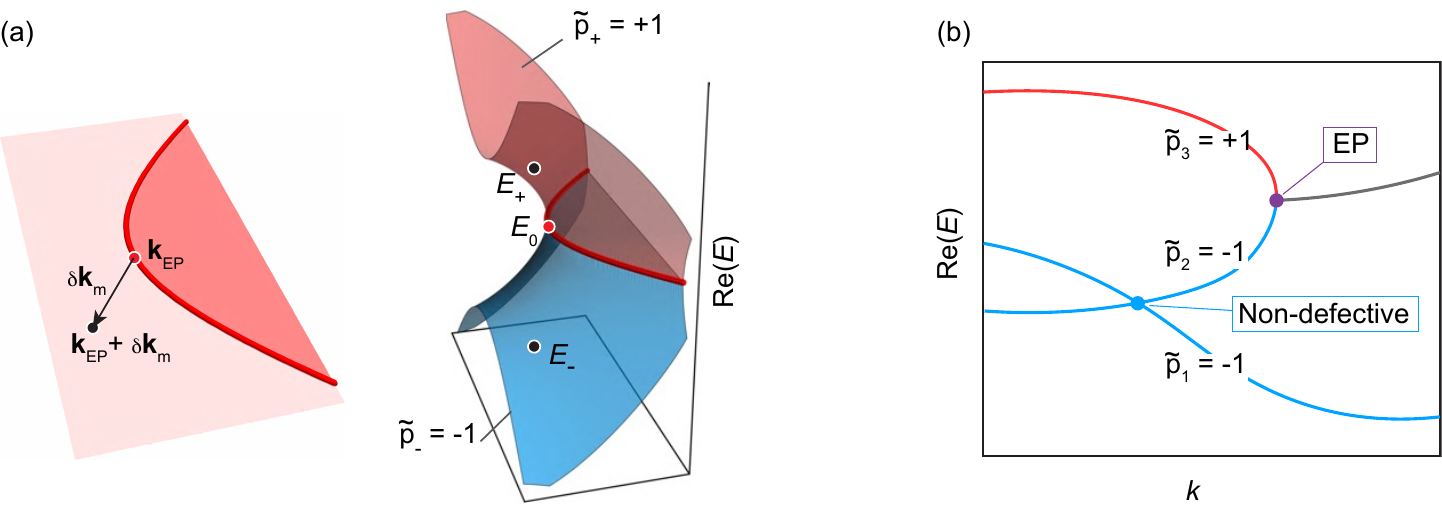}
\caption{\label{fig-Proof-MDagger} 
(a) Schematic of the perturbation near a \mdagger-protected real EP on the mirror plane. (b) Schematic of the \mdagger-protected band structure along a line on the mirror plane, where two upper bands with opposite \mdagger-parities in the exact phase are degenerate at an EP, while lower two bands with the same \mdagger-parity can be only degenerate at a non-defective point. Note that this non-defective degeneracy is generally accidental in the absence of other symmetry protection. }
\end{figure*}

As shown in Fig. \ref{fig-Proof-MDagger} (a), if $\delta\mathbf{k}=\delta\mathbf{k}_m$ is in the mirror plane, we have
\begin{equation}
    \langle\chi_0^L|\nabla_{\mathbf{k}}\mathcal{H}(\mathbf{k}_\mathrm{EP})|\chi_0^R\rangle=\langle\chi_0^L|\hat{M}\nabla_{\mathbf{k}}\mathcal{H}(\mathbf{k}_\mathrm{EP})^\dagger \hat{M}|\chi_0^R\rangle=\langle\chi_0^R|\nabla_{\mathbf{k}}\mathcal{H}(\mathbf{k}_\mathrm{EP})^\dagger|\chi_0^L\rangle=\langle\chi_0^L|\nabla_{\mathbf{k}}\mathcal{H}(\mathbf{k}_\mathrm{EP})|\chi_0^R\rangle^*\in\mathbb{R}.
\end{equation}
Hence, $\mu(\delta\hat{\mathbf{k}}_m)\in\mathbb{R}$ indicates that the eigenvalues $E_\pm(\mathbf{k}_\mathrm{EP}+\delta\mathbf{k}_m)$ near the EP2 should either be real if $\mu(\delta\hat{\mathbf{k}}_m)>0$ or form complex conjugate pairs if $\mu(\delta\hat{\mathbf{k}}_m)<0$.

Furthermore, we arrive at the following theorem and corollary:
\begin{theorem}[\bf \mdagger-symmetry protected ELs]\label{theorem-mdagger}
A \mdagger-symmetry protected order-2 exceptional line on the mirror plane $\Pi_M$ can only be formed by two bands with \textbf{opposite \mdagger-parities} in the nearby exact phase on $\Pi_M$.
\end{theorem}

The contrapositive of the theorem leads to
\begin{corollary}\label{corollary-mdagger}
In a \mdagger-symmetric system, if two bands have the same \mdagger-parity in the exact phase on the mirror plane, their degeneracies adjacent to the exact phase must be \textbf{non-defective}.
\end{corollary}

\begin{proof}
In terms of the expansion in Eqs.~(\ref{expansion1},\ref{expansion2}), we can directly compute the \mdagger-parities of the two branches of eigenstates in the exact phase in the vicinity of the EP:
\begin{equation*}
    \begin{split}
&\langle\psi^R_\pm(\mathbf{k}_\mathrm{EP}+\delta\mathbf{k}_m)|\hat{M}|\psi^R_\pm(\mathbf{k}_\mathrm{EP}+\delta\mathbf{k}_m)\rangle\\
=&\qty(\langle\chi_0^R|\pm\sqrt{\mu(\delta\hat{\mathbf{k}}_m)}\delta k^{1/2}_m\langle\chi_1^R|)\hat{M}\qty(|\chi_0^R\rangle\pm\sqrt{\mu(\delta\hat{\mathbf{k}}_m)}\,\delta k^{1/2}_m|\chi_1^R\rangle)+\mathcal{O}(\delta k_m)\\
=&\qty(\langle\chi_0^R|\pm\sqrt{\mu(\delta\hat{\mathbf{k}}_m)}\delta k^{1/2}_m\langle\chi_1^R|)\qty[\qty(\rho_0\pm c\rho_0\sqrt{\mu(\delta\hat{\mathbf{k}}_m)}\delta k^{1/2}_m)|\chi_0^L\rangle\pm\rho_0\sqrt{\mu(\delta\hat{\mathbf{k}}_m)}\,\delta k^{1/2}_m|\chi_1^L\rangle]+\mathcal{O}(\delta k_m)\\
=& \pm\rho_0\sqrt{\mu(\delta\hat{\mathbf{k}}_m)}\,\delta k^{1/2}_m\qty(\langle\chi_0^R|\chi_1^L\rangle+\langle\chi_1^R|\chi_0^L\rangle)+\mathcal{O}(\delta k_m)\\
=& \pm2\rho_0\sqrt{\mu(\delta\hat{\mathbf{k}}_m)}\,\delta k^{1/2}_m+\mathcal{O}(\delta k_m),
\end{split}\nonumber
\end{equation*}
where $\rho_0\in\mathbb{R}_{\neq0}$ and $\mu(\delta\hat{\mathbf{k}}_m)>0$ in the exact phase have been used.
Therefore, the two branches of the eigenstates coalescing at the EP2 always have opposite \mdagger-parities, $\tilde{p}_\pm=\mathrm{sign}\qty[\langle\psi^R_\pm(\mathbf{k}_\mathrm{EP}+\delta\mathbf{k}_m)|\hat{M}|\psi^R_\pm(\mathbf{k}_\mathrm{EP}+\delta\mathbf{k}_m)\rangle]=\pm\mathrm{sign}[\rho_0]$. And according to \textbf{Property}~\ref{property}, we know that $\mathrm{sign}[\rho_0]$ and hence the mirror parities of the two bands are irrespective to the gauge of the eigenvectors and generalized eigenvectors $|\chi^{R/L}_i\rangle$ ($i=1,2$).
\end{proof}

As shown in Fig.~\ref{fig-Proof-MDagger} (b), \textbf{Theorem}~\ref{theorem-mdagger} reveals that \textbf{an orthogonal EC protected by two perpendicular \mdagger\ symmetries can only be formed by two bands with opposite \mdagger-parties on both mirror planes.} 
And for a planar EC protected by two perpendicular \mdagger\ symmetries (see Fig.~2(c) in the main text), since the two bands forming the chain have opposite \mdagger-parities on the horizontal plane ($\Pi_1$) while they have an identical \mdagger-parity on the vertical plane ($\Pi_2$),  \textbf{Corollary~\ref{corollary-mdagger} ensures that the non-defective chain point cannot expand to an EP ring in $\Pi_2$}.



\vspace{30pt}

\section{Proof of quantized Berry phases protected by non-Hermitian spatiotemperal symmetries}

In this section, we will prove the quantization of different kinds of Berry phases protected by four types of two-fold DAS point-group symmetries, $G$, \gdagger, $G\mathcal{T}$ and \gtdagger,
which are summarized in Table II of the main text.  
Note that to guarantee the Berry phases are well-defined, namely, the results are identical in the sense of modulo $2\pi$ for any continuous gauge of the eigenvectors along $\Gamma$, we require that the continuous band concerned is \textbf{self-closed}, namely the eigenstate returns to the initial one after travelling on the concerned band along the loop one turn.

\vspace{-10pt}
\subsection{Quantized Berry phases protected by $G$ symmetry}
\begin{figure*}[t!]
\includegraphics[width=0.87\textwidth]{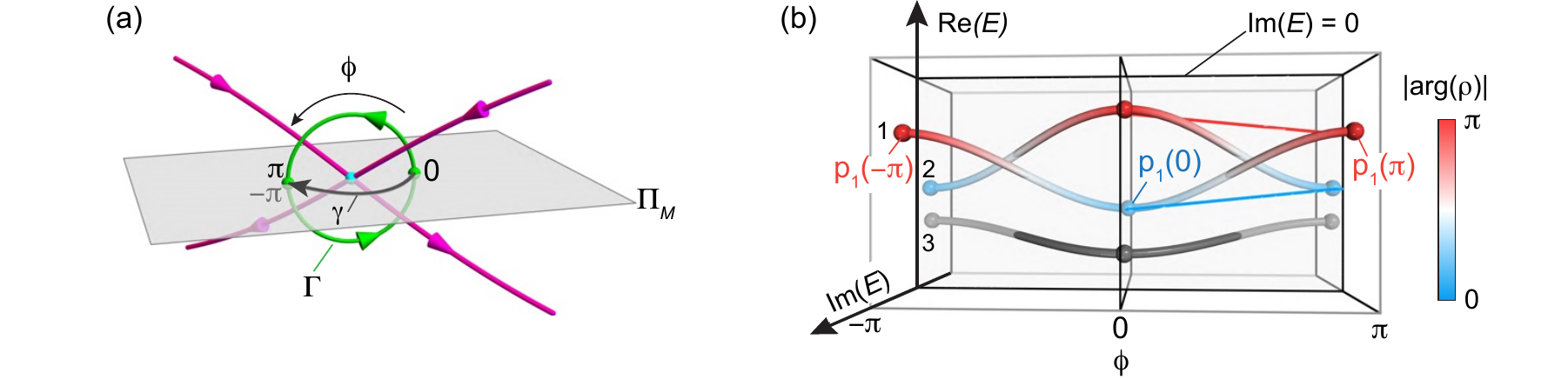}
\caption{\label{fig-MirrorBP} 
(a) Schematic of a mirror-symmetric (green) loop $\Gamma=\hat{m}\Gamma^{-1}$ that intersects with the mirror plane $\Pi_M$ at two $M$-invariant points $\phi=0,\pm \pi$ and encircles a  mirror symmetry protected EC. (b) Schematic of the non-degenerate band structure along the mirror-symmetric loop $\Gamma$, where the bands $1,2,3$ are ordered by their real parts according to their real parts at $\phi=-\pi$. The presence of mirror ($M$) symmetry ensures that the complex bands are mirror-symmetric about the planes at $\phi=0,\pi$. This spectral symmetry guarantees every non-degenerate band along $\Gamma$ to be self-closed. The color of the tubes represents the norm of the phase of $\rho(\phi)$, which are also mirror-symmetric according to Eq.~\eqref{G constraint 1}. The red and blue thin lines in (b) denote the two bands with constant $M$-parities along the in-plane path $\gamma$ in (a)}
\end{figure*}

\vspace{-5pt}We consider a Hamiltonian $\mathcal{H}(\mathbf{k})$ with a twofold point-group symmetry $G$, i.e., $\mathcal{H}(\mathbf{k})=\hat{G}\mathcal{H}(\hat{g}\mathbf{k})\hat{G}^{\dagger}$, and a $G$-symmetric loop 
\begin{equation}\label{G-symmetric loop}
    \Gamma=\{\mathbf{k}(\phi)\,|\,\hat{g}\mathbf{k}(\phi)=\mathbf{k}(-\phi)\ \text{and}\ \mathbf{k}(\pi)=\mathbf{k}(-\pi), -\pi<\phi\leq \pi\},
\end{equation}
which intersects with the $G$-invariant subspace $\Pi_G=\qty{\mathbf{k}_g\,|\,\hat{g}\mathbf{k}_g=\mathbf{k}_g}$ at $\phi=0$ and $\pi$. Obviously, transforming the loop with a $G$ operation, the shape of the loop remains unchanged, while the direction of transformed loop $\hat{g}\Gamma = \qty{\mathbf{k}'(\phi)=\hat{g}\mathbf{k}(\phi)\,|\,\mathbf{k}(\phi)\in\Gamma,\ -\pi<\phi\leq\pi}$ is reversed:
    $\hat{g} \Gamma = \Gamma^{-1}$,
where the superscript $-1$ denotes the direction reversal.
In the $G$-invariant subspace $\Pi_G$, any non-degenerate eigenstate $\ket{\psi^{R/L}(\mathbf{k}_g)}$ of the Hamiltonian is also an eigenstate of $\hat{G}$, whose eigenvalues are denoted as the $G$-parity of the states,
\begin{equation}\label{G-parity}
    \hat{G}\ket{\psi^{R/L}(\mathbf{k}_g)}= p_G(\mathbf{k}_g)\, \ket{\psi^{R/L}(\mathbf{k}_g)}.
\end{equation}
where $p_G=\pm1$ because of $\hat{G}^2=I$, and the left and right eigenvectors possess the same $G$-parity since $p_G^L(\mathbf{k}_g)=\bra{\psi^L(\mathbf{k}_g)}\hat{G}\ket{\psi^R(\mathbf{k}_g)}/\braket{\psi^L(\mathbf{k}_g)}{\psi^R(\mathbf{k}_g)}=p_G^R(\mathbf{k}_g)$.
For an eigenvector $\ket{\psi^{R/L}(\mathbf{k})}$ of the $G$-symmetric Hamiltonian, 
\begin{equation}\label{G correlation}
    \hat{G}\ket{\psi^{R/L}(\mathbf{k})}= \rho^{R/L}(\mathbf{k}) \ket{\psi^{R/L}(\hat{g}\mathbf{k})}.
\end{equation}
is proportional to the eigenvector $\ket{\psi^{R/L}(\hat{g}\mathbf{k})}$ at the $G$-partner point $\hat{g}\mathbf{k}$ of $\mathbf{k}$, where the coefficients $\rho^{R/L}(\mathbf{k})\neq0$ satisfy 
\begin{equation}\label{G constraint 1}
    \rho^{R/L}(\mathbf{k})=1/\rho^{R/L}(\hat{g}\mathbf{k}).
\end{equation}
Inside $\Pi_G$, Eqs.~(\ref{G-parity},\ref{G correlation}) lead to 
\begin{equation}\label{G constraint 2}
    \rho^R(\mathbf{k}_g)=\rho^L(\mathbf{k}_g)=p_G(\mathbf{k}_g)\in\qty{\pm1}.
\end{equation}
Importantly, we know that (see the example of a mirror-symmetric loop encircling a mirror-protected EC in Fig.~\ref{fig-MirrorBP}.)
\begin{property}\label{property of G}
Along a $G$-symmetric loop satisfying $\Gamma=\hat{g}\Gamma^{-1}$, provided that all bands  $E_n(\phi)$ are non-degenerate everywhere along $\Gamma$, the $G$ symmetry ensures that
\begin{enumerate}
    \item[(1)] the eigenenergies $E_n(\phi)$ along the loop $\Gamma$ are symmetric about the two  $G$-invariant points $\mathbf{k}(0)$ and $\mathbf{k}(\pi)$, \vspace{-6pt}
    \item[(2)] every continuous band of $E_n(\phi)$ with eigenvectors $\ket{\psi^{R/L}_n(\phi)}$ is \textbf{self-closed} for $-\pi<\phi\leq\pi$.
\end{enumerate}
\end{property}

The second property immediately implies that \textbf{any pair of states $\ket{\psi^{R/L}(\mathbf{k})}$ and $\ket{\psi^{R/L}(\hat{g}\mathbf{k})}\propto\hat{G}\ket{\psi^{R/L}(\mathbf{k})}$ at two $G$-partner points on $\Gamma$ are on the same self-closed band.} The self-closeness of non-degenerate bands guarantees the single-band Berry phase along a $G$-symmetric loop is well-defined. Indeed, we can prove that


\begin{theorem}[\bf $G$-protected quantized Berry phase]\label{Thm-G-protected Berry}
For a non-degenerate band along a $G$-symmetric loop $\Gamma$ satisfying $\hat{g}\Gamma=\Gamma^{-1}$, the twofold point-group symmetry $G$ of the Hamiltonian guarantees that all the four types of Berry phases of the band are consistently quantized modulo $2\pi$:
\begin{equation}
    \theta^{LL}(\Gamma)=\theta^{RR}(\Gamma)=\theta^{LR}(\Gamma)=\theta^{RL}(\Gamma)=(0\ \mathrm{or}\ \pi)\ \mathrm{mod}\ 2\pi.
\end{equation}
The quantized value is determined by the $G$-parities, $p_G(0),p_G(\pi)$, of the two eigenstates at the two $G$-invariant points $\mathbf{k}(0)$ and $\mathbf{k}(\pi)$ on $\Gamma$:
\begin{equation}\label{G-protected berry phase factor}
    \exp[i\theta^{LL}(\Gamma)]=\exp[i\theta^{RR}(\Gamma)]=\exp[i\theta^{LR}(\Gamma)]=\exp[i\theta^{RL}(\Gamma)]={p}_G(0){p}_G(\pi)\in\qty{\pm1}.
\end{equation}
\end{theorem}

\begin{proof}
The eigenvectors for calculating a specific type of Berry connection $A^{\alpha\beta}(\mathbf{k})={-i\bra{ {\psi}_n^\alpha(\mathbf{k})}\nabla_\mathbf{k}\ket{{\psi}_n^\beta(\mathbf{k})}},\qquad (\alpha,\beta\in\qty{L,R})$, should be normalized by the condition $\braket{{\psi}^{\alpha}_n(\mathbf{k})}{{\psi}^{\beta}_n(\mathbf{k})}=1$, yielding that
\begin{equation}\label{G constraint 3}
 1=\braket{\psi_n^\alpha(\mathbf{k})}{\psi_n^\beta(\mathbf{k})}=\bra{\psi_n^\alpha(\mathbf{k})}\hat{G}\hat{G}\ket{\psi_n^\beta(\mathbf{k})}=\rho_n^{\alpha}(\mathbf{k})^*\rho_n^\beta(\mathbf{k})\quad\Rightarrow\quad
 {\rho}_n^{\alpha}(\mathbf{k})=\frac{1}{{\rho}_n^{\beta}(\mathbf{k})^*}.
\end{equation}
Equations~(\ref{G constraint 1},\ref{G constraint 2},\ref{G constraint 3}) give the three constraints on $\rho_n^{\alpha/\beta}(\mathbf{k})$ along the $G$-symmetric self-closed band.
The the $G$-symmetry also connects the Berry connections at two $G$-partner points on $\Gamma$: 
\begin{equation*}
\begin{split}
A^{\alpha\beta}(\hat{g}\mathbf{k})=&-i\langle \psi_n^\alpha(\hat{g}\mathbf{k})|\nabla_{\hat{g}\mathbf{k}}|\psi_n^\beta(\hat{g}\mathbf{k})\rangle\\
=&-i\langle \psi_n^\alpha(\mathbf{k})|\hat{G}\frac{1}{\rho^\alpha_n(\mathbf{k})^*}\nabla_{\hat{g}\mathbf{k}}\,\frac{1}{\rho^\beta_n(\mathbf{k})}\hat{G}|\psi_n^\beta(\mathbf{k})\rangle\\
=&-i\langle \psi_n^\alpha(\mathbf{k})|\frac{1}{\rho^\alpha_n(\mathbf{k})^*}\Big(\underbrace{\frac{\partial\mathbf{k}}{\partial(\hat{g}\mathbf{k})}}_{=\hat{g}^{-1}=\hat{g}}\cdot\nabla_{\mathbf{k}}\Big)\rho^\alpha_n(\mathbf{k})^*|\psi_n^\beta(\mathbf{k})\rangle\\
=&\ \hat{g}\Big[-i\langle\psi^\alpha_n(\mathbf{k})|\nabla_\mathbf{k}|\psi_n^\beta(\mathbf{k})\rangle-i\nabla_\mathbf{k}\ln\rho^\alpha_n(\mathbf{k})^*\Big]\\
=&\ \hat{g}\Big[A^{\alpha\beta}(\mathbf{k})-i\nabla_\mathbf{k}\ln\rho_n^\alpha(\mathbf{k})^*\Big].
\end{split}
\end{equation*}

And the Berry phases along $\Gamma$ can be calculated as 

\begin{equation*}
\begin{split}
\theta^{\alpha\beta}(\Gamma)=\oint_{\Gamma}A^{\alpha\beta}(\mathbf{k})\cdot\,d\mathbf{k}=&\oint_{\Gamma}\big[\hat{g}A^{\alpha\beta}(\hat{g}\mathbf{k})+i\nabla_\mathbf{k}\ln(\rho^\alpha_n(\mathbf{k})^*)\big]\cdot\,d\mathbf{k}\\
=& \oint_{\Gamma} A^{\alpha\beta}(\hat{g}\mathbf{k})\cdot d(\hat{g}\mathbf{k})+i\oint_{\Gamma} d\ln(\rho_n^\alpha(\mathbf{k})^*)\\
=&\oint_{\hat{g}\Gamma=\Gamma^{-1}}A^{\alpha\beta}(\mathbf{k})\cdot\,d\mathbf{k}
+i\oint_{\Gamma}d\big(\underbrace{\ln|\rho^\alpha_n(\mathbf{k})|}_{\text{single valued}}-i\arg(\rho^\alpha_n(\mathbf{k}))\big)\\[-10pt]
=&-\theta^{\alpha\beta}(\Gamma)+\oint_{\Gamma} d\arg(\rho^\alpha_n(\mathbf{k})),
\end{split}
\end{equation*}
where $\oint_\Gamma d\ln\abs{\rho^\alpha_n(\mathbf{k})}=0$  in the penultimate step is due to the self-closeness of the non-degenerate band along $\Gamma$ (see Property~\ref{property of G}).  Therefore, we obtain
\begin{equation*}
    \theta^{\alpha\beta}(\Gamma)= \frac{1}{2}\oint_{\Gamma}d\arg(\rho^\alpha_n(\mathbf{k}))= \frac{1}{2} \int^{\pi}_{-\pi}d\arg(\rho^\alpha_n(\phi))=\int_0^{\pi}d\arg(\rho^\alpha_n(\phi))=\left[\arg(\rho^\alpha_n(\pi))-\arg(\rho^\alpha_n(0))\right] \bmod 2\pi,
\end{equation*}
where the third equality is because $\arg(\rho^\alpha_n(\phi))=-\arg(\rho^\alpha_n(-\phi)\ \mathrm{mod}\ 2\pi$ is an odd function of $\phi$ according to Eq.~\eqref{G constraint 1}.
Meanwhile, Eq.~\eqref{G constraint 2} shows that $\rho^\alpha_n(0)\equiv p_G(0)$ and $\rho^\alpha_n(\pi)\equiv p_G(\pi)$ always hold for any $\alpha,\beta\in\qty{R,L}$. As a result, the four types of Berry phases are consistently equal to the same quantized value determined by the $G$-parities, $p_G(0),p_G(\pi)\in\qty{\pm1}$, of the two states at the two $G$-invariant points $\mathbf{k}(0),\mathbf{k}(\pi)$:
\begin{equation*}
    \theta^{LL}(\Gamma)=\theta^{RR}(\Gamma)=\theta^{LR}(\Gamma)=\theta^{RL}(\Gamma)=\left[\arg(p_G(\pi))-\arg(p_G(0))\right] \bmod 2\pi=(0\ \mathrm{or}\ \pi)\ \mathrm{mod}\ 2\pi,
\end{equation*}
which can be equivalently expressed as Eq.~\eqref{G-protected berry phase factor}.
\end{proof}

\subsubsection{Example of mirror-symmetry protected Berry phases}
As a direct application of \textbf{Theorem}~\ref{Thm-G-protected Berry}, a mirror symmetry ($M$) can protect the quantization of the Berry phases of a band along a  $M$-symmetric loop traversing the mirror plane $\Pi_M$, $\Gamma=\hat{m}\Gamma^{-1}$, as depicted in Fig.~\ref{fig-MirrorBP}(a):
\begin{equation}
    \exp\qty[i\theta^{LL}(\Gamma)]=\exp\qty[i\theta^{RR}(\Gamma)]=\exp\qty[i\theta^{LR}(\Gamma)]=\exp\qty[i\theta^{RL}(\Gamma)]=p_1(0)p_1(\pi),
\end{equation}
where $p_1(0),\,p_1(\pi)$ denotes the mirror parities of the first band (in Fig.~\ref{fig-MirrorBP}(b)) at the intersections of the loop and $\Pi_M$ (green spheres in Fig.~\ref{fig-MirrorBP}(b)). 
If the loop encloses a $M$-symmetric EC (magenta tubes in Fig.~\ref{fig-MirrorBP}(a)), the two bands (gradient color tubes in Fig.~\ref{fig-MirrorBP}(b)) forming the EC must have opposite $M$-parities on the plane $\Pi_M$, i.e., $p_1(\phi)p_2(\phi)<0$ for $\phi=0,\pi$ (see Section ``Mirror-symmetric exceptional chains'' of the main text). Along an in-plane path $\gamma$, the parities of the bands are well-defined constants, as shown by the red ($p(\phi)=-1, \phi\in[0, \pi]$) and blue ($p(\phi)=1, \phi\in[0, \pi]$) thin lines in Fig.~\ref{fig-MirrorBP}(b). 
Furthermore, we consider the closed loop $c=\gamma\circ\Gamma_\mathrm{semi-low}$ combined by $\gamma$ and the lower-half of the green loop $\Gamma_\mathrm{semi-low}$, which encircles the EL directed to bottom right. Thus, starting at $E_{1,2}(-\pi)$,  the two bands along $c=\gamma\circ\Gamma_\mathrm{semi-low}$ braid once and undergoes the evolutions $E_1(-\pi)\rightarrow E_1(0)\rightarrow E_2(\pi)$ and $E_2(-\pi)\rightarrow E_2(0)\rightarrow E_1(\pi)$, respectively.  And since the evolution along $\gamma$ do not change the $M$-parities of the bands, we acquire that
\begin{equation}
    p_1(0)=p_2(\pi),\quad p_2(0)=p_1(\pi)\qquad\Rightarrow\qquad p_1(0)p_1(\pi)=p_2(0)p_1(\pi)=p_1(0)p_2(0)=-1,
\end{equation}
therefore bringing  about nontrivial Berry phases $\exp\qty[i\theta^{\alpha\beta}(\Gamma)]=-1$ of the two bands.

\subsection{Quantized Berry phase protected by $G\mbox{-}\dagger$ symmetry}

In this section, we consider a Hamiltonian respecting a twofold \gdagger-symmetry: $\hat{G}\mathcal{H}(\hat{g}\mathbf{k})^\dagger\hat{G}=\mathcal{H}(\mathbf{k})$. Indeed, most properties of a mirror-$\dagger$ symmetry derived in Section~\ref{section-mirror-dagger} can be directly generalized to any twofold \gdagger-symmetric system, which are listed as follows:
A Hamiltonian with a twofold \gdagger-symmetry has properties:
\begin{enumerate}
    \item[(1)] $\hat{G}$ maps a right/left eigenvector $\ket{\psi^{R/L}(\mathbf{k})}$ of $\mathcal{H}(\mathbf{k})$ at $\mathbf{k}$ to a left/right eigenvector $\ket{\psi^{L/G}(\hat{g}\mathbf{k})}$  of $\mathcal{H}(\hat{g}\mathbf{k})$ at $\hat{g}\mathbf{k}$:
    \begin{equation}\label{G-dagger map}
    \hat{G}\ket{\psi^{R/L}(\mathbf{k})}=\rho^{R/L}(\mathbf{k}) \ket{\psi^{L/G}(\hat{g}\mathbf{k})},
\end{equation}
whose eigenvalues form complex conjugate pair. Note that two $G$-partner states may \textbf{not} be on the same band.
    \item[(2)] In the $G$-invariant subspace $\Pi_G$, $\mathcal{H}(\mathbf{k}_g)$ is pseudo-Hermitian, and hence the eigenstates are either in the exact phases with real eigenvalues or in the broken with wherein the eigenvalues form complex conjugate pairs. In the exact phase in $\Pi_G$, the right and left eigenvectors $\ket{\psi^{R/L}(\mathbf{k}_g)}$ of the same state satisfies
    \begin{equation}
        \hat{G}\ket{\psi^{R/L}(\mathbf{k}_g)}=\rho^{R/L}(\mathbf{k}_g) \ket{\psi^{L/G}(\mathbf{k}_g)},
    \end{equation}
with $\rho^R(\mathbf{k}_g)=1/\rho^L(\mathbf{k}_g)\neq 0\ \text{or}\ \infty$.
\item[(3)] In the exact phase in $\Pi_G$, the \gdagger-parity of an eigenstate can be defined as
\begin{equation}
    \tilde{p}_G(\mathbf{k}_g)=\mathrm{sign}\qty[\bra{\psi^R(\mathbf{k}_g)}\hat{G}\ket{\psi^R(\mathbf{k}_g)}]=\mathrm{sign}\qty[\bra{\psi^L(\mathbf{k}_g)}\hat{G}\ket{\psi^L(\mathbf{k}_g)}]\in\qty{\pm1},
\end{equation}
which forms a quantized invariant of a band in a continuous region of exact phase.
\end{enumerate}

\begin{figure*}[t!]
\includegraphics[width=0.9\textwidth]{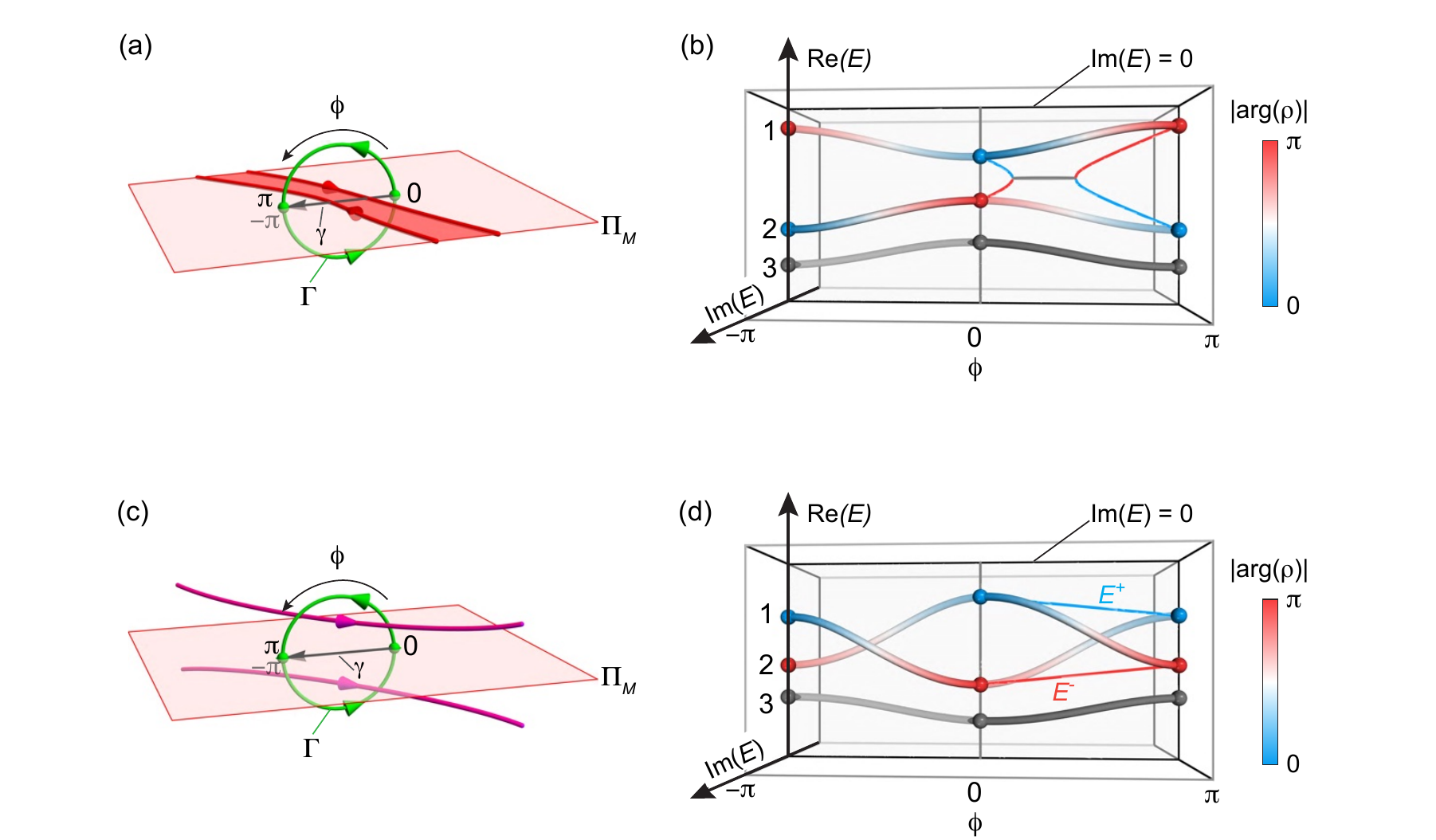}
\caption{\label{fig-MDagger-QuantizedBP} 
(a,c) Schematics of a mirror-symmetric (green) loop $\Gamma=\hat{m}\Gamma^{-1}$ that intersects with the mirror plane $\Pi_M$ at two $M$-invariant points $\phi=0,\pm \pi$ and encircles (a) two in-plane ELs in $\Pi_M$ or (c) two \mdagger-partner ELs at the two sides of $\pi_M$. (b,d) Schematics of the non-degenerate band structure along the loop $\Gamma$ in (a,c), respectively, where the bands $1,2,3$ are ordered by their real parts at $\phi=-\pi$. The presence of \mdagger\ symmetry ensures that the complex bands are $C_2$-symmetric about the lines at $(\phi=0,\Im[E]=0)$ and $(\phi=\pi,\Im[E]=0)$. The colors on bands 1,2 represent the norm of the phase of $\rho(\phi)$, which determines the \mdagger-parities $\tilde{p}(\phi_m)=\rho(\phi_m)=\pm1$ at $\phi_m=0,\pi$. The red and blue thin lines in (b,d) denote the bands $E^\pm$ along the in-plane path $\gamma$ in (a,c), whose colors in exact phases denote their \mdagger-parities. }
\end{figure*}

In addition, \gdagger\ symmetry may also ensure the self-closeness of a band along a loop (see examples in Fig.~\ref{fig-MDagger-QuantizedBP}):
\begin{property}\label{property of Gdagger self-closed band}
    For a continuous  band of  eigenstates $\ket{\psi^{R/L}_n(\phi)}$ ($-\pi\le\phi\le\pi$) along a $G$-symmetric loop $\Gamma=\hat{g}\Gamma^{-1}$ given by Eq.~\eqref{G-symmetric loop}, 
    provided that
\begin{enumerate}
    \item[(1)] the band remains non-degenerate along the loop $\Gamma$ ,\vspace{-4pt}
    \item[(2)] the eigenstates at the two $G$-invariant points $\mathbf{k}(0),\mathbf{k}(\pi)\in\Pi_G$ on $\Gamma$ both lie in the exact phase, namely, $E_n(0),E_n(\pm\pi)\in\mathbb{R}$ and $\hat{G}|\psi_n^{R}(\phi_g)\rangle=\rho(\phi_g)|\psi_n^{L}(\phi_g)\rangle$  ($\phi_g\in\{0,\pm\pi\}$),
\end{enumerate}
the \gdagger\ symmetry can guarantee the band is \textbf{self-closed}, i.e., $E_n(-\pi)=E_n(\pi)$ and $\ket{\psi^{R/L}_n(-\pi)}=\ket{\psi^{R/L}_n(\pi)}$.
\end{property}

Thanks to the self-closeness of the band along $\Gamma$, Eq.~\eqref{G-dagger map} is applicable to any pair of eigenstates on the band at two $G$-partner points $\mathbf{k}(\phi)$ and $\hat{g}\mathbf{k}(\phi)=\mathbf{k}(-\phi)$:
$
    \hat{G}\left|\psi^{R/L}_n(\phi)\right\rangle=\rho^{R/L}_n(\phi)\,\left|\psi^{L/R}_{n}(-\phi)\right\rangle
$, which leads to
\begin{equation}\label{Gdagger relaton 1}
    \rho_n^{R}(\mathbf{k})=1/\rho_n^{L}(\hat{g}\mathbf{k})\quad\Rightarrow\quad\rho_n^{R}(\phi)=1/\rho_n^{L}(-\phi).
\end{equation}


We have the following theorem for the quantized Berry phase protected by \gdagger\ symmetry:
\begin{theorem}[\bf \gdagger-protected quantized Berry phase]\label{thm_Gdagger Berry phase}
Provided that a band of eigenstates observe the above two conditions in Property~\ref{property of Gdagger self-closed band} along a $G$-symmetric loop $\Gamma=\hat{g}\Gamma^{-1}$, the \gdagger \ symmetry guarantees the quantization for the Berry phases of the self-closed band:
\begin{equation}
    \mathrm{Re}\qty[\theta^{LR}(\Gamma)]=\mathrm{Re}\qty[\theta^{RL}(\Gamma)]=\frac{1}{2}\qty[\theta^{LL}(\Gamma)+\theta^{RR}(\Gamma)]=(0\ \mathrm{or}\ \pi)\ \mathrm{mod}\ 2\pi,
\end{equation}
where the eigenvectors used to calculate $\frac{1}{2}(\theta^{LL}(\Gamma)+\theta^{RR}(\Gamma))$ should satisfy the gauge constraint $\braket{\psi^L_n(\mathbf{k})}{\psi^R_n(\mathbf{k})}\in\mathbb{R}_{>0}$. The quantized Berry phases are determined by the \gdagger-parities of the two eigenstates at the intersection points $\mathbf{k}(0)$ and $\mathbf{k}(\pi)$ of $\Gamma_G$ with the plane $\Pi_G$:
\begin{equation}
    \exp\qty[i\mathrm{Re}(\theta^{RL}(\Gamma))]=\exp\qty[i\mathrm{Re}(\theta^{LR}(\Gamma))]=\exp\qty[i\frac{1}{2}\qty(\theta^{LL}(\Gamma)+\theta^{RR}(\Gamma))]=\tilde{p}_G(0)\tilde{p}_G(\pi)=\pm1.
\end{equation}
\end{theorem}

\begin{proof}
In computing any type of Berry connection $A^{\alpha\beta}(\mathbf{k})={-i\bra{ {\psi}_n^\alpha(\mathbf{k})}\nabla_\mathbf{k}\ket{{\psi}_n^\beta(\mathbf{k})}},\ (\alpha,\beta\in\qty{L,R})$, the eigenstates should be normalized by
\begin{equation*}
1=\braket{\psi_n^\alpha(\mathbf{k})}{\psi_n^\beta(\mathbf{k})}=\bra{\psi_n^\alpha(\mathbf{k})}\hat{G}\hat{G}\ket{\psi_n^\beta(\mathbf{k})}=\rho_n^{\alpha}(\mathbf{k})^*\rho_n^\beta(\mathbf{k})\braket{\psi^{\bar\alpha}_n(\hat{g}\mathbf{k})}{\psi^{\bar\beta}_n(\hat{g}\mathbf{k})},
\end{equation*}
which combining with Eq.~\eqref{Gdagger relaton 1} gives
\begin{equation}\label{Gdagger relaton 2}
    \rho_n^{\bar\alpha}(\mathbf{k})^*\rho^{\bar\beta}_n(\mathbf{k})=\braket{\psi^{\bar\alpha}_n(\mathbf{k})}{\psi^{\bar\beta}_n(\mathbf{k})}\in\mathbb{R}_{>0}.
\end{equation}
Here, $\bar\alpha,\bar\beta$ represent the opposite indices to $\alpha,\beta$ respectively, say if $\alpha=R$,  $\bar\alpha=L$, and vice versa. Equations.~\eqref{Gdagger relaton 1}~and~\eqref{Gdagger relaton 2} also imply the equality in the sense of modulo $2\pi$:
\begin{equation}\label{argument relation}
    -\arg(\rho_n^{\bar\alpha}(-\phi))=\arg(\rho_n^\alpha(\phi))=\arg(\rho_n^\beta(\phi))=-\arg(\rho_n^{\bar\beta}(-\phi)).
\end{equation}

Then, we consider the effect of \gdagger\ symmetry on the Berry connections:
\begin{equation*}
\begin{split}
A^{\alpha\beta}(\hat{g}\mathbf{k})=&-i\bra{ \psi_n^\alpha(\hat{g}\mathbf{k})}\nabla_{\hat{g}\mathbf{k}}\ket{\psi_n^\beta(\hat{g}\mathbf{k})}\\
=&-i\bra{ \psi_n^{\bar\alpha}(\mathbf{k})}\hat{G}\frac{1}{\rho_n^{\bar\alpha}(\mathbf{k})^*}\nabla_{\hat{g}\mathbf{k}}\,\frac{1}{\rho_n^{\bar\beta}(\mathbf{k})}\hat{G}\ket{\psi_n^{\bar\beta}(\mathbf{k})}\\[-10pt]
=&\hat{g}\qty[-i\bra{ \psi_n^{\bar\alpha}(\mathbf{k})}\frac{1}{\rho_n^{\bar\alpha}(\mathbf{k})^*\rho^{\bar\beta}_n(\mathbf{k})}\nabla_{\mathbf{k}}\ket{\psi_n^{\bar\beta}(\mathbf{k})}+i\frac{\braket{\psi^{\bar\alpha}_n(\mathbf{k})}{\psi^{\bar\beta}_n(\mathbf{k})}}{\rho_n^{\bar\alpha}(\mathbf{k})^*\rho^{\bar\beta}_n(\mathbf{k})}\nabla_\mathbf{k}\ln\rho^{\bar\beta}_n(\mathbf{k})]\\
=&\hat{g}\qty[-i\bra{ \tilde{\psi}_n^{\bar\alpha}(\mathbf{k})}\nabla_{\mathbf{k}}\ket{\tilde{\psi}_n^{\bar\beta}(\mathbf{k})}-\frac{1}{2}i\nabla_{\mathbf{k}}\ln\qty(\braket{\psi^{\bar\alpha}_n(\mathbf{k})}{\psi^{\bar\beta}_n(\mathbf{k})})+i\nabla_\mathbf{k}\ln\rho^{\bar\beta}_n(\mathbf{k})]\\
=&\hat{g}\qty[A^{\bar\alpha \bar\beta}(\mathbf{k})-\frac{1}{2}i\nabla_{\mathbf{k}}\ln\qty(\braket{\psi^{\bar\alpha}_n(\mathbf{k})}{\psi^{\bar\beta}_n(\mathbf{k})})+i\nabla_\mathbf{k}\ln\rho^{\bar\beta}_n(\mathbf{k})],
\end{split}
\end{equation*}
where $\ket{\tilde\psi^{\bar\alpha/\bar\beta}_n(\mathbf{k})}=\ket{\psi^{\bar\alpha/\bar\beta}_n(\mathbf{k})}/\sqrt{\braket{\psi^{\bar\alpha}_n(\mathbf{k})}{\psi^{\bar\beta}_n(\mathbf{k})}}$ denotes the eigenvectors that naturally satisfy $\braket{\tilde\psi^{\bar\alpha}_n(\mathbf{k})}{\tilde\psi^{\bar\beta}_n(\mathbf{k})}=1$, hence  $A^{\bar\alpha\bar\beta}(\mathbf{k})=-i\bra{ \tilde{\psi}_n^{\bar\alpha}(\mathbf{k})}\nabla_{\mathbf{k}}\ket{\tilde{\psi}_n^{\bar\beta}(\mathbf{k})}$. It is worth noting that since $\braket{\psi^{\bar\alpha}_n(\mathbf{k})}{\psi^{\bar\beta}_n(\mathbf{k})}$ for any $\alpha,\beta$ always takes real positve values  along the loop $\Gamma$ according to Eq.\eqref{Gdagger relaton 2}, a continuous gauge of the original eigenvectors $\ket{\psi^{\alpha\beta}_n(\mathbf{k})}$ along $\Gamma$ guarantees that the renormalized eigenvectors $\ket{\tilde\psi^{\bar\alpha/\bar\beta}_n(\mathbf{k})}$  also form a continuous gauge along $\Gamma$. Hence, the two types of Berry phases $\theta^{\alpha\beta}(\Gamma)$ and $\theta^{\bar\alpha\bar\beta}(\Gamma)$ are both well-defined and obey the relation:

\vspace{-7pt}
\begin{equation}
\begin{split}
\theta^{\bar\alpha\bar\beta}(\Gamma)=\oint_{\Gamma}A^{\bar\alpha\bar\beta}(\mathbf{k})\cdot\,d\mathbf{k}=&\oint_{\Gamma}\big[\hat{g}A^{\alpha\beta}(\hat{g}\mathbf{k})+\frac{1}{2}i\nabla_{\mathbf{k}}\ln\qty(\braket{\psi^{\bar\alpha}_n(\mathbf{k})}{\psi^{\bar\beta}_n(\mathbf{k})})-i\nabla_\mathbf{k}\ln(\rho_n^{\bar\beta}(\mathbf{k}))\big]\cdot\,d\mathbf{k}\\
=& \oint_{\Gamma} A^{\alpha\beta}(\hat{g}\mathbf{k})\cdot d(\hat{g}\mathbf{k})+\oint_{\Gamma} d\arg(\rho_n^{\bar\beta}(\mathbf{k}))+i\oint_\Gamma d\Bigg(\underbrace{\frac{1}{2}\ln\qty(\braket{\psi^{\bar\alpha}_n(\mathbf{k})}{\psi^{\bar\beta}_n(\mathbf{k})})-\ln\abs{\rho^{\bar\beta}_n(\mathbf{k})}}_{\text{single valued}}\Bigg)\\[-5pt]
=&-\theta^{\alpha\beta}(\Gamma)+\oint_{\Gamma} d\arg(\rho^{\bar\beta}_n(\mathbf{k})).
\end{split}
\end{equation}
Therefore, from Eq.~\eqref{argument relation}, we obtain
\begin{equation}\label{Gdagger Berry phase}
 \frac{1}{2}[\theta^{\alpha\beta}(\Gamma)+\theta^{\bar\alpha\bar\beta}(\Gamma)]=\frac{1}{2}\oint_{\Gamma} d\arg(\rho^{\bar\beta}_n(\mathbf{k}))
 =\frac{1}{2} \int^{\pi}_{-\pi}d\arg(\rho^{\bar\beta}_n(\phi))=\frac{1}{2}\qty[ \int^{\pi}_{0}d\arg(\rho^{\bar\beta}_n(\phi))+\int^{\pi}_{0}d\arg(\rho^{\beta}_n(\phi))].
\end{equation}
In order to acquire quantized Berry phases, we introduce  \textbf{a gauge constraint on the eigenvectors $\ket{\psi^\beta_n(\mathbf{k})}$ and $\ket{\psi^{\bar\beta}(\mathbf{k})}$ along the loop:}
\begin{equation}\label{additional constrant}
    \braket{\psi^\beta_n(\mathbf{k})}{\psi^{\bar\beta}_n(\mathbf{k})}=\bra{\psi^\beta_n(\mathbf{k})}\hat{G}\hat{G}\ket{\psi^{\bar\beta}_n(\mathbf{k})}=\rho^\beta_n(\mathbf{k})^*\rho^{\bar\beta}_n(\mathbf{k})\braket{\psi^{\bar\beta}_n(\mathbf{k})}{\psi^{\beta}_n(\mathbf{k})} \in \mathbb{R}_{>0}.
\end{equation}
This constraint is automatically satisfied by the binormalized eigenvectors used to compute the biorthogonal Berry phases $\theta^{\alpha\beta}(\Gamma)=\theta^{LR}(\Gamma)$ and $\theta^{\bar\alpha\bar\beta}(\Gamma)=\theta^{RL}(\Gamma)$. In calculating $\theta^{\alpha\beta}(\Gamma)=\theta^{LL}(\Gamma)$ and $\theta^{\bar\alpha\bar\beta}(\Gamma)=\theta^{RR}(\Gamma)$, $\ket{\psi^L_n(\mathbf{k})}$ and $\ket{\psi^R_n(\mathbf{k})}$ are normalized separately, and their phases are two gauge degrees of freedom. But the constraint~\eqref{additional constrant} locks the relative phase between $\ket{\psi^L_n(\mathbf{k})}$ and $\ket{\psi^R_n(\mathbf{k})}$, hence leaving only one gauge degree of freedom. Moreover, $\braket{\psi^\beta_n(\mathbf{k})}{\psi^{\bar\beta}_n(\mathbf{k})}=\braket{\psi^L_n(\mathbf{k})}{\psi^{R}_n(\mathbf{k})}\neq0$ always holds, then plugging Eq.~\eqref{G-dagger map} into Eq.~\eqref{additional constrant} leads to
\begin{equation}\label{additional constrant 2}
    \rho^\beta_n(\mathbf{k})^*\rho^{\bar\beta}_n(\mathbf{k})=\frac{\braket{\psi^\beta_n(\mathbf{k})}{\psi^{\bar\beta}_n(\mathbf{k})}}{\braket{\psi^{\bar\beta}_n(\mathbf{k})}{\psi^{\beta}_n(\mathbf{k})}}=1\quad\Rightarrow\quad
    \arg(\rho^\beta_n(\mathbf{k}))=\arg(\rho^{\bar\beta}_n(\mathbf{k}))\bmod 2\pi.
\end{equation}
Combining Eqs.~\eqref{Gdagger relaton 1} and \eqref{additional constrant 2}, we obtain 
\begin{equation}
    \arg(\rho_n^\beta(\phi)) =\arg(\rho_n^{\bar\beta}(\phi))= -\arg(\rho_n^\beta(-\phi)), 
\end{equation}
indicating $\rho_n^\beta(\phi)$ is an odd function about the $G$-invariant point $\phi_g=0,\pm \pi$.

In the $G$-invariant subspace $\Pi_G$, the constraint~\eqref{additional constrant} also guarantees 
\begin{equation}\label{Gdagger-parity 2}
    \rho^\beta_n(\mathbf{k}_g)=\frac{\bra{\psi^\beta_n(\mathbf{k}_g)}\hat{G}\ket{\psi^\beta_n(\mathbf{k}_g)}}{\braket{\psi^\beta_n(\mathbf{k}_g)}{\psi^{\bar\beta}_n(\mathbf{k}_g)}}\in\mathbb{R}_{\neq 0}\quad\Rightarrow\quad
    \mathrm{sign}(\rho^\beta_n(\mathbf{k}_g))=\mathrm{sign}\qty[\bra{\psi^\beta_n(\mathbf{k}_g)}\hat{G}\ket{\psi^\beta_n(\mathbf{k}_g)}]=\tilde{p}_G(\mathbf{k}_g)
\end{equation}
Substitution of Eqs.~\eqref{additional constrant 2} and~\eqref{Gdagger-parity 2} into Eq.~\eqref{Gdagger Berry phase} yields
\begin{equation}
    \frac{1}{2}[\theta^{\alpha\beta}(\Gamma)+\theta^{\bar\alpha\bar\beta}(\Gamma)]=\frac{1}{2}\int_{-\pi}^{\pi} d\arg(\rho^{\beta}_n(\phi))
 =\qty[\arg(\rho^\beta_n(\pi))-\arg(\rho^\beta_n(0))]\bmod 2\pi=\mathrm{arg}\qty(\tilde{p}_G(0)\tilde{p}_G(\pi))\bmod 2\pi.
\end{equation}
which shows that \gdagger\ symmetry can protect the quantization of Berry phases for any $\alpha,\beta$ in the above form and the quantized value is uniquely dictated by the \gdagger-parities of the two states at the two $G$-invariant points $\mathbf{k}(0),\mathbf{k}(\pi)$ on $\Gamma$.
\end{proof}

\subsubsection{Examples of \mdagger\ protected Berry phases}
As shown in Fig.~\ref{fig-MDagger-QuantizedBP}, we consider the application of \textbf{Theorem}~\ref{thm_Gdagger Berry phase} to a mirror-symmetric loop $\Gamma=\hat{m}\Gamma^{-1}$ in a \mdagger-symmetric system. The loop crosses the mirror plane $\Pi_M$ at $\phi=0$ and $\pi$. For a non-degenerate band along $\Gamma$, provided that two states at $\phi=0,\pi$ on the band are in the exact phases, the Berry phases of the band is quantized as
\begin{equation}
    \exp\qty[i\Re[\theta^{LR}(\Gamma)]]=\exp\qty[i\Re[\theta^{RL}(\Gamma)]]=\exp\qty[i\theta^{LL}(\Gamma)]=\exp\qty[i\theta^{RR}(\Gamma)]=\tilde{p}(0)\tilde{p}(\pi),
\end{equation}
where $\tilde{p}(0)$ and $\tilde{p}(\pi)$ denote the \mdagger-parities of the two states at the $M$-invairant points. 

Supposing all ELs enclosed by the loop $\Gamma$ are formed by two bands which are well separated from other bands by real line gaps, Figs.~\ref{fig-MDagger-QuantizedBP}(a,b) and Figs.~\ref{fig-MDagger-QuantizedBP}(c,d) illustrate two different scenarios supporting nontrivial \mdagger-protected Berry phases. In both scenarios, the eigenenergies of the two bands are fixed on the real axis at $\phi=0,\pi$ where the order of the two real eigenvalues are definite, hence the energy braiding of the two bands along each semi-loop (i.e, $\phi\in[-\pi,0]$ and  $\phi\in[0,\pi]$) is unambiguous. This fact together with the spectral symmetry $E_i(\phi)=E_i(-\phi)^*$ ensures that \textbf{the energy braidings along the upper and lower semi-loops are always identical, so the DN along $\Gamma$, twice as big as the semi-loop DN $\mathcal{D}(\Gamma_\mathrm{semi})\in\mathbb{Z}$, should be an even number}
\begin{equation}
    \mathcal{D}(\Gamma)=2\mathcal{D}(\Gamma_\mathrm{semi})\in2\mathbb{Z},
\end{equation}
indicating the loop $\Gamma$ must enclose an even number of ELs.
If we connect the two terminals of $\Gamma_\mathrm{semi}$ by a path $\gamma=\qty{\mathbf{k}_m(\varphi)\in\Pi_M\,|\,\mathbf{k}_m(0)=\Gamma(0)\text{ and }\mathbf{k}_m(\pi)=\Gamma(\pi),\varphi\in[0,\pi]}$ inside the mirror plane (see Fig.~\ref{fig-MDagger-QuantizedBP}(a,c)) and order the eigenenergies $E^\pm(\varphi)$ along the path $\gamma$ be their real parts, $\Re[E^+(\varphi)]>\Re[E^-(\varphi)]$ (see the red and blue thin lines in Figs.~\ref{fig-MDagger-QuantizedBP}(b,d)), we should have the following permutation relation:
\begin{equation}\label{semi-loop permutation}
    \begin{pmatrix}
    \ket{\psi_1(\pi)}\\\ket{\psi_2(\pi)}
    \end{pmatrix}=\begin{pmatrix}
    \ket{\psi^+(\pi)}\\\ket{\psi^-(\pi)}
    \end{pmatrix}\quad\Rightarrow\quad
    \begin{pmatrix}
    \ket{\psi_1(0)}\\\ket{\psi_2(0)}
    \end{pmatrix}=\underbrace{\pmqty{0 & 1 \\ 1 & 0}^{\mathcal{D}(\Gamma_\mathrm{semi})} }_{\hat{\mathcal{P}}(\Gamma_\mathrm{semi})}
    \begin{pmatrix}
    \ket{\psi^+(0)}\\\ket{\psi^-(0)}
    \end{pmatrix}.
\end{equation}

For the first scenario in Figs.~\ref{fig-MDagger-QuantizedBP}(a,b), the loop $\Gamma$ encloses two \mdagger-protected ELs lying inside $\Pi_M$, which implies that the two considered bands possess opposite \mdagger-parities at the two terminals, i.e., $\tilde{p}_1(0)\tilde{p}_2(0)=\tilde{p}_1(\pi)\tilde{p}_1(\pi)=-1$ according to \textbf{Theorem}~\ref{theorem-mdagger} (see bands 1 and 2 in Fig.~\ref{fig-MDagger-QuantizedBP}(b)). In particular, since the two ELs pass through the loop from opposite directions, the two energy bands do not braid along $\Gamma$ ($\mathcal{D}(\Gamma)=0$) and hence \textbf{the orders of the two eigenenergies are identical at two terminals}, i.e. the two terminals of band 1 $\ket{\psi_1(0)},\ket{\psi_1(\pi)}$ (band 2 $\ket{\psi_2(0)},\ket{\psi_2(\pi)}$) are on the in-plane band $E^+$ ($E^-$).
When band 1,2 take nontrivial Berry phases, the two terminals of each band $\ket{\psi_{1,2}(0)}=\ket{\psi^{+,-}(0)}$ and $\ket{\psi_{1,2}(\pi)}=\ket{\psi^{+,-}(\pi)}$ should have opposite \mdagger-parities, 
\[
\tilde{p}_{1,2}(0)\tilde{p}_{1,2}(\pi)=\tilde{p}^{+,-}(0)\tilde{p}^{+,-}(\pi)=-1.
\]
However, since the \mdagger-parity along a continuous in-plane band should be a constant unless encountering degeneracies, the opposite terminal \mdagger-parities indicates the in-plane bands $E^\pm$ have to be degenerate at some points along the path $\gamma$ (i.e., the points encountering the ELs), in other words, \textbf{the quantized Berry phases along $\Gamma$ forbid the annihilation of the two in-plane ELs in spite of their opposite directions.}

For the second scenario in Figs.~\ref{fig-MDagger-QuantizedBP}(c,d), the loop encloses two \mdagger-partner ELs symmetrically located at the two sides of $\Pi_M$. From Table~\ref{DN table}, we know that the two \mdagger-partner ELs must thread the loop from the same direction, so the DN is quantized to $\mathcal{D}(\Gamma)=2$ and the two bands braid once along each semi-loop, as shown in Fig.~\ref{fig-MDagger-QuantizedBP}(d).  Seemingly, the Berry phase along $\Gamma$ would always be trivial in this case as the two terminals $\phi=0,\pi$ are in a continuous exact phase inside the plane $\Pi_M$: $\tilde{p}^{\pm}(0)=\tilde{p}^{\pm}(\pi)$. However, thanks to the mode braiding $\mathcal{D}(\Gamma_\mathrm{semi})=1$, the two states $\ket{\psi_{1}(0)},\ket{\psi_{1}(\pi)}$ on band 1 along $\Gamma$ fall on two different in-plane bands $E^\pm$ along $\gamma$  in light of Eq.~\eqref{semi-loop permutation}. As a consequence, provided the two in-plane bands have opposite \mdagger-parities $\tilde{p}^+\tilde{p}^-=-1$ (e.g., the red and blue thin lines in Fig.~\ref{fig-MDagger-QuantizedBP}(d)), we have 
\begin{equation}
    \tilde{p}_1(0)\tilde{p}_1(\pi)=\tilde{p}_2(0)\tilde{p}_2(\pi)=\tilde{p}^+\tilde{p}^-=-1,
\end{equation}
giving rise to the nontrivial Berry phases of the bands 1,2 along $\Gamma$. 
\textbf{This scenario manifests that the single-band Berry phases can also be quantized even if the band undergoes nontrivial energy braiding, which therefore goes beyond the previous result requiring a real line gap and has no Hermitian counterpart.}

\subsection{Quantized Berry phases protected by $G\mathcal{T}$ symmetry}
Let $\mathcal{H}(\mathbf{k})$ now be a Hamiltonian with a twofold antiunitary symmetry $G\mathcal{T}$ combined by the point group operation $G$ and time-reversal $\mathcal{T}$, i.e., $\mathcal{H}(\mathbf{k})=\hat{\tau}\mathcal{H}(-\hat{g}\mathbf{k})^*\hat{\tau}^{\dagger}$, where $\tau=\tau^\intercal=(\tau^{-1})^*$ is a symmetric unitary operator acting on the internal degrees of freedom. And we consider a $G\mathcal{T}$-symmetric loop
\begin{equation}\label{GT-symmetric loop}
    \Gamma=G\mathcal{T}(\Gamma)=\qty{\mathbf{k}(\phi)\,|\,\hat{g}\mathbf{k}=-\mathbf{k}\ \text{and}\ \mathbf{k}(\pi)=\mathbf{k}(-\pi),-\pi\le\phi\le\pi}=\in\Pi_{G\mathcal{T}}
\end{equation}
lying in the $G\mathcal{T}$-invariant subspace $\Pi_{G\mathcal{T}}=\qty{\mathbf{k}\,|\,\hat{g}\mathbf{k}=-\mathbf{k}}$.
In the subspace $\Pi_{G\mathcal{T}}$, the $G\mathcal{T}$ symmetry ensures the eigenstates are either in the exact or broken phases where the eigenvalues are purely real or form complex conjugate pairs. And in the exact phase in $\Pi_{G\mathcal{T}}$, the $G\mathcal{T}$ operation maps a non-degenerate eigenvector to itself:
\begin{gather}\label{GT map}
    \mathcal{H}(\mathbf{k})\hat{\tau}\ket{\psi^{R}(\mathbf{k})}^*= \hat{\tau}\mathcal{H}(\mathbf{k})^*\ket{\psi^{R}(\mathbf{k})}^*=E^*\hat{\tau}\ket{\psi^{R}(\mathbf{k})}^*= E\hat{\tau}\ket{\psi^{R}(\mathbf{k})}^*\nonumber\\
    \Rightarrow\quad G\mathcal{T}\ket{\psi^{R/L}(\mathbf{k})}=\hat{\tau}\ket{\psi^{R/L}(\mathbf{k})}^*=\rho^{R/L}(\mathbf{k})\ket{\psi^{R/L}(\mathbf{k})},
\end{gather}
where $\rho^{R/L}(\mathbf{k})=\exp\qty[i\varphi^{R/L}(\mathbf{k})]$ is a pure phase due to $(G\mathcal{T})^2=1$.

Importantly, we know that
\begin{property}\label{property of GT}
$G\mathcal{T}$ symmetry can ensure a band of continuous eigenstates along a $G\mathcal{T}$-symmetric loop $\Gamma$ is \textbf{self-closed}, provided that
\begin{enumerate}
    \item[(1)] the band concerned is in the exact phase, \vspace{-4pt}
    \item[(2)] the band concerned is isolated from other bands by real line gaps.
\end{enumerate}
\end{property}

Then we have the following theorem of $G\mathcal{T}$ symmetry protected quantized Berry phases along such loops: 
\begin{theorem}[\bf $G\mathcal{T}$-protected quantized Berry phase]
Along a $G\mathcal{T}$-symmetric loop $\Gamma\in\Pi_{G\mathcal{T}}$, provided that a band of exact eigenstates satisfies the Property~\ref{property of GT}, $G\mathcal{T}$ symmetry ensures that the real parts of all four types of Berry phases of the band are consistently quantized modulo $2\pi$:
\begin{equation}
    \theta^{LL}(\Gamma)=\theta^{RR}(\Gamma)=\mathrm{Re}\qty[\theta^{LR}(\Gamma)]=\mathrm{Re}\qty[\theta^{RL}(\Gamma)]=(0\ \mathrm{or}\ \pi)\ \mathrm{mod}\ 2\pi.
\end{equation}
\end{theorem}

\begin{proof}
For a specific type of Berry connection $A^{\alpha\beta}$ ($\alpha,\beta\in\qty{R,L}$) of eigenstates in the exact phase in $\Pi_{G\mathcal{T}}$, the normalization condition requires that
\begin{equation*}
 1=\braket{\psi_n^\alpha(\mathbf{k})}{\psi_n^\beta(\mathbf{k})}=\bra{\psi_n^\alpha(\mathbf{k})^*}\hat{\tau}^*\hat{\tau}\ket{\psi_n^\beta(\mathbf{k})^*}=\rho_n^{\alpha}(\mathbf{k})^*\rho_n^\beta(\mathbf{k})\quad\Rightarrow\quad
 {\rho}_n^{\alpha}(\mathbf{k})=\frac{1}{{\rho}_n^{\beta}(\mathbf{k})^*}=\rho^\beta_n(\mathbf{k}).
\end{equation*}
And the Berry connections satisfy 
\begin{equation*}
\begin{split}
A^{\alpha\beta}(\mathbf{k})=&-i\bra{ \psi_n^\alpha(\mathbf{k})}\nabla_{\mathbf{k}}\ket{\psi_n^\beta(\mathbf{k})}\\
=&-i\bra{\psi_n^\alpha(\mathbf{k})}\hat{\tau}\hat{\tau}^*\nabla_{\mathbf{k}}\ket{\psi_n^\beta(\mathbf{k})}\\
=&-i\bra{\psi_n^\alpha(\mathbf{k})^*}\rho^\alpha_n(\mathbf{k})\nabla_{\mathbf{k}}\rho^\beta_n(\mathbf{k})^*\ket{\psi_n^\beta(\mathbf{k})^*}\\
=& -i\qty(\bra{\psi^\alpha_n(\mathbf{k})}\nabla_\mathbf{k}\ket{\psi_n^\beta(\mathbf{k})})^*-i\nabla_\mathbf{k}\ln\rho^\beta_n(\mathbf{k})^*\\
=&\ -A^{\alpha\beta}(\mathbf{k})^*-i\nabla_\mathbf{k}\ln\rho_n^\beta(\mathbf{k})^*.
\end{split}
\end{equation*}
Therefore, the real parts of all types of Berry phases along $\Gamma$ are quantized:
\begin{equation*}
\begin{split}
\mathrm{Re}\qty[\theta^{\alpha\beta}(\Gamma)]=\frac{1}{2}\qty[\theta^{\alpha\beta}(\Gamma)+\theta^{\alpha\beta}(\Gamma)^*]=\frac{1}{2}\oint_{\Gamma}\qty[A^{\alpha\beta}(\mathbf{k})+A^{\alpha\beta}(\mathbf{k})^*]\cdot\,d\mathbf{k}
=&-\frac{1}{2}\oint_\Gamma d\,\arg(\rho^\beta_n(\mathbf{k}))\big)=0\ \text{or}\ \pi\bmod 2\pi.
\end{split}
\end{equation*}
Under a gauge transformation $\ket{\psi^\beta_n(\phi)}\rightarrow\ket{\psi'^\beta_n(\phi)}=e^{i\vartheta(\phi)}\ket{\psi^\beta_n(\phi)}$ with $\oint_\Gamma d\vartheta(\pi)=2m\pi$ ($m\in\mathbb{Z}$), we have $\hat{\tau}=\ket{\psi'^\beta_n(\phi)}^*=e^{2i\vartheta(\phi)}\ket{\psi'^\beta_n(\phi)}$ and hence $\arg\qty[\rho'_n(\phi)]=\arg\qty[\rho_n(\phi)]+2\vartheta(\phi)$. As a result, we see that any kind of the Berry phases are gauge invariant in the sense of modulo $2\pi$:
\begin{equation*}
    \mathrm{Re}\qty[\theta'^{\alpha\beta}(\Gamma)]=\mathrm{Re}\qty[\theta^{\alpha\beta}(\Gamma)]+\oint_\Gamma d\vartheta(\phi)=\mathrm{Re}\qty[\theta^{\alpha\beta}(\Gamma)]+2m\pi.
\end{equation*}
Therefore, as long as the four types of Berry phases take the same quantized value in any special gauge, they will be always identical modulo $2\pi$ in all gauges. This is indeed the case if we select a gauge of $\ket{\psi_n^{L/R}(\mathbf{k})}$
satisfying $\braket{\psi^L_n(\mathbf{k})}{\psi^R_n(\mathbf{k})}=1$ to compute $\theta^{LR}$,$\theta^{RL}$ and in the mean time take $\ket{\psi'^{L/R}_n(\mathbf{k})}=\ket{\psi^{L/R}_n(\mathbf{k})}/|\psi^{L/R}_n(\mathbf{k})|$ to calculate $\theta^{LL}$ and $\theta^{RR}$, respectively. On account that $\theta^{LL}$, $\theta^{RR}$ always take real values, we thus complete the proof.
\end{proof}

\begin{figure*}[b!]
\includegraphics[width=0.9\textwidth]{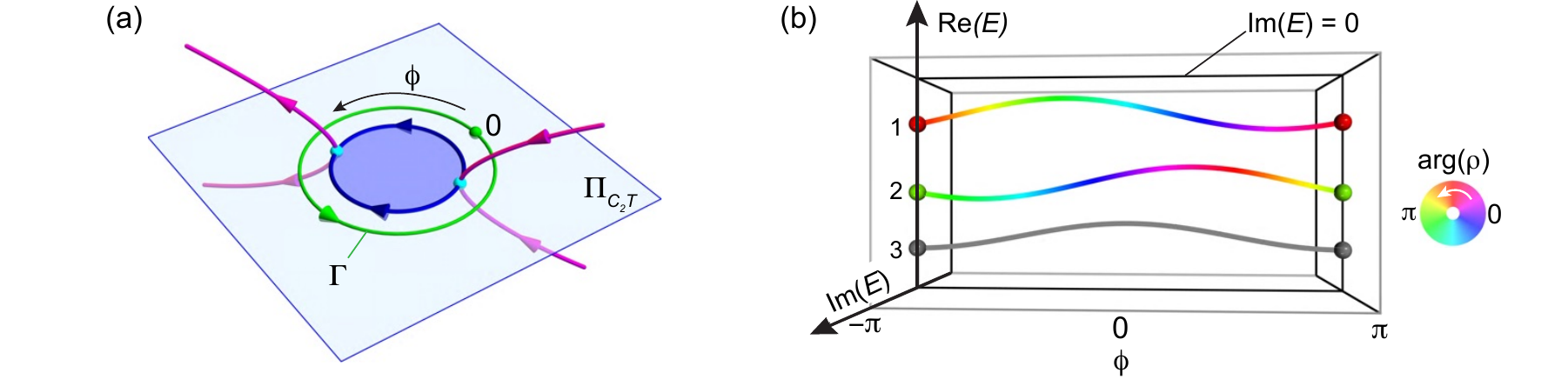}
\caption{\label{fig-C2T-QuantizedBP} 
(a) Schematic of a $C_2\mathcal{T}$-symmetric (green) loop $\Gamma=-\hat{c}_2\Gamma$ lying in the exact phase of $C_2\mathcal{T}$-invariant plane $\Pi_{C_2\mathcal{T}}$ encircling an exceptional ring (blue tube). (b) Schematic of the non-degenerate multi-band structure along the loop $\Gamma$ in (a), where the bands $1,2,3$ are purely real. The colors on bands 1 and 2 represent $\arg[\rho_{1,2}]$ of the bands.}
\end{figure*}

In Fig.~\ref{fig-C2T-QuantizedBP}(a), we illustrate an example of $C_2\mathcal{T}$-protected quantized Berry phases along a loop $\Gamma=-\hat{c}_2\Gamma$ in the exact phase of the $C_2\mathcal{T}$-invariant plane $\Pi_{C_2\mathcal{T}}$. The loop $\Gamma$ encircles an in-plane EP ring (blue) which is chained with the other two out-of-plane ELs (magenta). Figure~\ref{fig-C2T-QuantizedBP}(b) shows the schematic of the band structure along $\Gamma$, where the two colored lines denote the two bands 1,2 forming the EC in Fig.~\ref{fig-C2T-QuantizedBP}(a). Since the two bands are in the exact phase, their eigenenergies are fixed in the real plane ($\Im[E]=0$), the colors on the bands represent the phase $\arg[\rho^\beta_{1,2}]$ associated with the two bands, each of which changes a cycle as $\phi$ travels from $-\pi$ to $\pi$, corresponding to $\pi$-quantized Berry phases of each band.

\subsection{Quantized Berry phases protected by \gtdagger\ symmetry}
We consider a $\mathcal{H}(\mathbf{k})$ respecting with a  \gtdagger\ symmetry , i.e., $\mathcal{H}(\mathbf{k})^\dagger=\hat{\tau}\mathcal{H}(-\hat{g}\mathbf{k})^*\hat{\tau}^{\dagger}$ with $\tau=\tau^\intercal=(\tau^{-1})^*$ and a $G\mathcal{T}$-symmetric loop $\Gamma$ (Eq.~\eqref{GT-symmetric loop}) lying in the $G\mathcal{T}$-invariant subspace $\Pi_{G\mathcal{T}}$ (i.e., $\hat{g}\mathbf{k}=-\mathbf{k}$).
In the subspace $\Pi_{G\mathcal{T}}$, the $G\mathcal{T}$ operation maps a non-degenerate eigenvector to itself:
\begin{gather}\label{GT map}
    \mathcal{H}(\mathbf{k})^\dagger\hat{\tau}\ket{\psi^{R}(\mathbf{k})}^*= \hat{\tau}\mathcal{H}(\mathbf{k})^*\ket{\psi^{R}(\mathbf{k})}^*=E^*\hat{\tau}\ket{\psi^{R}(\mathbf{k})}^*\nonumber\\
    \Rightarrow\quad G\mathcal{T}\ket{\psi^{R/L}(\mathbf{k})}=\hat{\tau}\ket{\psi^{R/L}(\mathbf{k})}^*=\rho^{R/L}(\mathbf{k})\ket{\psi^{L/R}(\mathbf{k})},
\end{gather}
where $\rho^{L}(\mathbf{k})=1/\rho^{R}(\mathbf{k})^*$ as a result of $(G\mathcal{T})^2=1$.


Then we have the following theorem of \gtdagger\ symmetry protected quantized Berry phases: 

\begin{theorem}[\bf $G\mathcal{T}$-protected quantized Berry phase]
For a \textbf{self-closed} band along a $G\mathcal{T}$-symmetric loop $\Gamma\in\Pi_{G\mathcal{T}}$, \gtdagger\ symmetry can protect the quantization of Berry phases in the following forms
\begin{equation}
    \theta^{LR}(\Gamma)=\theta^{RL}(\Gamma)=\frac{1}{2}\qty[\theta^{LL}(\Gamma)+\theta^{RR}(\Gamma)]=(0\ \mathrm{or}\ \pi)\ \mathrm{mod}\ 2\pi,
\end{equation}
where the eigenvectors used to calculate $\frac{1}{2}[\theta^{LL}(\Gamma)+\theta^{RR}(\Gamma)]$ should obey the gauge constraint $\braket{\psi^L_n(\mathbf{k})}{\psi^R_n(\mathbf{k})}\in\mathbb{R}_{>0}$.
\end{theorem}

\begin{proof}
For a specific type of Berry connection $A^{\alpha\beta}$ ($\alpha,\beta\in\qty{R,L}$) of eigenstates in the exact phase in $\Pi_{G\mathcal{T}}$, the normalization condition requires that
\begin{gather}\label{GTdagger constraint 3}
 1=\braket{\psi_n^\alpha(\mathbf{k})}{\psi_n^\beta(\mathbf{k})}=\bra{\psi_n^\alpha(\mathbf{k})^*}\hat{\tau}^*\hat{\tau}\ket{\psi_n^\beta(\mathbf{k})^*}=\rho_n^{\alpha}(\mathbf{k})^*\rho_n^\beta(\mathbf{k})\braket{\psi^{\bar\alpha}_n(\mathbf{k})}{\psi^{\bar\beta}_n(\mathbf{k})}
 =\frac{\braket{\psi^{\bar\alpha}_n(\mathbf{k})}{\psi^{\bar\beta}_n(\mathbf{k})}}{\rho^{\bar\alpha}_n(\mathbf{k})\rho^{\bar\beta}_n(\mathbf{k})^*}\nonumber\\
 \Rightarrow\quad
 \rho^{\bar\alpha}_n(\mathbf{k})\rho^{\bar\beta}_n(\mathbf{k})^*=\braket{\psi^{\bar\alpha}_n(\mathbf{k})}{\psi^{\bar\beta}_n(\mathbf{k})}\in\mathbb{R}_{>0}.
\end{gather}
And the Berry connections satisfy 
\begin{equation*}
\begin{split}
A^{\alpha\beta}(\mathbf{k})=&-i\bra{ \psi_n^\alpha(\mathbf{k})}\nabla_{\mathbf{k}}\ket{\psi_n^\beta(\mathbf{k})}\\
=&-i\bra{\psi_n^\alpha(\mathbf{k})}\hat{\tau}\hat{\tau}^*\nabla_{\mathbf{k}}\ket{\psi_n^\beta(\mathbf{k})}\\
=&-i\bra{\psi_n^{\bar\alpha}(\mathbf{k})^*}\rho^\alpha_n(\mathbf{k})\nabla_{\mathbf{k}}\rho^\beta_n(\mathbf{k})^*\ket{\psi_n^{\bar\beta}(\mathbf{k})^*}\\
=& -i\bra{\psi^{\bar\alpha}_n(\mathbf{k})^*}\rho_n^{\alpha}(\mathbf{k})^*\rho_n^{\beta}(\mathbf{k})\nabla_\mathbf{k}\ket{\psi_n^{\bar\beta}(\mathbf{k})^*}-\qty[\rho_n^{\alpha}(\mathbf{k})^*\rho_n^\beta(\mathbf{k})\braket{\psi^{\bar\alpha}_n(\mathbf{k})^*}{\psi^{\bar\beta}_n(\mathbf{k})^*}]i\nabla_\mathbf{k}\ln\rho^\beta_n(\mathbf{k})^*\\
=& -i\bra{\tilde{\psi}^{\bar\alpha}_n(\mathbf{k})^*}\nabla_\mathbf{k}\ket{\tilde{\psi}_n^{\bar\beta}(\mathbf{k})^*}-\frac{i}{2}\nabla_\mathbf{k}\ln\qty(\braket{\psi^{\bar\beta}_n(\mathbf{k})}{\psi^{\bar\alpha}_n(\mathbf{k})})-i\nabla_\mathbf{k}\ln\rho^\beta_n(\mathbf{k})^*\\
=&\ -A^{\bar\alpha\bar\beta}(\mathbf{k})^*-i\nabla_\mathbf{k}\qty[\frac{1}{2}\ln\qty(\braket{\psi^{\bar\beta}_n(\mathbf{k})}{\psi^{\bar\alpha}_n(\mathbf{k})})+\ln\rho_n^\beta(\mathbf{k})^*],
\end{split}
\end{equation*}
where $\ket{\tilde\psi^{\bar\alpha/\bar\beta}_n(\mathbf{k})}=\ket{\psi^{\bar\alpha/\bar\beta}_n(\mathbf{k})}/\sqrt{\braket{\psi^{\bar\alpha}_n(\mathbf{k})}{\psi^{\bar\beta}_n(\mathbf{k})}}$ denote the eigenvectors used for calculating $A^{\bar\alpha\bar\beta}(\mathbf{k})$. 
Therefore, the Berry phases along $\Gamma$ satisfy
\begin{equation*}
\begin{split}
\frac{1}{2}\qty[\theta^{\alpha\beta}(\Gamma)+\theta^{\bar\alpha\bar\beta}(\Gamma)^*]=\frac{1}{2}\oint_{\Gamma}\qty[A^{\alpha\beta}(\mathbf{k})+A^{\bar\alpha\bar\beta}(\mathbf{k})^*]\cdot\,d\mathbf{k}
=&-\frac{1}{2}\oint_\Gamma d\,\arg(\rho^\beta_n(\mathbf{k}))\big)=0\ \text{or}\ \pi\bmod 2\pi.
\end{split}
\end{equation*}
When $\alpha,\beta=R,L$ or $L,R$, the above result indicates $\theta^{RL}=\theta^{LR}\in\mathbb{R}$ are quantized. When $\alpha=\beta=R$ or $L$, to guarantee $\frac{1}{2}(\theta^{LL}+{\theta^{RR}}^*)=\frac{1}{2}(\theta^{LL}+{\theta^{RR}})$ being gauge invariant up to modulo $2\pi$, we should demand the left and right eigenvectors used to compute $\theta^{LL}$ and $\theta^{RR}$, respectively, satisfy the \textbf{gauge constraint} $\braket{\psi^L_n(\mathbf{k})}{\psi^R_n(\mathbf{k})}\in\mathbb{R}_{>0}$ (see the analysis below Eq.~\eqref{additional constrant}).
Moreover, as long as we select a gauge of $\ket{\psi_n^{L/R}(\mathbf{k})}$
satisfying $\braket{\psi^L_n(\mathbf{k})}{\psi^R_n(\mathbf{k})}=1$ to compute $\theta^{LR}$,$\theta^{RL}$ and in the mean time take $\ket{\psi'^{L/R}_n(\mathbf{k})}=\ket{\psi^{L/R}_n(\mathbf{k})}/|\psi^{L/R}_n(\mathbf{k})|$ to calculate $\frac{1}{2}(\theta^{LL}+\theta^{RR})$, the obtained three types of Berry phases take the same value in this gauge, and hence they must be identical up to modulo $2\pi$ in all gauges satisfying Eq.~\eqref{additional constrant}.
\end{proof}

\section{Determinisc realization of ECs arising from Hermitian nodal lines}
In this section, we analyze the typical local EC configurations 
evolving from a $z$-mirror-symmetry ($M_z$) protected Hermitian nodal line from a non-Hermitian perturbation perspective. To \textbf{deterministically} obtain ECs, we suppose there exists a second symmetry $R_2\in\qty{M_x,C_{2x}\mathcal{T}}$, apart from $M_z$, in the Hermitian nodal line system.
Hence, there are two symmetry-invariant planes $\Pi_z=\qty{\mathbf{k}_1\,|\,\mathbf{k}_1=\hat{m}_z\mathbf{k}_1}$ and  $\Pi_x=\qty{\mathbf{k}_2\,|\,\mathbf{k}_2=\hat{R}_2\mathbf{k}_2}$. The nodal line lies in $\Pi_z$ and intersects with $\Pi_x$ at $\mathbf{K}$ (see the nodal structure in the second lines of Tables~\ref{perturbation table MzMx} and \ref{perturbation table 2xMz}, respectively).

\begin{table}[b!]
\begin{adjustwidth}{-.5in}{-.5in}  
\centering
\setlength\cellspacetoplimit{3pt}
\setlength\cellspacebottomlimit{3pt}
\addtolength\tabcolsep{1pt}
\vspace{-10pt}\caption{Deterministic schemes for realizing ECs by thresholdless non-Hermitian perturbations from a Hermitian nodal point (the black point shown by the nodal structure in the second line) protected by $m_zm_x2_y\blue{1^\dagger}$ point group at the intersection $\mathbf{K}$ of a nodal line and a high-symmetry line (the crossing line of the red and blue planes). The first part (the first two lines) shows the properties of Hermitian nodal point. The second part displays different non-Hermitian perturbations near the nodal point classified by non-Hermitian little-group symmetries (the first column) of the perturbations, wherein the second column shows the non-Hermitian evolution path of the little group in each case that can guarantee the formation of ECs. The third column gives the perturbative $\mathbf{k}\cdot\mathbf{p}$ Hamiltonians near the nodal point $\mathbf{k}=\mathbf{K}+\mathbf{q}$, where all coefficients are real numbers. \label{perturbation table MzMx}}
\begin{threeparttable}
\begin{tabular}{Sc|Sc|Sc|Sc|Sc|Sc}\hline\hline
\textbf{\makecell{Hermitian\\ point group}} & \textbf{\makecell{non-Hermitian\\ subgroups~\tnote{1}}} & \textbf{Corep} & \textbf{Generators} & \textbf{$\mathbf{k}\cdot\mathbf{p}$ Hamiltonian} & \textbf{nodal structure}\\\hline

$m_zm_x2_y\blue{1^\dagger}$ & \makecell{$\underline{\blue{m_z^\dagger m_x^\dagger}2_y}$, $\blue{m_z^\dagger }m_x\blue{2_y^\dagger}$, $\underline{m_z\blue{m_x^\dagger 2_y^\dagger}}$,\\[8pt] $m_zm_x2_y$, $\underline{\underline{\blue{m_z^\dagger}}}$,
$\blue{m_x^\dagger}$,
$\underline{\underline{m_z}}$, $m_x$,\\[8pt]  $\blue{2_y^\dagger}$, $2_y$}  & 
$A_1\oplus B_2$ & 
\makecell{$\hat{M}_z=\sigma_z$\\[5pt]$\hat{M}_{x}=\sigma_0$} &
$\begin{aligned}\mathcal{H}_0 &=(h_{3}^xq_z+h^x_{23} q_yq_z)\sigma_x\\&+(h^y_3 q_z+h_{23}^y q_yq_z)\sigma_y\\&\quad+(h^z_2q_y+h^x_{11}q_x^2+h^y_{11}q_y^2\\&+h^z_{11}q_z^2)\sigma_z\end{aligned}$ &
\begin{minipage}{.2\textwidth}
      \includegraphics[height=21.4mm]{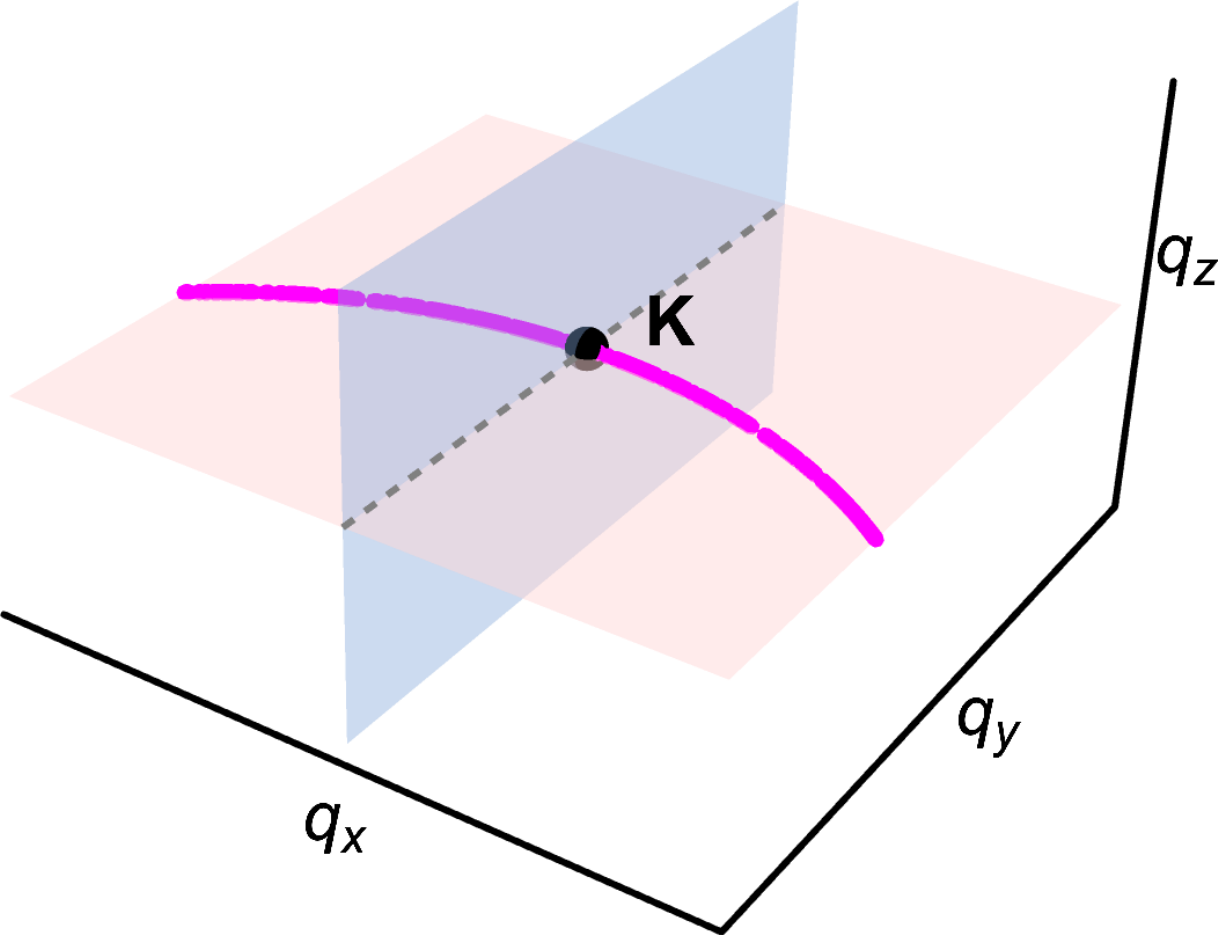}
    \end{minipage}\\\specialrule{.15em}{.05em}{.05em} 

\textbf{subgroup} & \textbf{evolution path} & \multicolumn{3}{Sc|}{\textbf{$\mathbf{k}\cdot\mathbf{p}$ Hamiltonian}} & \textbf{EC structure}\\\hline

$\blue{m_z^\dagger m_x^\dagger}2_y$ & $m_zm_x 2_y\blue{1^\dagger}\rightarrow\blue{m_z^\dagger m_x^\dagger}2_y$ & \multicolumn{3}{Sc|}{\makecell{$\mathcal{H}_1=\mathcal{H}_0+i\mathcal{H}^\mathrm{ah}_1$\\[5pt] $\begin{aligned}\mathcal{H}^\mathrm{ah}_1 =&(a^x_1q_x+a^x_{12}q_xq_y)\sigma_x+(a^y_{1}q_x+a^y_{12}q_xq_y)\sigma_y\\&+a^z_{13}q_xq_z\sigma_z\end{aligned}$}} & 
\begin{minipage}{.2\textwidth}
      \includegraphics[height=21.4mm]{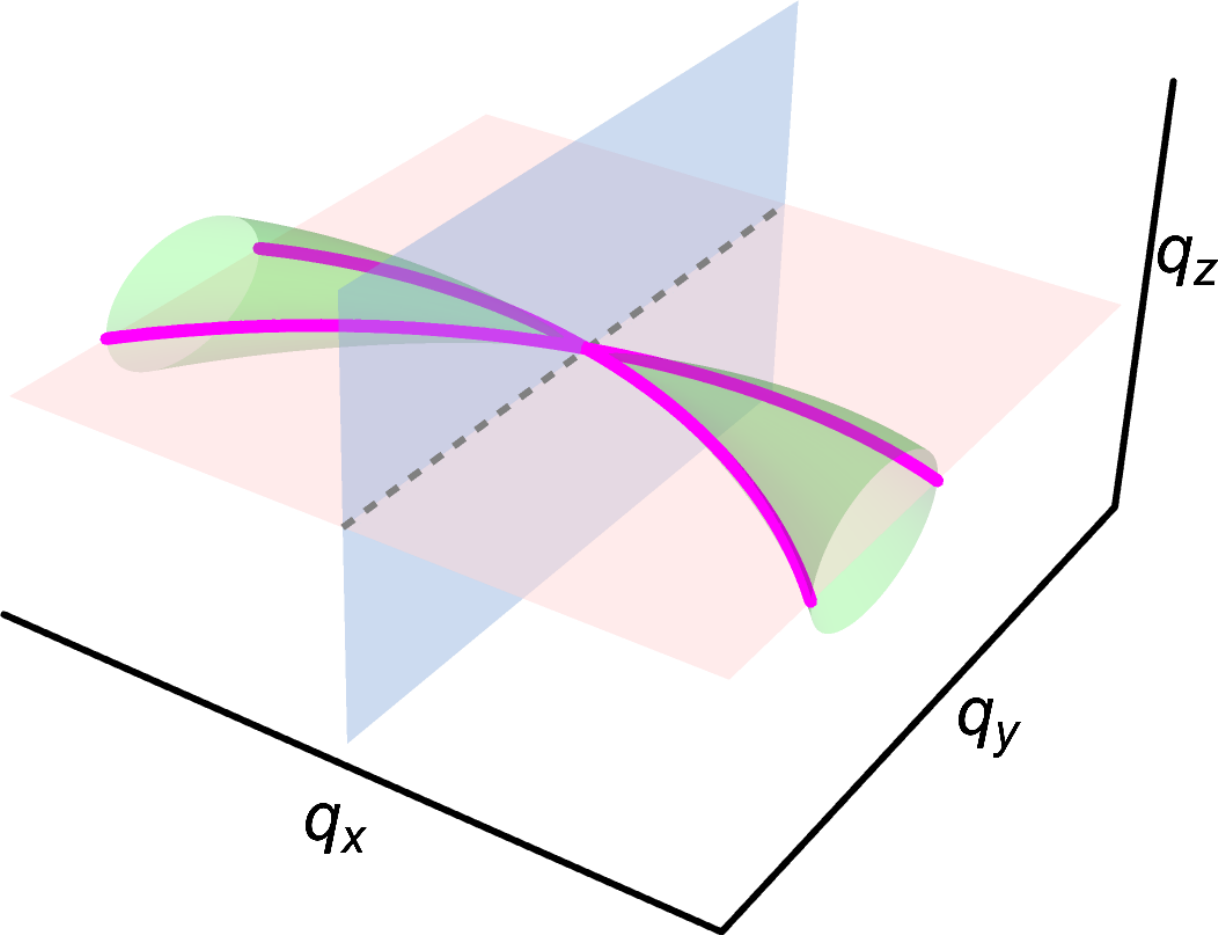}
    \end{minipage}\\\hline

$m_z\blue{m_x^\dagger }\blue{2_y^\dagger}$ & $m_zm_x 2_y\blue{1^\dagger}\rightarrow m_z\blue{m_x^\dagger }\blue{2_y^\dagger}$ & \multicolumn{3}{Sc|}{\makecell{$\mathcal{H}_2=\mathcal{H}_0+i\mathcal{H}^\mathrm{ah}_2$\\[5pt] $\begin{aligned}\mathcal{H}^\mathrm{ah}_2 =&a^x_{13}q_xq_z\sigma_x+a^y_{13}q_xq_z\sigma_y+(a^z_1q_x+a^z_{12}q_xq_y)\sigma_z\end{aligned}$}} & 
\begin{minipage}{.2\textwidth}
      \includegraphics[height=21.4mm]{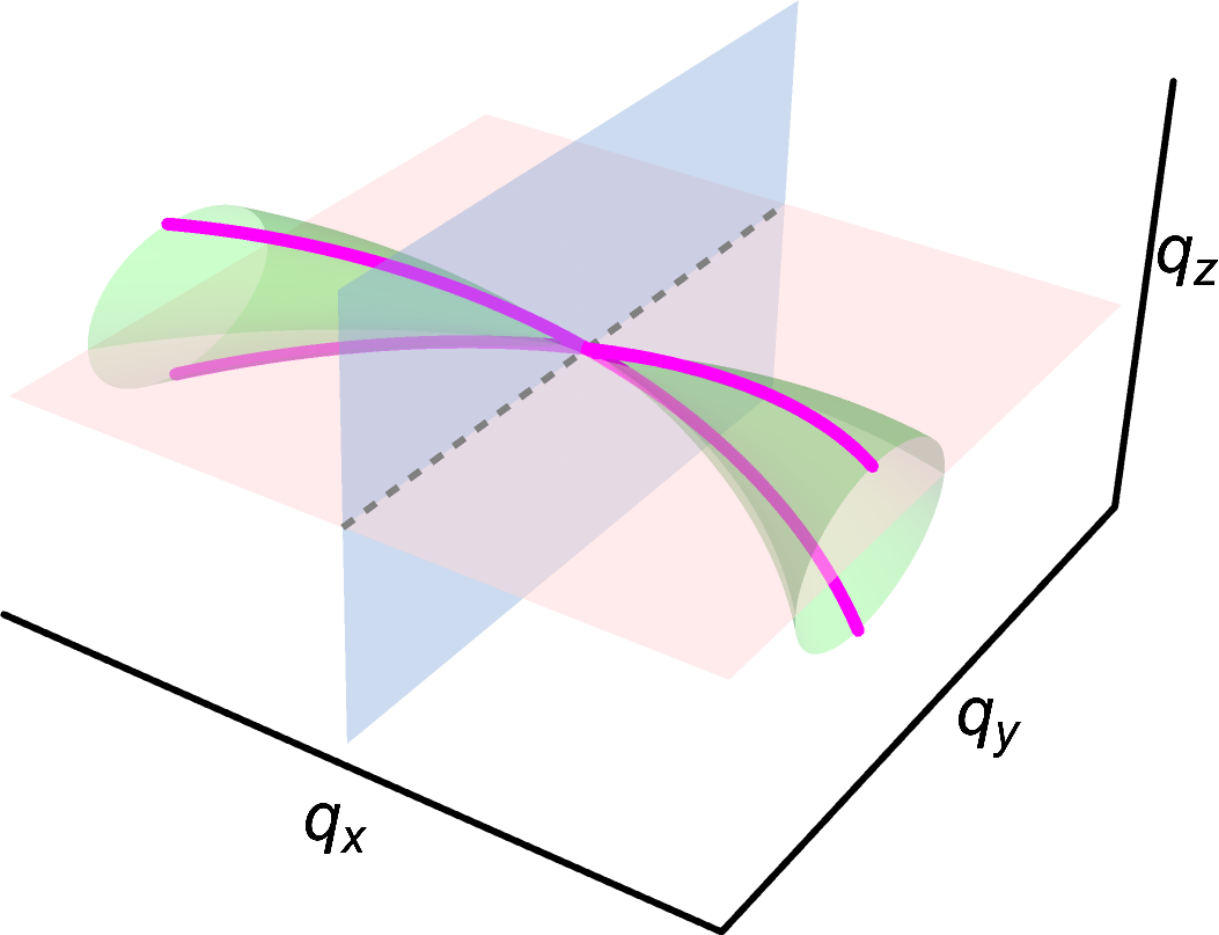}
    \end{minipage}\\\hline

$\blue{m_z^\dagger}$ & $m_zm_x 2_y\blue{1^\dagger}\rightarrow\blue{m_z^\dagger m_x^\dagger}2_y\rightarrow\blue{m_z^\dagger}$ & \multicolumn{3}{Sc|}{\makecell{$\mathcal{H}_3=\mathcal{H}_1+i\mathcal{H}^\mathrm{ah}_3$\\[5pt] $\begin{aligned}\mathcal{H}^\mathrm{ah}_3 =&(a^x_0+a^x_2q_y+a^x_{11}q_x^2+a^x_{22}q_y^2+a^x_{33}q_z^2)\sigma_x\\&+(a^y_0+a^y_2q_y+a^y_{11}q_x^2+a^y_{22}q_y^2+a^y_{33}q_z^2)\sigma_y\\&+(a^z_{3}q_z+a^z_{23}q_yq_z)\sigma_z\end{aligned}$}} & 
\begin{minipage}{.2\textwidth}
      \includegraphics[height=21.4mm]{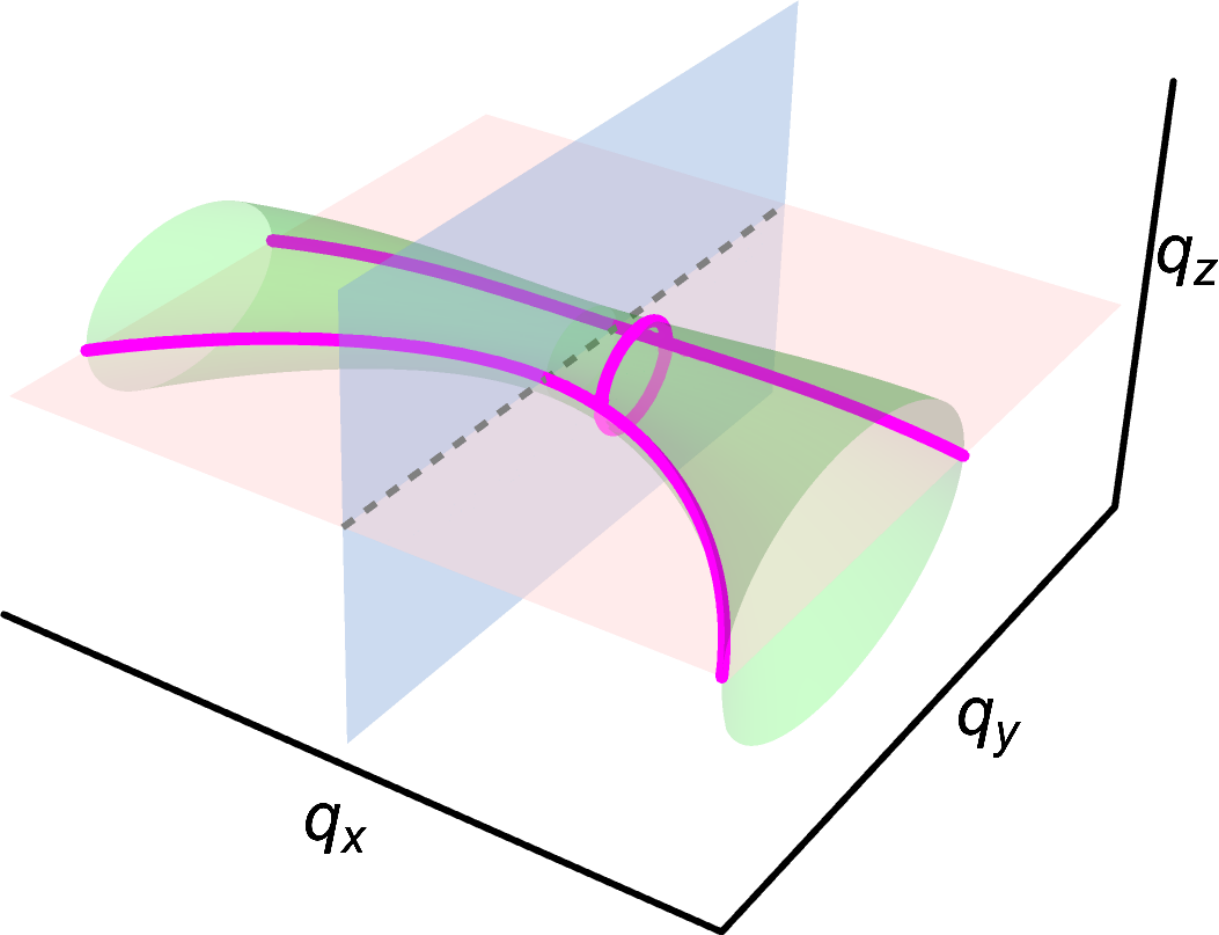}
    \end{minipage}\\\hline

${m_z}$ & $m_zm_x 2_y\blue{1^\dagger}\rightarrow m_z\blue{m_x^\dagger }\blue{2_y^\dagger}\rightarrow {m_z}$ & \multicolumn{3}{Sc|}{\makecell{$\mathcal{H}_4=\mathcal{H}_2+i\mathcal{H}^\mathrm{ah}_4$\\[5pt] $\begin{aligned}\mathcal{H}^\mathrm{ah}_4 =&(a^x_3q_z+a^x_{23}q_yq_z)\sigma_x+(a^y_3q_z+a^y_{23}q_yq_z)\sigma_y\\&+(a^z_0+a^z_2q_y+a^z_{11}q_x^2+a^z_{22}q_y^2+a^z_{33}q_z^2)\sigma_z\end{aligned}$}} & 
\begin{minipage}{.2\textwidth}
      \includegraphics[height=21.4mm]{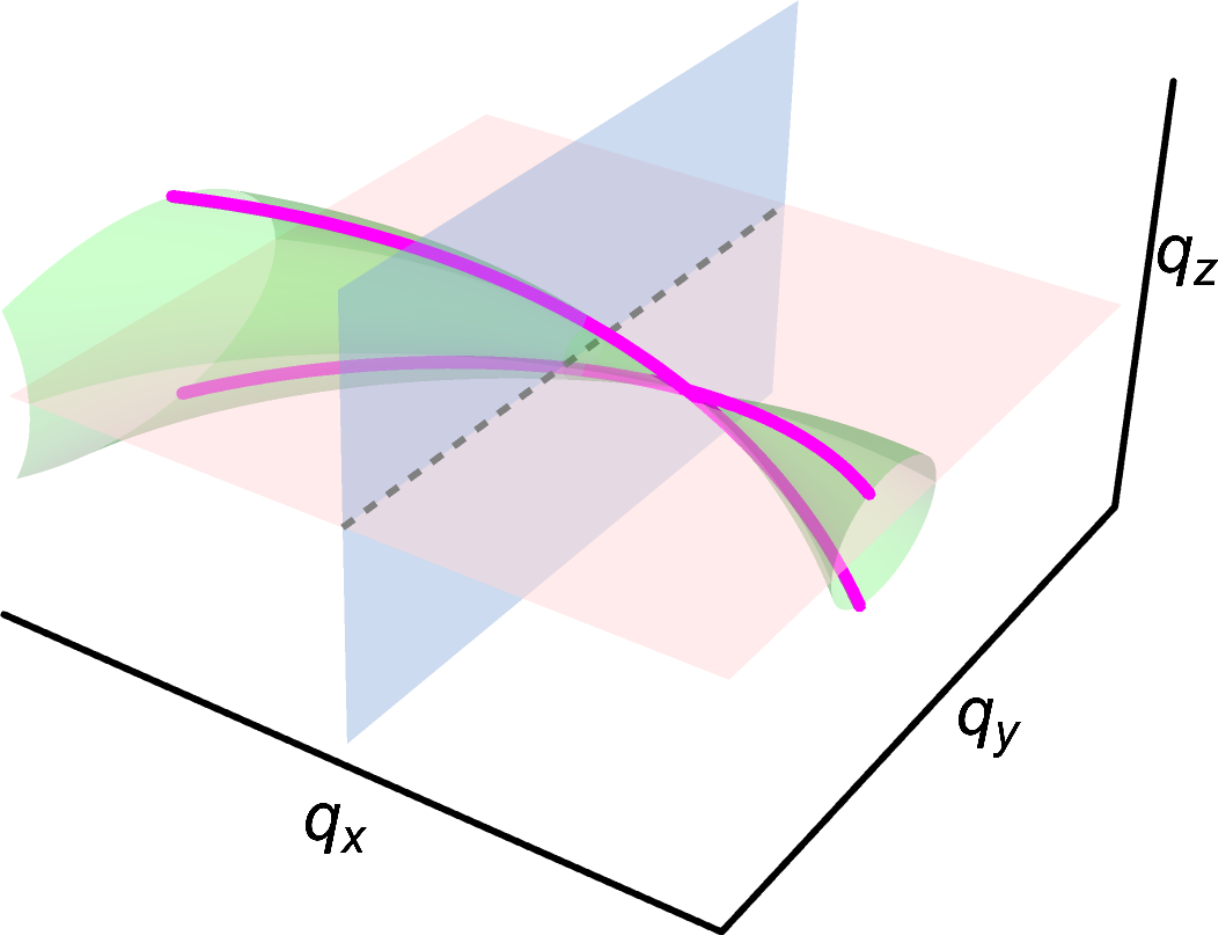}
    \end{minipage}\\\hline\hline

\end{tabular}
\begin{tablenotes}
\item[1] The underlines ``$\underline{\quad}$'' and ``$\underline{\underline{\quad}}$'' denote the non-Hermitian subgroups that can deterministically guarantee the generation of ECs in one-step and two-step perturbations, respectively.
\end{tablenotes}
\end{threeparttable}
\end{adjustwidth}
\end{table}

Now we investigate the situations of $R_2=M_x$ and $R_2=C_{2x}\mathcal{T}$, separately. For the case of $R_2=M_x$, the little group at the intersection $\mathbf{K}$ of the nodal line and $\Pi_x$ is isomorphic to the Hermitian magnetic point group $\mathcal{G}_1=m_zm_x2_y\blue{1^\dagger}$. And the corepresentation of the nodal point at $\mathbf{K}$ is given by $A_1\oplus B_2$ in Mulliken notation, which allows us express the unitary part of $M_z$ and $M_x$ operators as $\hat{M}_z=\sigma_z$ and $\hat{M}_x=\sigma_0$. Hence,the two-band $\mathbf{k}\cdot\mathbf{p}$ Hamiltonian $\mathcal{H}_0(\mathbf{q})$ near the nodal point $\mathbf{k}=\mathbf{K}+\mathbf{q}$ can be obtained (see Table~\ref{perturbation table MzMx}). Then we introduce purely non-Hermitian (i.e., anti-Hermitian) perturbations to the $\mathbf{k}\cdot\mathbf{p}$ Hamiltonian precise to the second order of $\mathbf{q}$:
\begin{equation}
    \mathcal{H}(\mathbf{q})=\mathcal{H}_0(\mathbf{q})+i\mathcal{H}^\mathrm{ah}(\mathbf{q})=\mathcal{H}_0(\mathbf{q})+i\sum_{\mu}\qty[a^\mu_0+\sum_{i}a^\mu_i q_i+\sum_{i,j}a^\mu_{ij}q_iq_j]\sigma_\mu,
\end{equation}
where the anti-Hermitian part $i\mathcal{H}^\mathrm{ah}$ should obey the symmetries of a \textbf{non-Hermitian double-antisymmetry point group} as a subgroup of $\mathcal{G}_1=m_z m_x2_y\blue{1^\dagger}$. Based on the three mechanisms of EC formation we proposed in the main text, we list in Table~\ref{perturbation table MzMx} all the possible evolution paths that can deterministically guarantee the generation of ECs under thresholdless perturbations in each step of the path. And the local EC configurations protected by the corresponding non-Hermitian subgroups are exhibited in the last column of the table. It shows that planar EC and mirror-symmetric EC can be realized by thresholdless perturbations obeying $\blue{m_z^\dagger m_x^\dagger}2_y$ and $m_z\blue{m_x^\dagger2_y^\dagger}$ subgroups, respectively, both of which have non-defective chain point fixing along the high-symmetry line of $q_x=q_z=0$. If we further impose $\blue{m_z^\dagger}$ and $m_z$ types of perturbations to the planar and mirror-symmetric ECs, respectively, the chains persist while the plannar EC becomes a earring EC and the chain point of the mirror-symmetric EC moves always from the high-symmetry line.

\begin{figure*}[b!]
\includegraphics[width=0.5\textwidth]{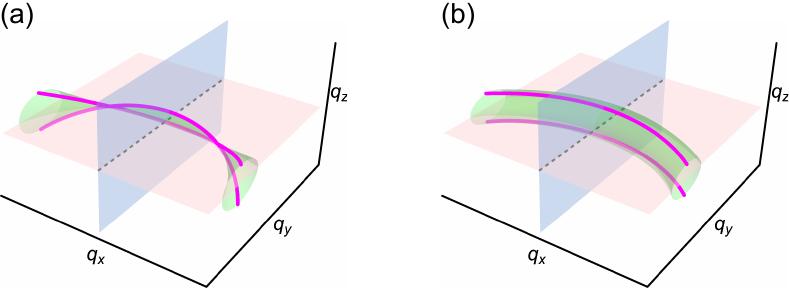}
\caption{\label{MzMx_nH} 
Possible configurations of ELs induced by the thresholdless non-Hermitian perturbations obeying $m_zm_x2_y$ from the Hermitian nodal line. (a) EC with two $M_x$-symmetric chain points, (b) two nonconnected ELs.}
\end{figure*}

It should be aware that stable ECs may also be generated by other types of non-Hermitian perturbations not listed in Table~\ref{perturbation table MzMx}, e.g.  the non-Hermitian perturbations respecting the subgroup $m_zm_x2_y$ as shown in Fig.~\ref{MzMx_nH}. However, in these cases, the formation of ECs are not deterministic but relies on the detailed parameters of the perturbations.

\begin{table}[b!]
\begin{adjustwidth}{-.5in}{-.5in}  
\centering
\setlength\cellspacetoplimit{3pt}
\setlength\cellspacebottomlimit{3pt}
\addtolength\tabcolsep{1pt}
\caption{Deterministic schemes for realizing ECs by thresholdless non-Hermitian perturbations from a Hermitian nodal point (the black point shown by the nodal structure in the first line) on a nodal line protected by $\red{2'_x}/m_z\blue{1^\dagger}$ point group at the intersection $\mathbf{K}$ of a nodal line and a high-symmetry line (the crossing line of the red and blue planes). \label{perturbation table 2xMz}}
\begin{threeparttable}
\begin{tabular}{Sc|Sc|Sc|Sc|Sc|Sc}\hline\hline
\textbf{\makecell{Hermitian\\ point group}} & \textbf{\makecell{non-Hermitian\\ subgroups~\tnote{1}}}
& \textbf{Corep} & \textbf{Generators} & \textbf{$\mathbf{k}\cdot\mathbf{p}$ Hamiltonian} & \textbf{nodal structure}\\\hline

$\red{2'_xm'_y}m_z\blue{1^\dagger}$ & \makecell{$\underline{\red{2'_x}\green{m_y^{\prime\dagger}\blue{m_z^\dagger}}}$, $\underline{\red{2'_xm'_y}{m_z}}$, $\green{2_x^{\prime\dagger}}\red{m'_y}\blue{m_z^\dagger}$,\\[8pt] $\green{2_x^{\prime\dagger}m_y^{\prime\dagger}}{m_z}$, $\underline{\underline{\blue{m_z^\dagger}}}$, $\underline{\underline{m_z}}$, $\underline{\underline{\red{2'_x}}}$, $\green{2_x^{\prime\dagger}}$,\\[8pt] $\red{m'_y}$, $\green{m_y^{\prime\dagger}}$} & 
$A'\oplus A''$ & 
\makecell{$\hat{M}_z=\sigma_z$\\[5pt]$\hat{C}_{2x}\hat{T}=\sigma_z$} &
$\begin{aligned}\mathcal{H}_0 &=h_{13}^x q_xq_z\sigma_x\\&+(h^y_3 q_z+h_{23}^y q_yq_z)\sigma_y\\&+(h^z_2q_y+h^x_{11}q_x^2+h^y_{11}q_y^2\\&\quad+h^z_{11}q_z^2)\sigma_z\end{aligned}$ &
\begin{minipage}{.2\textwidth}
      \includegraphics[height=21.4mm]{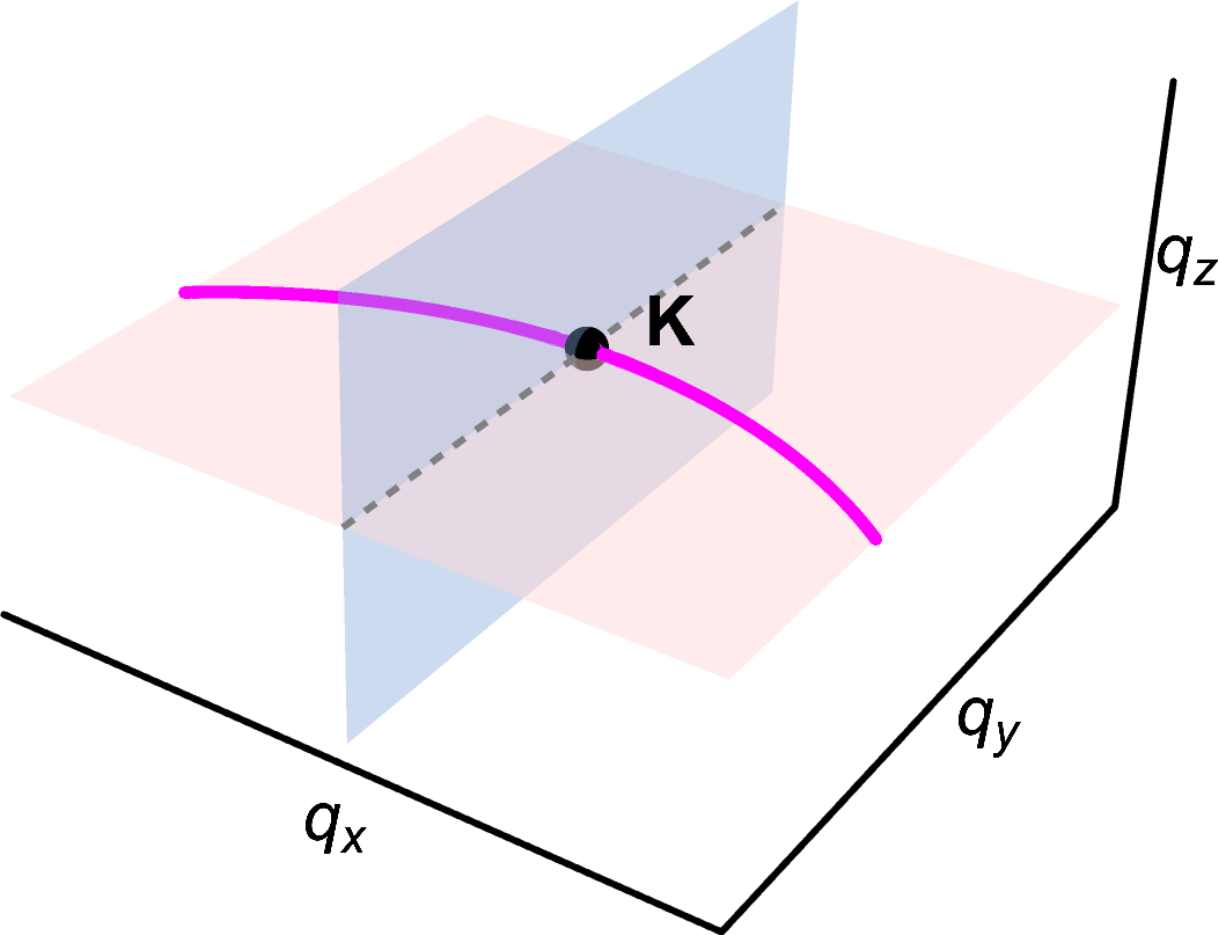}
    \end{minipage}\\\specialrule{.15em}{.05em}{.05em} 

\textbf{subgroup} & \textbf{evolution path } & \multicolumn{3}{Sc|}{ \textbf{$\mathbf{k}\cdot\mathbf{p}$ Hamiltonian}} & \textbf{EC structure}\\\hline

$\red{2'_x}\green{m_y^{\prime\dagger}\blue{m_z^\dagger}}$ & $\red{2'_xm'_y}m_z\blue{1^\dagger}\rightarrow\red{2'_x}\green{m_y^{\prime\dagger}\blue{m_z^\dagger}}$ & \multicolumn{3}{Sc|}{\makecell{$\mathcal{H}_1=\mathcal{H}_0+i\mathcal{H}^\mathrm{ah}_1$\\[5pt] $\begin{aligned}\mathcal{H}^\mathrm{ah}_1 =&(a^x_0+a^x_2q_y+a^x_{11}q_x^2+a^x_{22}q_y^2+a^x_{33}q_z^2)\sigma_x\\&+(a^y_{1}q_x+a^y_{12}q_xq_y)\sigma_y+a^z_{13}q_xq_z\sigma_z\end{aligned}$}} & 
\begin{minipage}{.2\textwidth}
      \includegraphics[height=21.4mm]{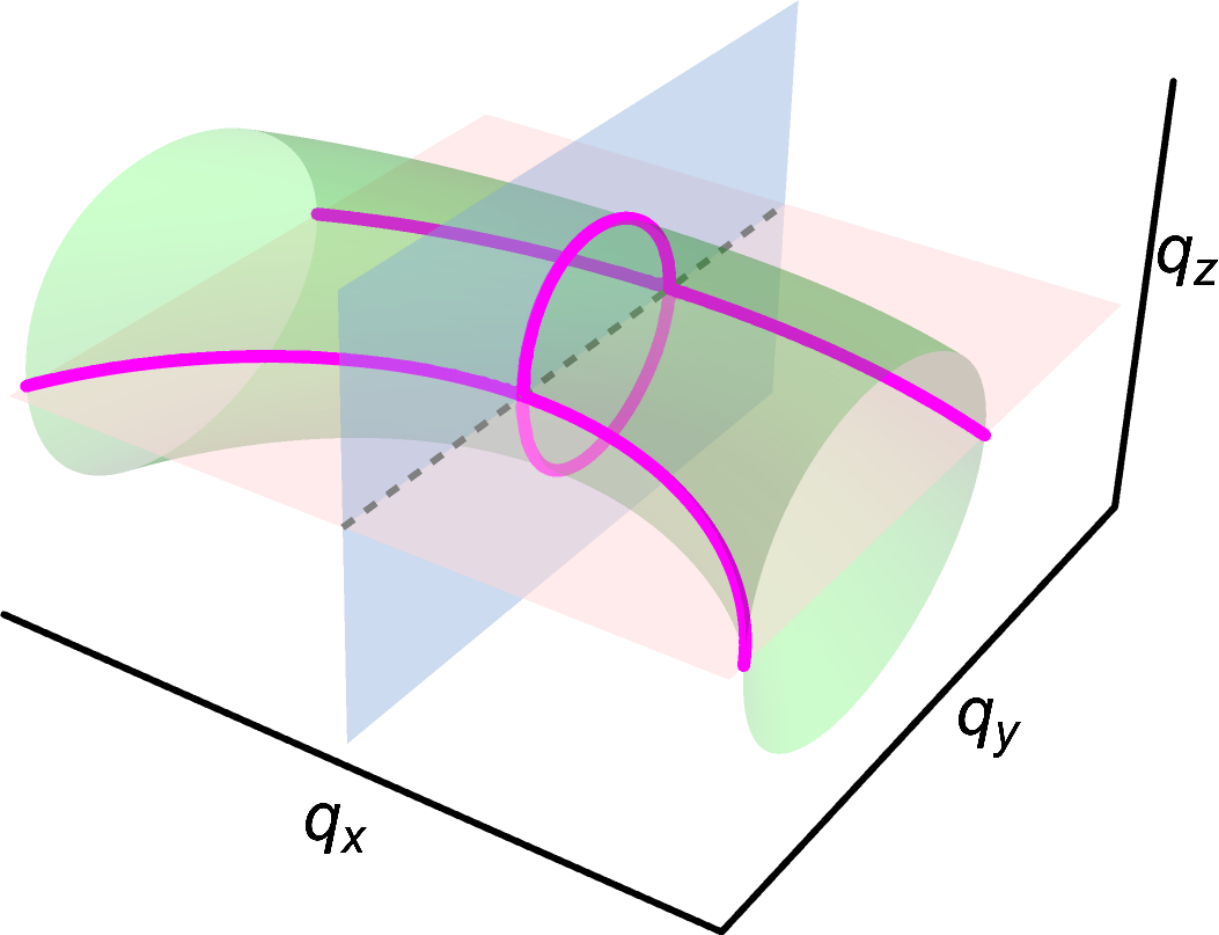}
    \end{minipage}\\\hline
$\red{2'_xm'_y}{m_z}$ & $\red{2'_xm'_y}m_z\blue{1^\dagger}\rightarrow\red{2'_xm'_y}{m_z}$ & \multicolumn{3}{Sc|}{\makecell{$\mathcal{H}_2=\mathcal{H}_0+i\mathcal{H}^\mathrm{ah}_2$\\[5pt] $\begin{aligned}\mathcal{H}^\mathrm{ah}_2 =&(a^x_3q_z+a^x_{23}q_yq_z)\sigma_x+a^y_{13}q_xq_z\sigma_y\\&+(a^z_1q_x+a^z_{12}q_xq_y)\sigma_z\end{aligned}$}} & 
\begin{minipage}{.2\textwidth}
      \includegraphics[height=21.4mm]{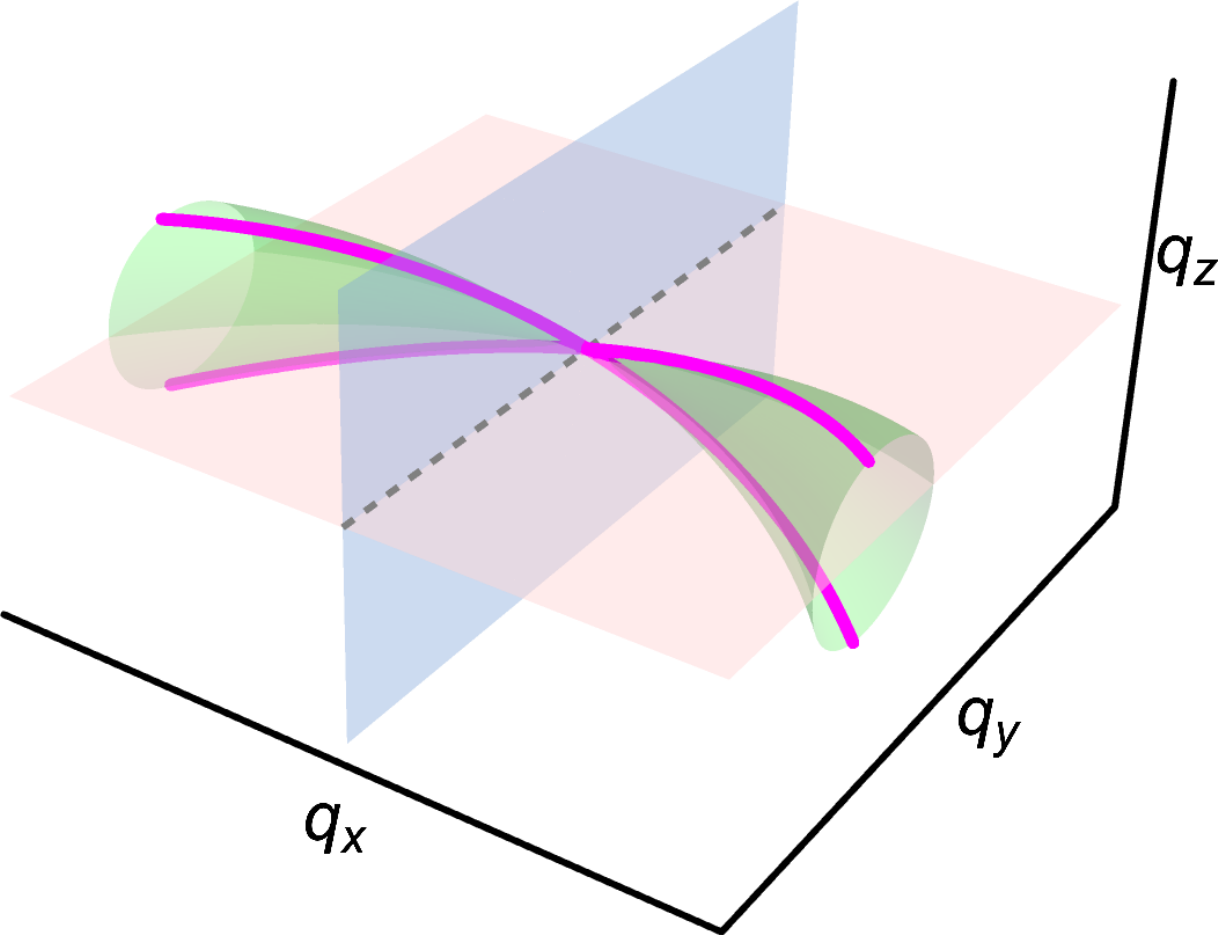}
    \end{minipage}\\\hline


$\blue{m_z^\dagger}$ & $\red{2'_xm'_y}m_z\blue{1^\dagger}\rightarrow\red{2'_x}\green{m_y^{\prime\dagger}\blue{m_z^\dagger}}\rightarrow\blue{m_z^\dagger}$ & \multicolumn{3}{Sc|}{\makecell{$\mathcal{H}_3=\mathcal{H}_1+i\mathcal{H}^\mathrm{ah}_3$\\[5pt] $\begin{aligned}\mathcal{H}^\mathrm{ah}_3 =&(a^x_1q_x+a^x_{12}q_xq_y)\sigma_x\\&+(a^y_0+a^y_2q_y+a^y_{11}q_x^2+a^y_{22}q_y^2+a^y_{33}q_z^2)\sigma_y\\&+(a^z_3q_z+a^z_{23}q_yq_z)\sigma_z\end{aligned}$}} & 
\begin{minipage}{.2\textwidth}
      \includegraphics[height=21.4mm]{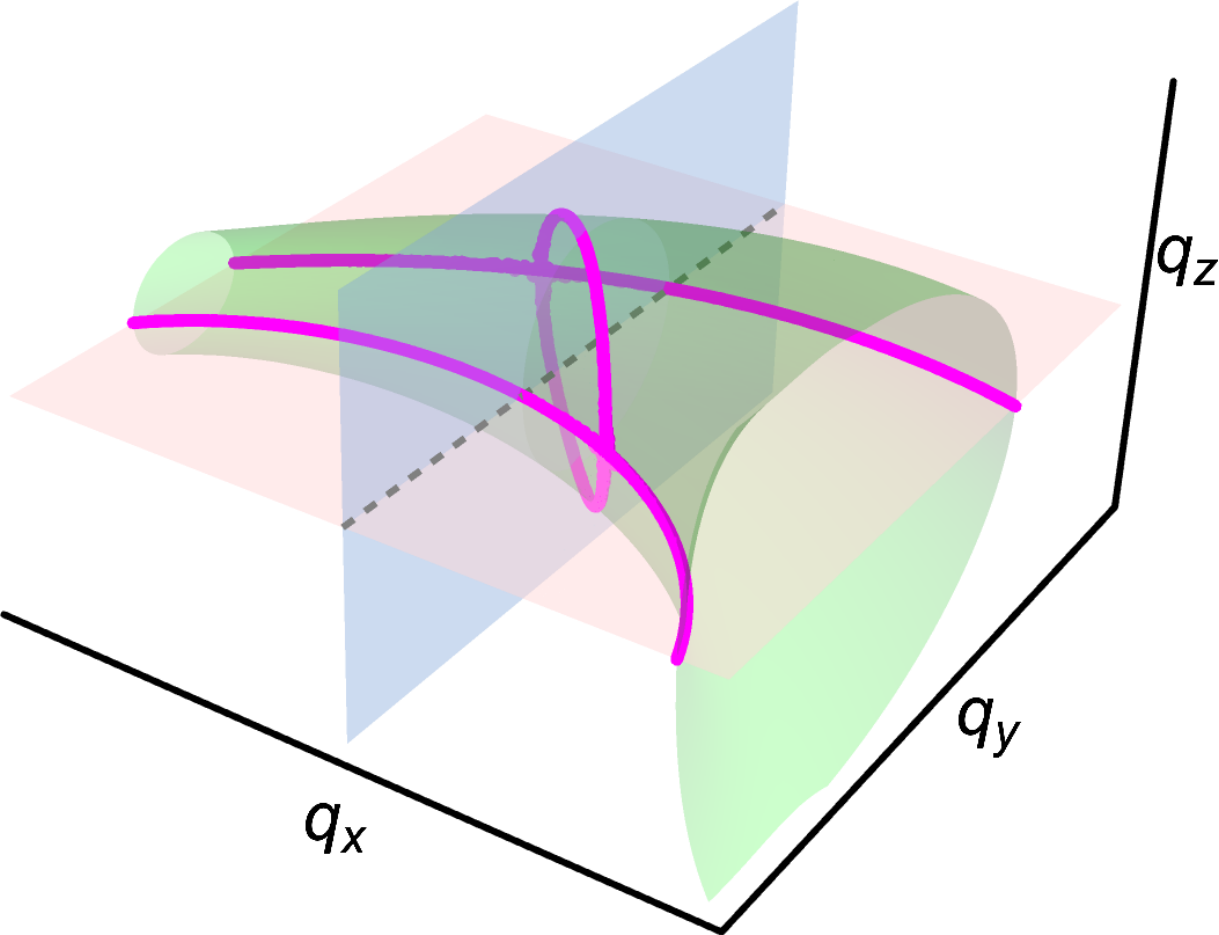}
    \end{minipage}\\\hline
    
${m_z}$ & $\red{2'_xm'_y}m_z\blue{1^\dagger}\rightarrow\red{2'_xm'_y}{m_z}\rightarrow {m_z}$ & \multicolumn{3}{Sc|}{\makecell{$\mathcal{H}_4=\mathcal{H}_2+i\mathcal{H}^\mathrm{ah}_4$\\[5pt] $\begin{aligned}\mathcal{H}^\mathrm{ah}_4 =&a^x_{13}q_xq_z\sigma_x+(a^y_3q_z+a^y_{23}q_yq_z)\sigma_y\\&+(a^z_0+a^z_2q_y+a^z_{11}q_x^2+a^z_{22}q_y^2+a^z_{33}q_z^2)\sigma_z\end{aligned}$}} & 
\begin{minipage}{.2\textwidth}
      \includegraphics[height=21.4mm]{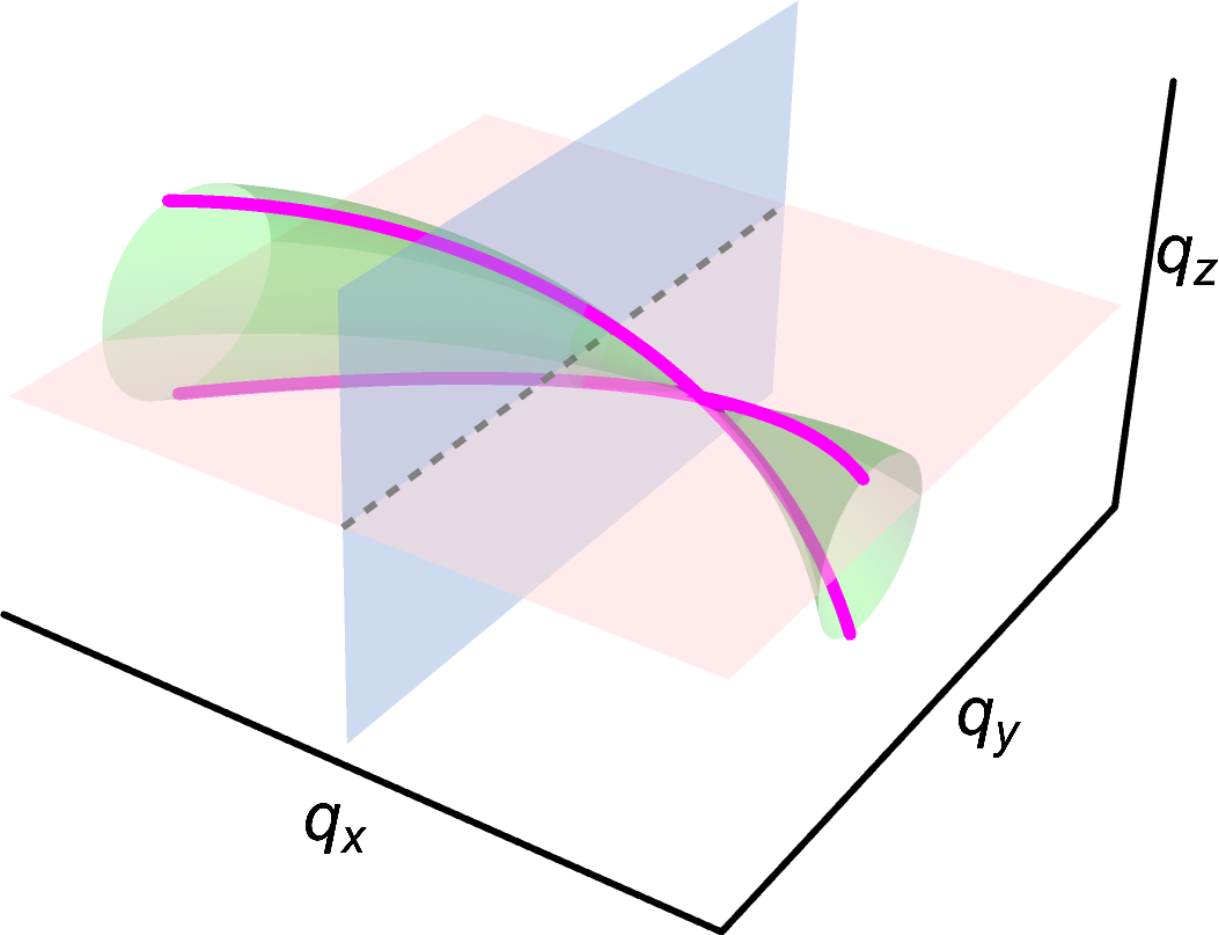}
    \end{minipage}\\\hline
    
\multirow{2}{*}[-18pt]{$\red{2'_x}$} & $\red{2'_xm'_y}m_z\blue{1^\dagger}\rightarrow\red{2'_x}\green{m_y^{\prime\dagger}}\blue{/m_z^\dagger}\rightarrow \red{2'_x}$ & \multicolumn{3}{Sc|}{\makecell{$\mathcal{H}_5=\mathcal{H}_1+i\mathcal{H}^\mathrm{ah}_5$\\[5pt] $\begin{aligned}\mathcal{H}^\mathrm{ah}_5 =&(a^x_3q_z+a^x_{23}q_yq_z)\sigma_x+a^y_{13}q_xq_z\sigma_y+(a^z_1q_x+a^z_{12}q_xq_y)\sigma_z\end{aligned}\hspace{-10pt}$}}& 
\multirow{2}{*}[5pt]{\begin{minipage}{.2\textwidth}
      \includegraphics[height=21.4mm]{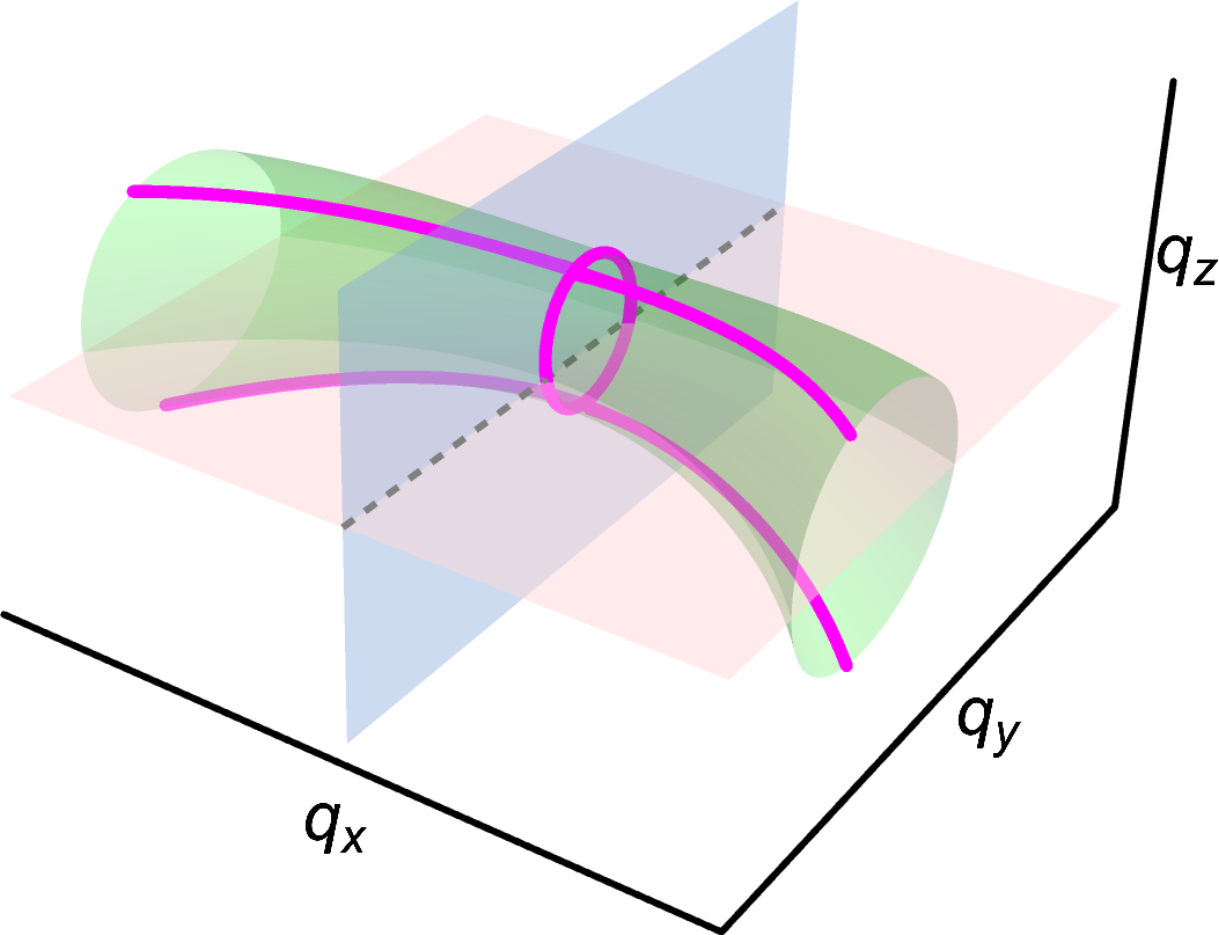}
    \end{minipage}}\\\cline{2-5}

& $\red{2'_xm'_y}m_z\blue{1^\dagger}\rightarrow\red{2'_xm'_y}{m_z}\rightarrow \red{2'_x}$ & \multicolumn{3}{Sc|}{\makecell{$\mathcal{H}_5'=\mathcal{H}_2+i\mathcal{H}^\mathrm{ah\prime}_5$\\[5pt] $\begin{aligned}\mathcal{H}^\mathrm{ah\prime}_5 =&(a^x_0+a^x_2q_y+a^x_{11}q_x^2+a^x_{22}q_y^2+a^x_{33}q_z^2)\sigma_x\\&+(a^y_1q_x+a^y_{12}q_xq_y)\sigma_y+a^z_{13}q_xq_z\sigma_z\end{aligned}$}} & \\\hline\hline

\end{tabular}
\begin{tablenotes}
\item[1] The underlines ``$\underline{\quad}$'' and ``$\underline{\underline{\quad}}$'' denote the non-Hermitian subgroups that can deterministically guarantee the generation of ECs in one-step and two-step perturbations, respectively.
\end{tablenotes}
\end{threeparttable}
\end{adjustwidth}
\end{table}

Similarly, for $R_2=C_{2x}\mathcal{T}$,  the Hermitian nodal point at $\mathcal{K}$ has a little group isomorphic to $\mathcal{G}_2=\red{2'_xm'_y}m_z\blue{1^\dagger}$ with the corepresentation $A'\oplus A''$. Writing the generators of $\mathcal{G}_2$ as $\hat{M}_z=\sigma_z$ and $\hat{C}_{2x}\hat{T}=\sigma_z$, we can obtain the $\mathbf{k}\cdot\mathbf{p}$ Hamiltonians near $\mathbf{K}$ and all possible pathways of thresholdless non-Hermitian perturbations that can deterministically induce the generation of ECs, as listed in Table~\ref{perturbation table 2xMz}.

Indeed, we can also generalize these local deterministic schemes for achieving ECs from a local nodal point on a Hermitian nodal lines to global 2-band models as shown by the schemtic in Fig.~4 of the main text. In Table~\ref{Hamiltonians for Fig4}, the detailed Hamiltonians utilized to produce the EC configurations are listed.

\begin{table}[t!]
\begin{adjustwidth}{-.5in}{-.5in}  
\centering
\setlength\cellspacetoplimit{3pt}
\setlength\cellspacebottomlimit{3pt}
\addtolength\tabcolsep{1pt}
\caption{List of the two-band Hamiltonians with certain symmetries used in Fig.~4 of the main text for generating the EC configurations perturbed from a Hermitian nodal ring. In all cases, the local symmetry operators are identically given by $\hat{M}_z=\sigma_z$, $\hat{M}_x=\sigma_0$, and $\hat{C}_{2x}\hat{T}=\sigma_z$.  \label{Hamiltonians for Fig4}}
\begin{tabular}{Sc|Sc|Sc}\hline\hline
\textbf{\makecell{Symmetries}} & \textbf{\makecell{2-band Hamiltonian}} & \textbf{nodal structure}\\\hline

\parbox[c][1.15cm]{3cm}{$(M_z,C_{2x}\mathcal{T},1^\dagger)$} & 
\makecell{$\begin{aligned}\mathcal{H}_0 &=(v_x\sin k_x\,\sigma_x+v_y\sigma_y)\sin{k_z}+v_z(\cos{k_x}+\cos{k_y}+\cos{k_z})\sigma_z\end{aligned}$\\[5pt] with $v_x=v_y=v_z=1$} &
\multirow{2}{*}[12pt]{\makecell{
      \includegraphics[height=21.4mm,valign=c]{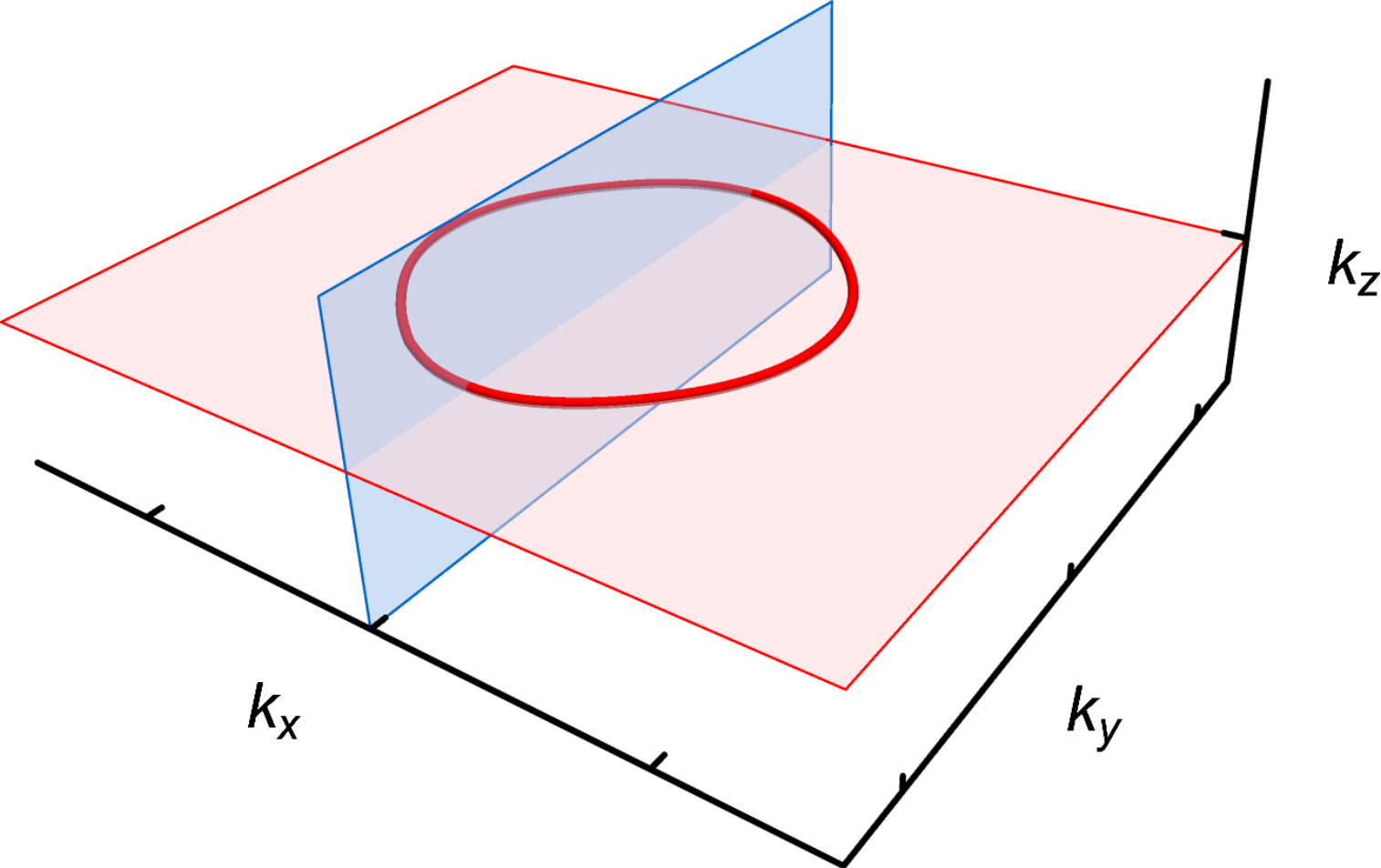}
    }}\\\cline{1-2}

\parbox[c][1.15cm]{3cm}{$(M_z,M_x,1^\dagger)$} & \makecell{$\begin{aligned}\mathcal{H}'_0 &=(v_x\sigma_x+v_y\sigma_y)\sin{k_z}+v_z(\cos{k_x}+\cos{k_y}+\cos{k_z})\sigma_z\end{aligned}$\\[5pt] with $v_x=v_y=v_z=1$} & \\\hline

(\mdagger[z],\,\mdagger[x]\,) & \makecell{$\mathcal{H}_1=\mathcal{H}'_0+i\qty(u_x\sigma_x+u_y\sigma_y)\sin{k_x}$\\[5pt]
with $u_x=u_y=-0.4$} & \begin{minipage}{.2\textwidth}
      \includegraphics[height=21.4mm]{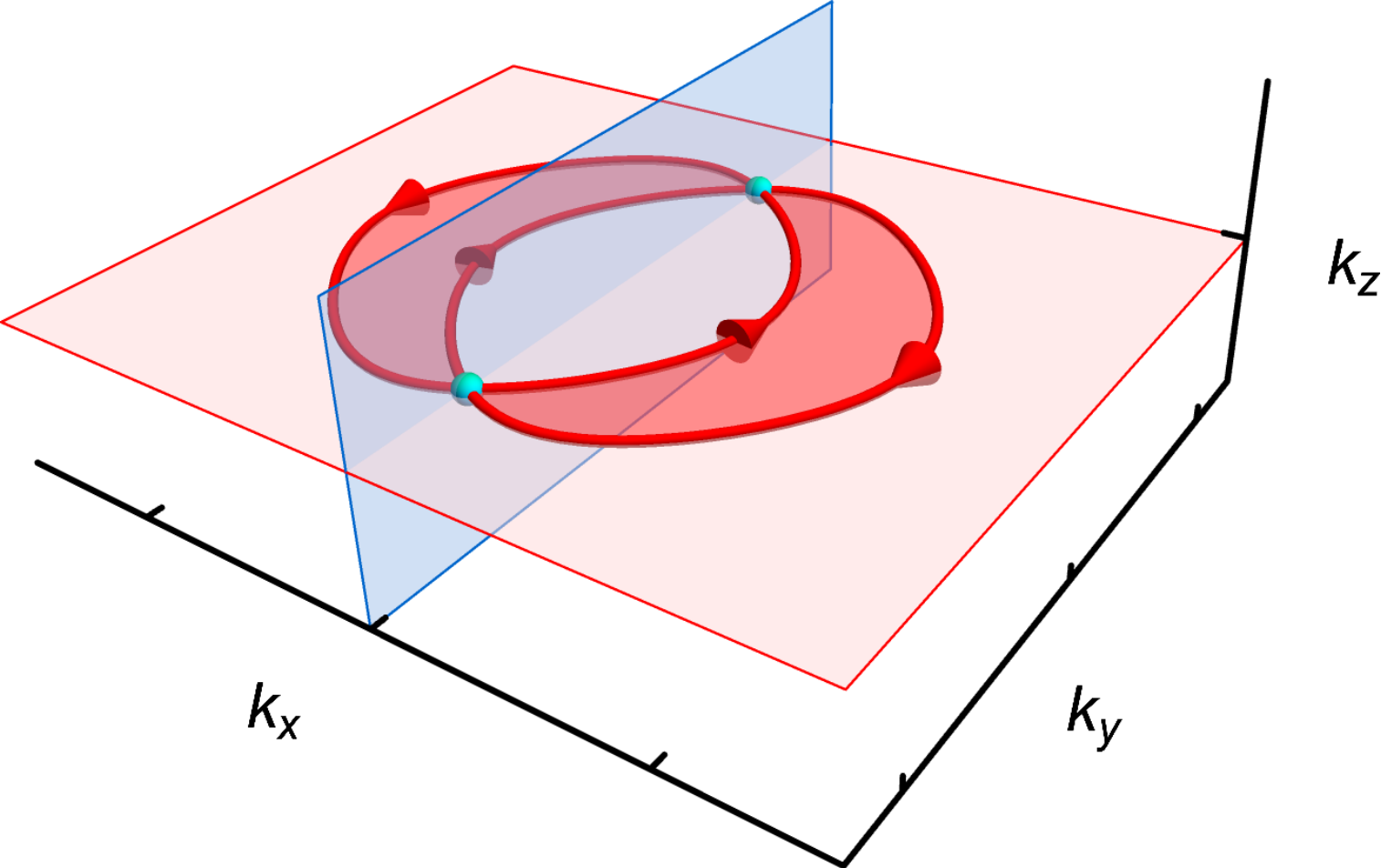}\end{minipage}\\\hline

(\mdagger[z],\,$C_{2x}\mathcal{T}$) & \makecell{$\mathcal{H}_2=\mathcal{H}_0+i\qty(u_x\sigma_x+u_y\sin{k_x}\sigma_y)$\\[5pt] with $u_x=u_y=-0.4$} & \begin{minipage}{.2\textwidth}
      \includegraphics[height=21.4mm]{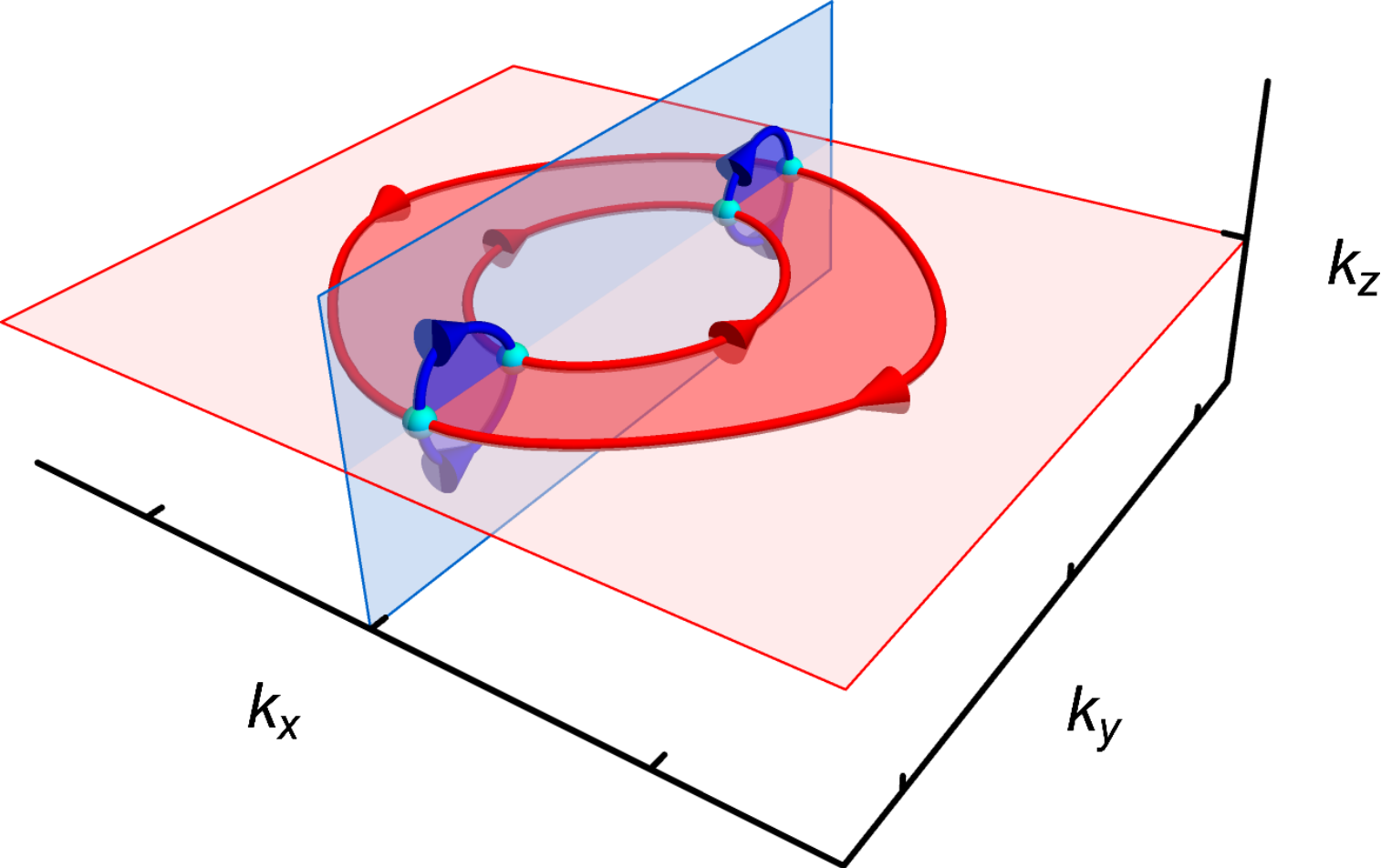}\end{minipage}\\\hline
      
\parbox[c][1.15cm]{3cm}{$(M_z,C_{2x}\mathcal{T})$} & 
\makecell{$\begin{aligned}\mathcal{H}_3 &=\mathcal{H}_0+i\qty[(u_x\sigma_x+u_y\sin{k_x}\sigma_y)\sin{k_z}+u_z\sin{k_x}\sigma_z]\end{aligned}$\\[5pt] with $u_x=u_y=u_z=0.4$} &
\multirow{2}{*}[12pt]{\makecell{
      \includegraphics[height=21.4mm,valign=c]{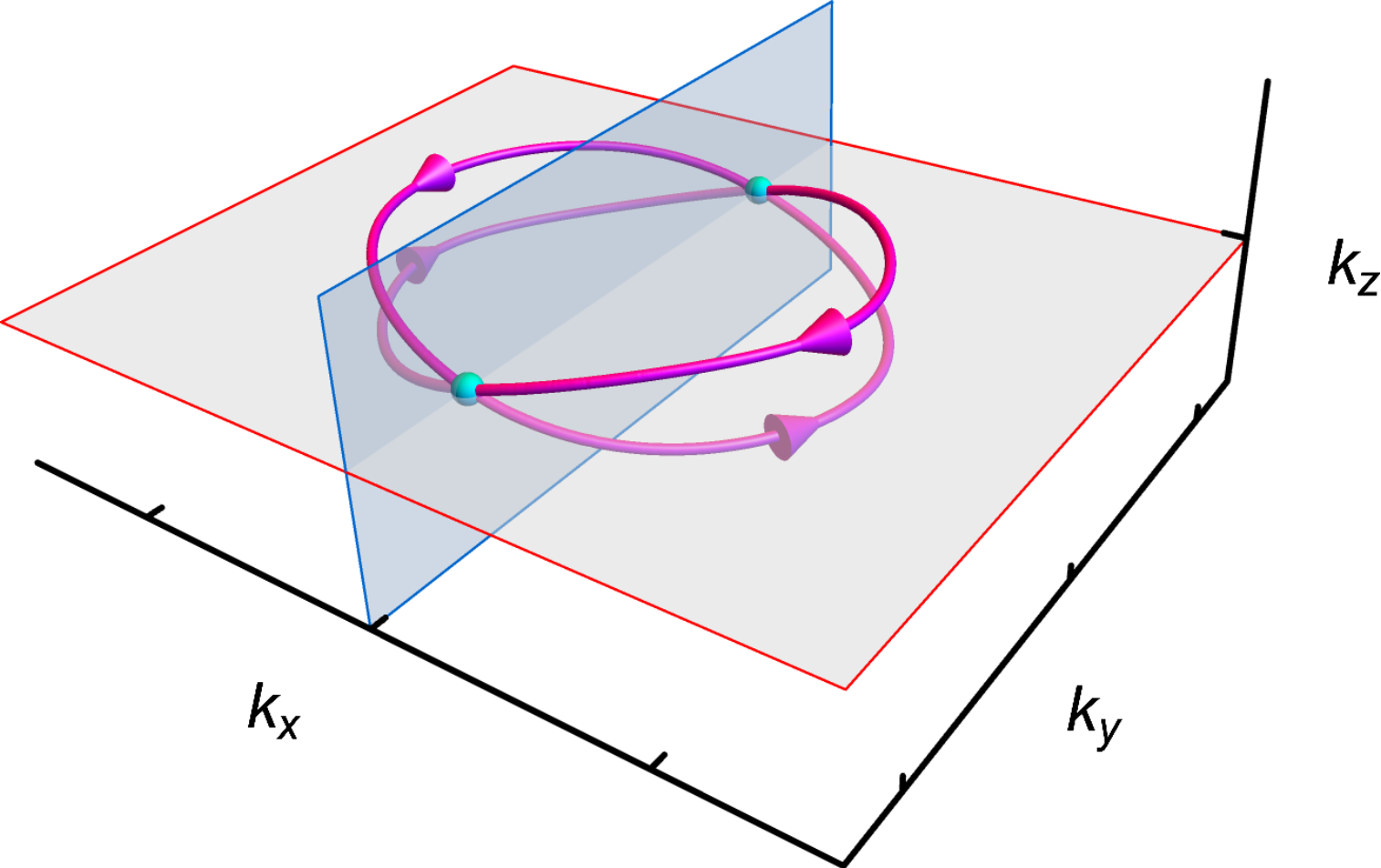}
    }}\\\cline{1-2}
\parbox[c][1.15cm]{3cm}{$(M_z,M_x\mbox{-}\dagger)$} & \makecell{$\begin{aligned}\mathcal{H}'_3 &=\mathcal{H}'_0+i\qty[(u_x\sigma_x+u_y\sigma_y)\sin{k_x}\sin{k_z}+u_z\sin{k_x}\sigma_z]\end{aligned}$\\[5pt] with $u_x=u_y=u_z=0.4$} & \\\hline

(\mdagger[z]) & \makecell{$\mathcal{H}_4=\mathcal{H}'_0+i\qty[u_x\sin{k_x}\sigma_x+u_y(\sin{k_x}+\cos{k_x})\sigma_y+u_z\sin{k_z}\sigma_z]$\\[5pt]
with $u_x=u_y=-0.4$ and $u_z=1.1$} & \begin{minipage}{.17\textwidth}
      \includegraphics[height=21.4mm]{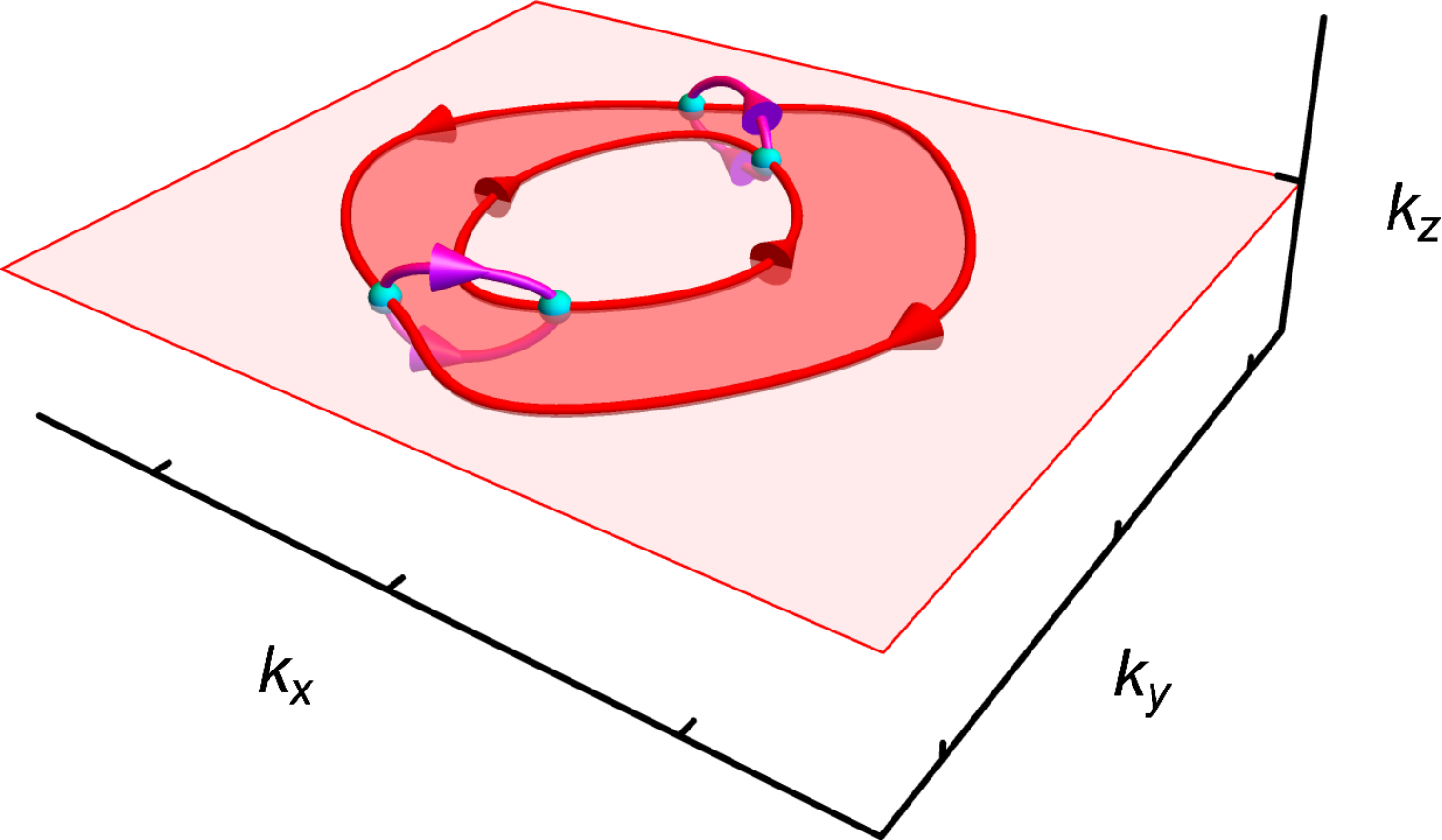}\end{minipage}\\\hline
      
($C_{2x}\mathcal{T}$) & \makecell{$\mathcal{H}_5=\mathcal{H}_0+v'_y\sigma_y+i\qty(u_x\sin{k_y}\sigma_x+u_y\sin{k_x}\sigma_y)$\\[5pt]
with $v'_y=0.02$ and $u_x=-0.4$, $u_y=-0.2$} & \begin{minipage}{.2\textwidth}
      \includegraphics[height=21.4mm]{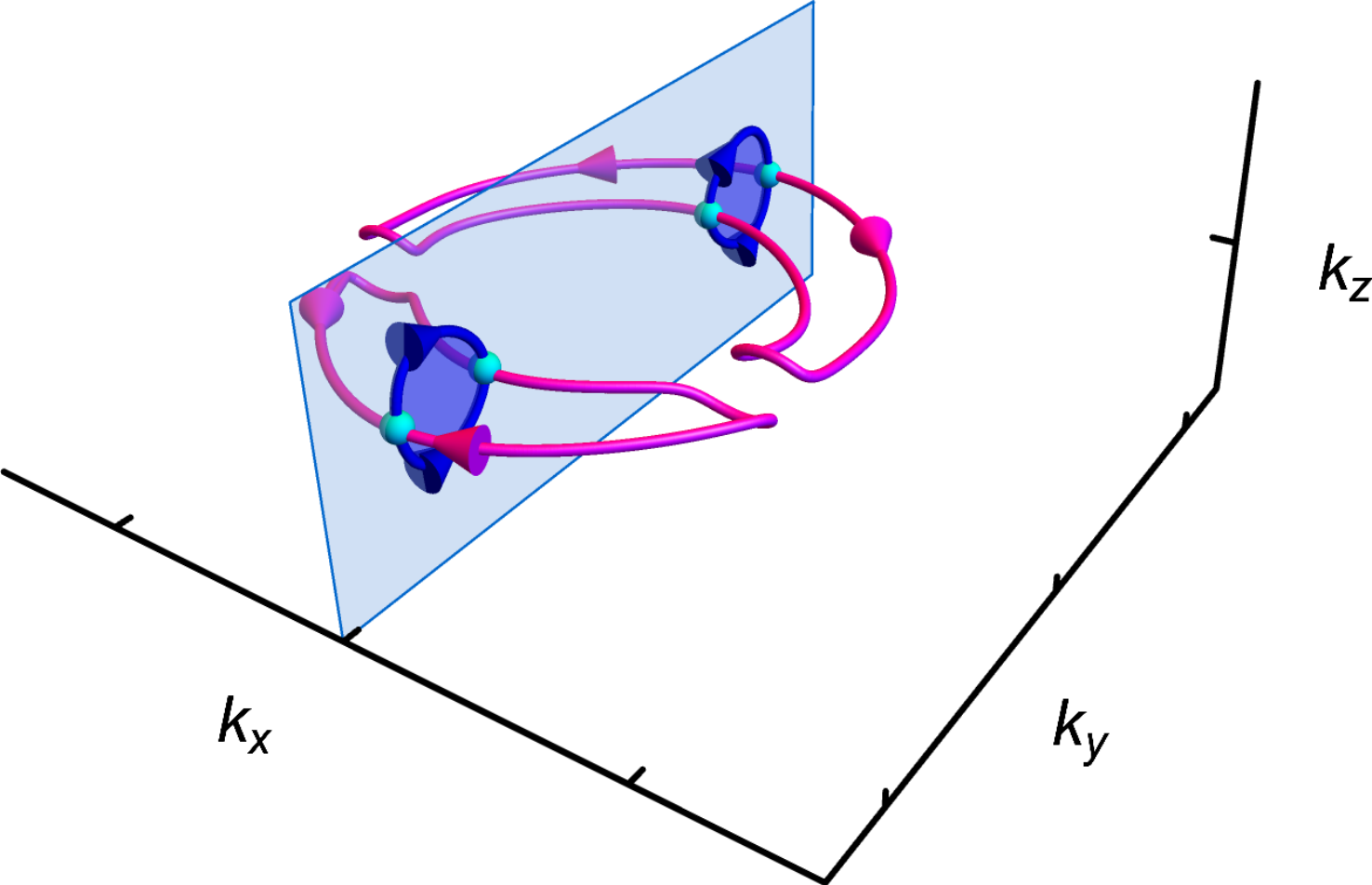}\end{minipage}\\\hline
      
($M_z$) & \makecell{$\mathcal{H}_6=\mathcal{H}'_0+i\qty[(u_x\sigma_x+u_y\sigma_y)\sin{k_x}\sin{k_z}+(u_z\sin{k_z}+u'_z\cos{k_x})\sigma_z]$\\[5pt]
with $u_x=u_z=0.5$, $u_y=0.4$ and $u'_z=-0.7$} & \begin{minipage}{.2\textwidth}
      \includegraphics[height=21.4mm]{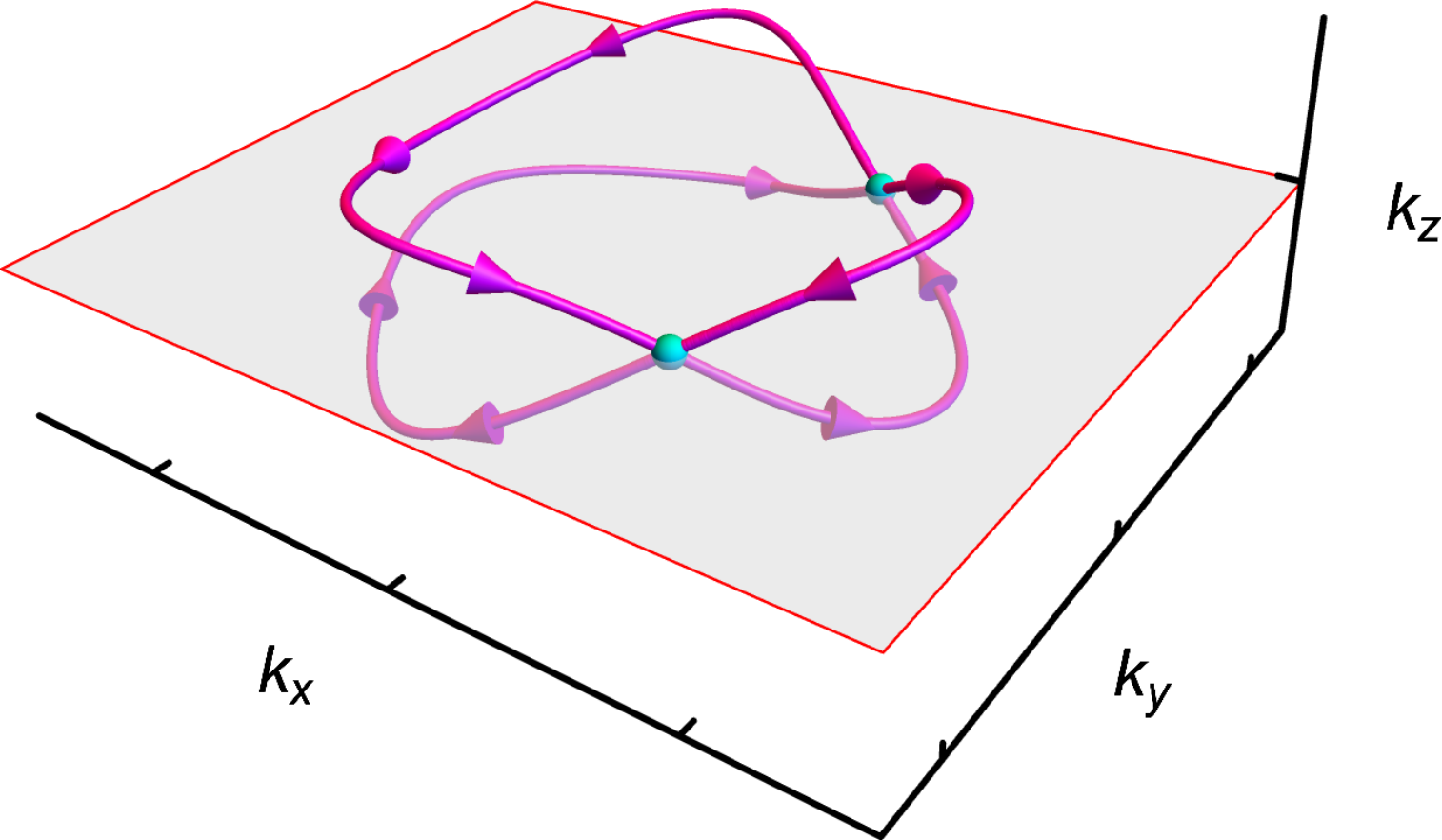}\end{minipage}\\\hline\hline

\end{tabular}
\end{adjustwidth}
\end{table}

\clearpage
\section{Exceptional links evolved from ECs}

\begin{figure*}[h!]
\includegraphics[width=0.6\textwidth]{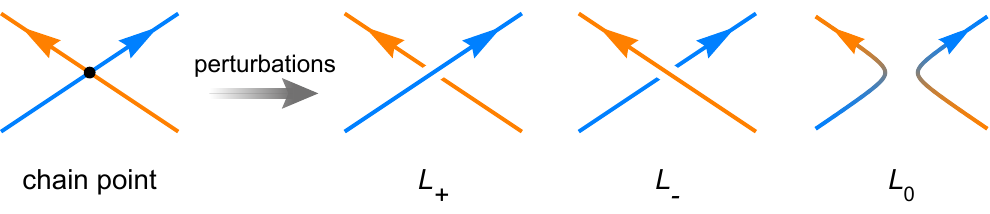}
\caption{\label{Chain point evolution} 
Three possible evolutions of a chain point intersected by two directed ELs induced by perturbations. The first two evolutions, $L_+$ and $L_-$, denote the two types of EL braidings. The third type of eovlution, $L_0$, denotes EL repelling.
}
\end{figure*} 

The strategies for realizing exceptional links (EP links) in previous works rely on the elaborately designed tight-binding models~\cite{carlstrom2018Exceptional,yang2019NonHermitiana, zhang2020Bulkboundary, he2020Double, wang2021Simulating}, hence hardly transplanted to real material systems beyond tight-binding models. However, our crystalline-symmetry-based EC schemes provide an opportunity to achieve EP links by simply controlling the crystalline symmetries of the systems, where the ECs serve as the critical phases for generating EP links.

We start with the local ECs  protected by $\blue{m_z^\dagger m_x^\dagger} 2_y$ and $m_z\blue{m_x^\dagger 2_y^\dagger}$ point groups, respectively, which can be determinately produced from a Hermitian nodal line (see Table~\ref{perturbation table MzMx}). The critical intersection of the directed ELs in both cases can be lifted by breaking the point group symmetries, leading to three possible local evolutions compatible with the DN conservation, i.e., EL braidings in two manners $L_+,L_-$ and EL repelling $L_0$,~\cite{yang2020Jones}, as shown in Fig.~\ref{Chain point evolution}. However, if the perturbations are restricted to be purely non-Hermitian, we have 
\begin{theorem}\label{no-go gap theorem}
For a local EC protected by $\blue{m_z^\dagger m_x^\dagger} 2_y$ or $m_z\blue{m_x^\dagger 2_y^\dagger}$ point group, a generic \textbf{thresholdless} and \textbf{purely non-Hermitian} (i.e., anti-Hermitian) perturbation can only make the EC evolve to the two types of EL braidings, $L_+$ and $L_-$, while the EL repelling, $L_0$, can never happen under such perturbations. 
\end{theorem}

\begin{proof}
 Since the EL repelling, $L_0$, requires the the chain point be gapped on some $q_x=\mathrm{const.}$ planes near the chain point (see figures in Table~\ref{perturbation table MzMx}), to prove the theorem, \textbf{we only need to show that the degeneracies on any $q_x=\mathrm{const.}(\ll1)$ plane cannot be lifted by sufficiently small anti-Hermitian perturbations}. Near the chain point $\mathbf{q}=0$,  the $\mathbf{k}\cdot\mathbf{p}$ Hamiltonian up to first order of $\mathbf{q}$ for both point groups can be expressed as (see $\mathcal{H}_1$ and $\mathcal{H}_2$ in Table~\ref{perturbation table MzMx})
 \begin{equation*}
     \mathcal{H}(\mathbf{q})=\underbrace{(h_x\sigma_x+h_y\sigma_y)q_z+h_zq_y\sigma_z}_{\mathcal{H}^\mathrm{h}(\mathbf{q})}+\underbrace{iq_x\sum_{\mu=x,y,z}a_\mu\sigma_\mu}_{\mathcal{H}^\mathrm{ah}(\mathbf{q})},
 \end{equation*}
where either one of the point groups protects the Hermitian part to be a gapless Dirac Hamiltonian $\mathcal{H}^\mathrm{h}(\mathbf{q})=\tilde{h}_xq_z\tilde{\sigma}_x+h_zq_y\sigma_z$  with $\tilde{\sigma}_x=(h_x\sigma_x+h_y\sigma_y)/\tilde{h}_x$ and $\tilde{h}_x=\sqrt{h_x^2+h_y^2}$, while the anti-Hermitian part is only dependent on $q_x$ ($a_z=0$ for $\blue{m_z^\dagger m_x^\dagger} 2_y$ point group and $a_x=a_y=0$ for $m_z\blue{m_x^\dagger 2_y^\dagger}$ point group). Introducing  a generic anti-Hermitian perturbation $\delta\mathcal{H}^\mathrm{ah}(\mathbf{q})=i\sum_{\mu}(\delta a^\mu_0+\sum_{\nu}\delta a^\mu_\nu q_\nu)\tilde{\sigma}_\mu$, the perturbed Hamiltonian reads
 \begin{equation*}
     \mathcal{H}'(\mathbf{q})=\mathcal{H}(\mathbf{q})+i\delta\mathcal{H}^\mathrm{ah}(\mathbf{q})=\underbrace{\tilde{h}_xq_z\tilde{\sigma}_x+h_zq_y\tilde{\sigma}_z}_{\mathcal{H}^\mathrm{h}}+\underbrace{i\sum_{\mu=x,y,z}\Big[\overbrace{(\delta a_0^\mu+\tilde{a}_\mu q_x+\delta a^\mu_x q_x)}^{=a'^\mu_0(q_x)}+\sum_{i=y,z}\delta a^\mu_i q_i\Big]\tilde{\sigma}_\mu}_{\mathcal{H}'^{\mathrm{ah}}=\mathcal{H}^{\mathrm{ah}}+\delta\mathcal{H}^{\mathrm{ah}}}  ,
 \end{equation*}
where $\qty{\tilde{\sigma}_x,\tilde{\sigma}_y,\tilde{\sigma}_z}=\qty{\tilde{\sigma}_x,i\tilde{\sigma}_x\sigma_z,\sigma_z}$ denotes a new basis of ${su}(2)$ Lie algebra and all terms dependent on $q_x$ have been collected into $a'^\mu_0(q_x)$. In a plane of $q_x=q_{x0}$, the discriminant of the perturbed Hamiltonian $\mathcal{H}'(q_x=q_{x0})=\mathcal{H}(q_x=q_{x0})+\delta\mathcal{H}(q_x=q_{x0})$ reads
\begin{equation*}
    \Delta_f(q_{x0},q_y,q_z)=\qty[\qty({\tilde{h}_x}^2q_z^2+{h_z}^2q_y^2)-\sum_{\mu}\qty(a'^\mu_0(q_{x0})+\sum_{i={y,z}}\delta a^\mu_i q_i)^2]+2i\qty[\tilde{h}_xq_z\Big(a'^x_0(q_{x0})+\sum_{i={y,z}}\delta a^x_i q_i\Big)+h_zq_y\Big( a'^z_0(q_{x0})+\sum_{i={y,z}}\delta a^z_i q_i\Big)].
\end{equation*}
The degeneracy of eigenenergies requires the real and imaginary parts of $\Delta_f(q_{x0},q_y,q_z)$ both equal zero:
\begin{gather}
    \hspace{-10pt}
    \begin{aligned}
        {\tilde{h}_x}^2q_z^2+{h_z}^2q_y^2&=
     \sum_{\mu}(a'^\mu_0+\sum_{i={y,z}}\delta a^\mu_i q_i)^2\\
     &=\underbrace{\Big(\sum_{\mu}{a'^\mu_0}^2\Big)}_{=A_0\geq0}+\underbrace{\Big(\sum_{\mu}{\delta a_z^\mu}^2\Big)}_{=A_z\geq0}q_z^2+\underbrace{\Big(\sum_{\mu}{\delta a_y^\mu}^2\Big)}_{=A_y\geq 0}q_y^2+\underbrace{\Big(2\sum_{\mu}a'^\mu_0 \delta a^\mu_z\Big)}_{=B_z}q_z+\underbrace{\Big(2\sum_{\mu}a'^\mu_0 \delta a^\mu_y\Big)}_{=B_y}q_y
     + \underbrace{\Big(2\sum_{\mu}\delta a^\mu_y\delta a^\mu_z\Big)}_{=C_{yz}}q_yq_z
    \end{aligned},\label{Re Dis}\\
     \tilde{h}_xq_z\Big(a'^x_0+\sum_{i}\delta a^x_i q_i\Big)+h_zq_y\Big(a'^z_0+\sum_{i}\delta a^z_i q_i\Big)=0.\label{Im Dis}
\end{gather}
From Eq.~\eqref{Im Dis}, we get
\begin{equation*}
    q_y=\underbrace{\frac{-\tilde{h}_xa'^x_0}{h_za'^z_0}}_{=c_1}q_z+\underbrace{\frac{-h_x({a'^x_0}^2\delta a^z_y h_x+{a'^z_0}^2 \delta a^x_z h_z-a'^x_0a'^z_0(\delta a^x_y\tilde{h}_x+\delta a^z_z h_z))}{{a'^z_0}^3h_z}}_{=c_2}q_z^2+\mathcal{O}(q_z^3),
\end{equation*}
Substituting this relation into Eq.~\eqref{Re Dis} and omitting higher order terms $\mathcal{O}(q_z^3)$, we obtain a quadratic equation of $q_z$
\begin{equation*}
\big[({\tilde{h}_x}^2+c_1^2h_z^2)-(A_z+c_1^2A_y+c_2B_y+c_1C_{yz})\big]q_z^2-(B_z+c_1B_y)q_z-A_0=0,
\end{equation*}
where the coefficients are functions of $q_{x0}$.
For the thresholdless anti-Hermitian perturbations, the perturbation coefficients $\delta a_{y,z}^\mu$ can take arbitrarily small values, and so do $A_y, A_z,B_y,B_z,C_{yz}$. Hence, as long as the perturbations are sufficiently small relative to $h_{x,y,z}$, we can guarantee $\big[({\tilde{h}_x}^2+c_1^2h_z^2)-(A_z+c_1^2A_y+c_2B_y+c_1C_{yz})\big]>0$, which together with the fact $A_0\geq 0$ leads to a nonnegative discriminant of the above quadratic equation of $q_z$: 
\begin{equation*}
    \Delta=(B_z+c_1B_y)^2+4A_0    \big[({\tilde{h}_x}^2+c_1^2h_z^2)-(A_z+c_1^2A_y+c_2B_y+c_1C_{yz})\big]\geq 0,
\end{equation*}
implying that the original chain point degeneracy cannot be gapped out by thresholdless non-Hermitian perturbations on any $q_x=\text{const.}$ plane but will generally split into two EPs (as long as one of $a'^\mu_0$ is nonzero) on that plane. This result demonstrates that the ELs must traverse the $q_x=\text{const.}$ planes, hence eliminating the possibility of the third type evolution in Fig. under anti-Hermitian and thresholdless perturbations.
\end{proof}

\begin{figure*}[t!]
\includegraphics[width=0.75\textwidth]{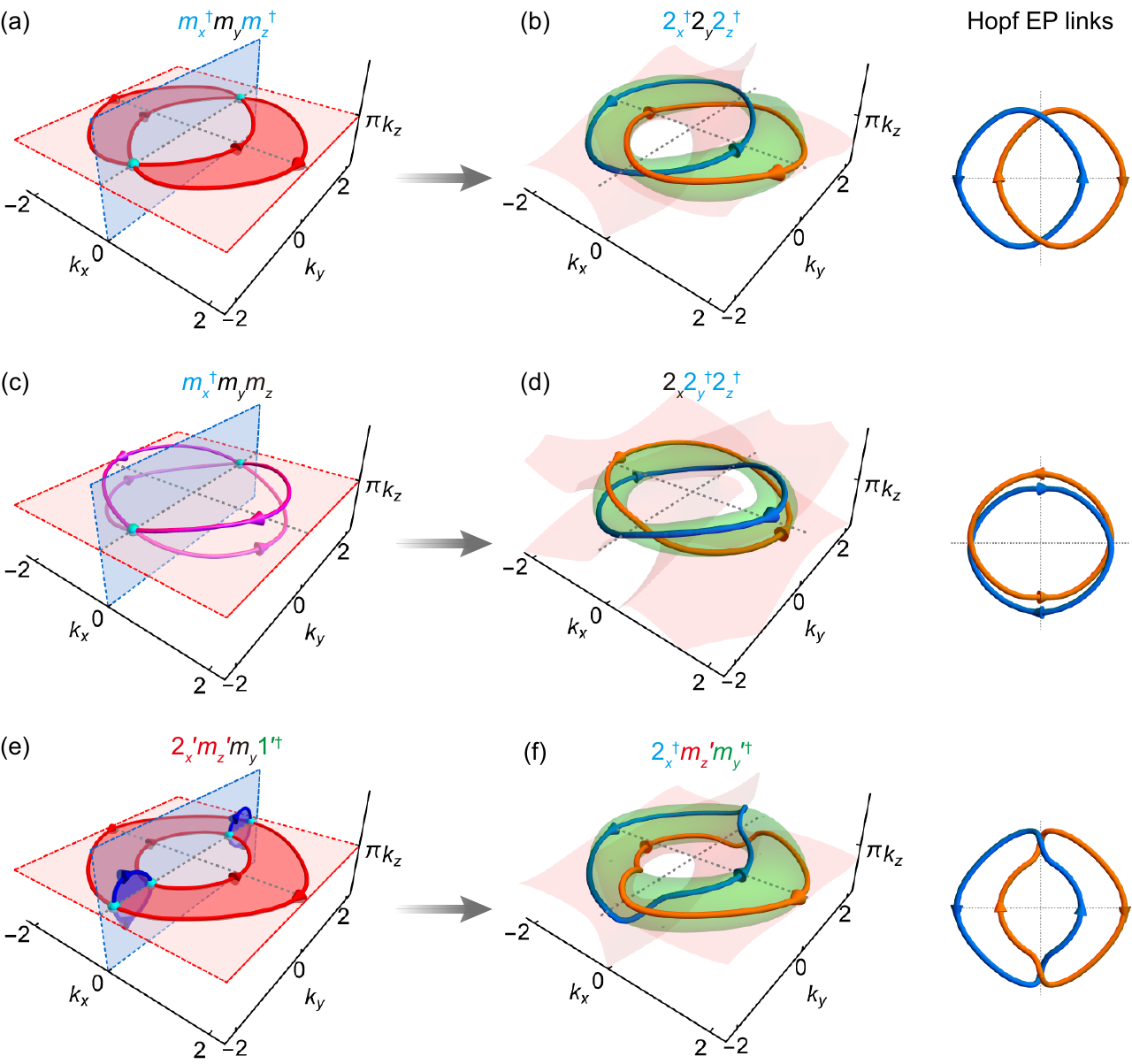}
\caption{\label{HopfEPlinks} 
Three schemes for realizing Hopf EP links ($2^2_1$) via evolutions from different types of ECs. The first and the second columns display the prototype ECs and the Hopf EP links evolved from the ECs in the same row, where the label above each panel denotes the double-antisymmetry point group of the corresponding exceptional structure. The third column exhibits the top view of each EP link. The red and green surfaces in the second column denote the zero surface of the real and imaginary parts of the discriminant, $\Re[\Delta_f(\mathbf{k})]=0$ and $\Im[\Delta_f(\mathbf{k})]=0$, respectively.
}
\end{figure*} 

Now, we turn to the global ECs with two non-defective chain points as shown by Figs.~\ref{HopfEPlinks}(a) and (c), which are locally stabilized by the little groups $\blue{m_z^\dagger m_x^\dagger} 2_y$ and $m_z\blue{m_x^\dagger 2_y^\dagger}$, respectively, along the line of $k_x=0$ and $k_z=\pi$. The Hamiltonians of the two cases are given by $\mathcal{H}_1(\mathbf{k})$ and $\mathcal{H}'_3(\mathbf{k})$ in Table~\ref{Hamiltonians for Fig4}. 
As we have shown that thresholdless purely non-Hermitian perturbations can only make the EL crossings at the chain points evolve to the ``EL braidings'' $L_+$ and $L_-$, \textbf{we can always produce EP links from the global ECs, providing that the ELs  at the two chain points braid in the same manner, namely either both $L_+$ or both $L_-$, which indeed can be guaranteed by imposing additional symmetry constraints to the non-Hermitian perturbations.} 

For the EC shown in Fig.~\ref{HopfEPlinks}(a), as long as we preserve either $C_{2x}\mbox{-}\dagger$ or $C_{2z}\mbox{-}\dagger$ symmetry in the exerted non-Hermitian perturbations, a Hopf EP link ($2_1^2$ in Alexander–Briggs notation) can be decisively generated. In Fig.~\ref{HopfEPlinks}(b), we present a concrete example  generated by the perturbed Hamiltonian 
\begin{gather}
    \mathcal{H}_\mathrm{HL1}(\mathbf{k})=\mathcal{H}_1(\mathbf{k})+{\color{red}i0.2\sin k_y\sigma_z},
\end{gather}
which respects both $C_{2x}\mbox{-}\dagger$ and $C_{2z}\mbox{-}\dagger$ symmetries, expressed as $\sigma_z\mathcal{H}_\mathrm{HL1}(\hat{c}_{2x}\mathbf{k})^\dagger\sigma_z=\mathcal{H}_\mathrm{HL1}(\mathbf{k})$ and $\mathcal{H}_\mathrm{HL1}(\hat{c}_{2z}\mathbf{k})^\dagger=\mathcal{H}_\mathrm{HL1}(\mathbf{k})$. The orientations of the ELs must respect the relations in Table~\ref{DN table}, as the ELs near the two chain points both braid as $L_-$, reasulting in a Hopf EP link.

Similarly, for the EC in Fig.~\ref{HopfEPlinks}(c), a thresholdless anti-Hermitian perturbation with either $C_{2x}$ or $C_{2z}\mbox{-}\dagger$ symmetry on the EC is also bound to deform the EC into a Hopf EP links, see the example in Fig.~\ref{HopfEPlinks}(d) generated by the Hamiltonian
\begin{equation}
    \mathcal{H}_\mathrm{HL2}(\mathbf{k})=\mathcal{H}'_3(\mathbf{k})+{\color{red}i0.2\sin k_y\sigma_x}
\end{equation}
which observes both $C_{2x}$ and $C_{2z}\mbox{-}\dagger$ symmetries, $\sigma_z\mathcal{H}_\mathrm{HL2}(\hat{c}_{2x}\mathbf{k})\sigma_z=\mathcal{H}_\mathrm{HL2}(\mathbf{k})$ and $\mathcal{H}_\mathrm{HL1}(\hat{c}_{2z}\mathbf{k})^\dagger=\mathcal{H}_\mathrm{HL1}(\mathbf{k})$.

In addition, we show that the earring EC in Fig.~~\ref{HopfEPlinks}(e) (generated by the Hamiltonian $\mathcal{H}_2(\mathbf{k})$ in Table~\ref{Hamiltonians for Fig4}) can also evolve into a Hopf EP links by introducing purely non-Hermitian perturbations that breaks both \mdagger[z]\ and $C_{2x}\mathcal{T}$ symmetries while preserves additional $C_{2x}\mbox{-}\dagger$ and $M_z\mathcal{T}$ symmetries, both of which can ensure the EL braidings at the two EP earrings are in the same manner. The perturbed Hamiltonian is given by
\begin{equation}
    \mathcal{H}_\mathrm{HL3}(\mathbf{k})=\mathcal{H}_2(\mathbf{k})+{\color{red}i0.1\sin k_y\sigma_z}.
\end{equation}
And the two additional symmetries are $\sigma_z\mathcal{H}_\mathrm{HL3}(\hat{c}_{2x}\mathbf{k})^\dagger\sigma_z=\mathcal{H}_\mathrm{HL3}(\mathbf{k})$ and $\sigma_z\mathcal{H}_\mathrm{HL3}(-\hat{m}_{z}\mathbf{k})^*\sigma_z=\mathcal{H}_\mathrm{HL3}(\mathbf{k})$.

\begin{figure*}[t!]
\includegraphics[width=0.75\textwidth]{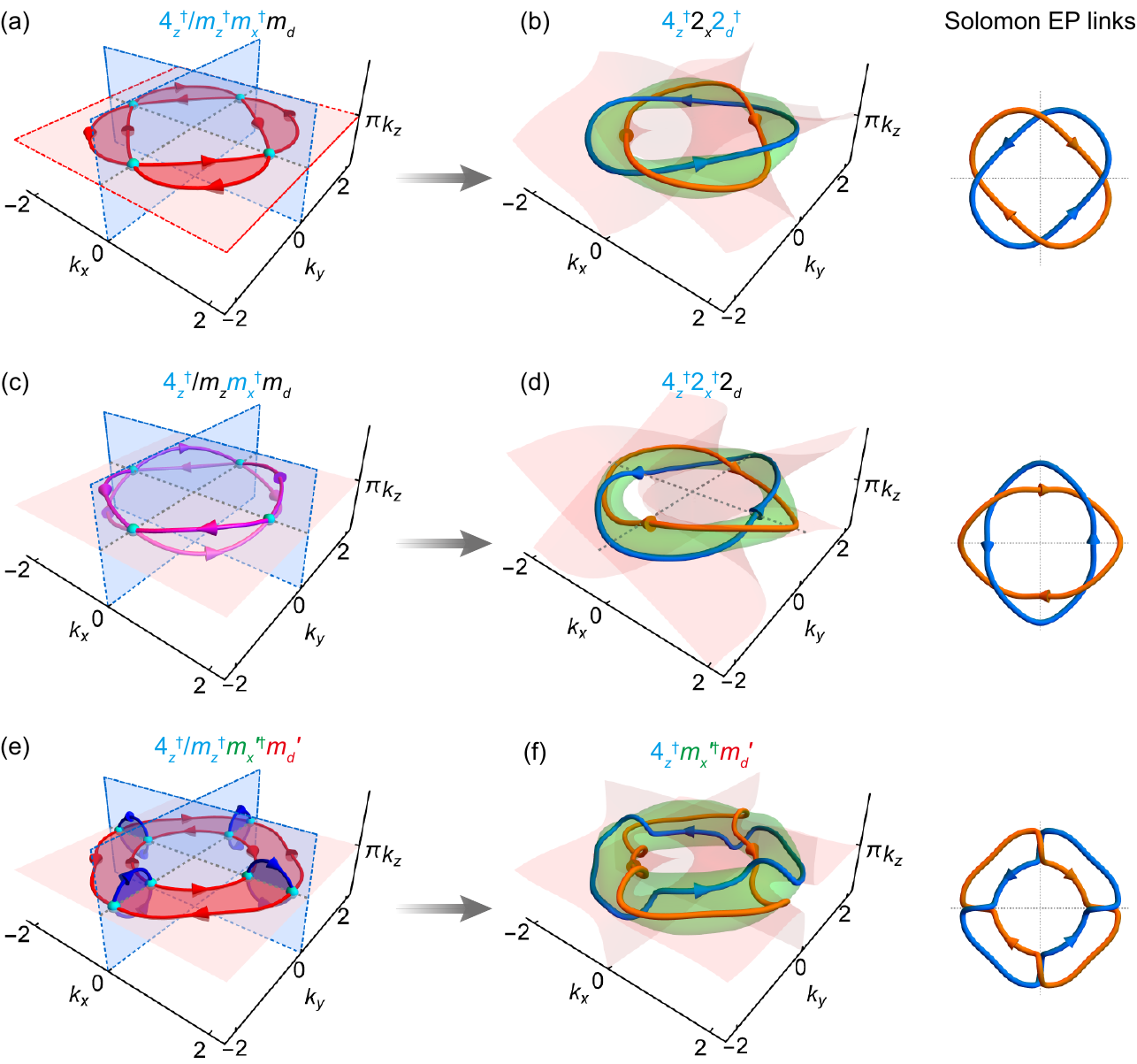}
\caption{\label{SolomonEPlinks} 
Three schemes for realizing Solomon EP links ($4^2_1$) via evolutions from different types of ECs. The first and the second columns display the prototype ECs and the Solomon EP links evolved from the ECs in the same row, where the label above each panel denotes the double-antisymmetry point group of the corresponding exceptional structure. The third column exhibits the top view of each EP link. The red and green surfaces in the second column denote the zero surface of the real and imaginary parts of the discriminant, $\Re[\Delta_f(\mathbf{k})]=0$ and $\Im[\Delta_f(\mathbf{k})]=0$, respectively.
}
\end{figure*}

Following the same strategy, more intriguing exceptional link configurations can be generated. In Fig.~\ref{SolomonEPlinks}, we provide three examples for achieving Solomon EP links ($4_1^2$) from three different types of ECs. Figures~\ref{SolomonEPlinks}(a) and (c) show a planar EC and a $M_z$-symmetric EC, respecitvely, both of which have four non-defective chain points. According to \textbf{Theorem}~\ref{no-go gap theorem}, thresholdless purely non-Hermitian perturbations distort the chain points into EL braidings. And  $C_{4z}\mbox{-}\dagger$-symmetric perturbations demand the ELs at each chain points braid in the same manner, therefore guaranteeing the formation of Solomon EP links, as shown in Figs.~\ref{SolomonEPlinks}(b) and (d). The solomon EP link can also evolve from a $C_{4z}\mbox{-}\dagger$-symmetric earring EC as illustrated in Figs.~\ref{SolomonEPlinks}(e) and (f), providing that $C_{4z}\mbox{-}\dagger$ symmetry is preserved during the perturbation. The concrete Hamiltonians for the ECs and links in Fig.~\ref{SolomonEPlinks} are listed as follows:
\begin{align}
    (a):&\quad\mathcal{H}_\mathrm{EC1}=\qty[(v_x\sigma_x+v_y\sigma_y)\sin{k_z}+v_z(\cos{k_x}+\cos{k_y}+\cos{k_z})\sigma_z]+i\qty[(u_x\sigma_x+u_y\sigma_y)\sin k_x\sin k_y]\\
    (b):&\quad\mathcal{H}_\mathrm{SL1}=\mathcal{H}_\mathrm{EC1}+{\color{red}i 0.2(\cos k_x-\cos k_y)\sigma_z}\\[5pt]
    (c):&\quad\mathcal{H}_\mathrm{EC2}=\qty[(v_x\sigma_x+v_y\sigma_y)\sin{k_z}+v_z(\cos{k_x}+\cos{k_y}+\cos{k_z})\sigma_z]+i\qty[(u_x\sigma_x+u_y\sigma_y)\sin k_z+u_z\sigma_z]\sin k_x\sin k_y\\
    (d):&\quad\mathcal{H}_\mathrm{SL2}=\mathcal{H}_\mathrm{EC2}+{\color{red}i 0.3(\cos k_x-\cos k_y)\sigma_x}\\[5pt]
    (e):&\quad\begin{aligned}\mathcal{H}_\mathrm{EC3} =& \qty[(v_x\sin k_x\sin k_y\sigma_x+v_y(\cos k_x-\cos k_y)\sigma_y)\sin{k_z}+v_z(\cos{k_x}+\cos{k_y}+\cos{k_z})\sigma_z]\\&+i\qty[u_x\sigma_x+u_z(\cos k_x-\cos k_y)\sin k_x\sin k_z\sigma_y]\end{aligned}\\
    (f):&\quad\mathcal{H}_\mathrm{SL3}=\mathcal{H}_\mathrm{EC3}+{\color{red}i 0.03(\cos k_x-\cos k_y)\sigma_z}
\end{align}
In all cases, we set $v_x=v_y=v_z=1$ and $u_x=u_y=u_z=0.4$. The $C_{4z}\mbox{-}\dagger$ symmetry for the Hamiltonians in (a-d) is expressed as $\mathcal{H}(k_y,-k_x,k_z)^\dagger=\mathcal{H}(k_x,k_y,k_z)$, while the $C_{4z}\mbox{-}\dagger$ symmetry for the Hamiltonians in (e-f) is $\sigma_z\mathcal{H}(k_y,-k_x,k_z)^\dagger\sigma_z=\mathcal{H}(k_x,k_y,k_z)$.  

Finally, we stress that our strategy can deterministically generate EP links from ECs by perturbatively controlling the symmetry breaking. With this advantage, our method is a pragmatic approach to realizing EP links in real crystalline materials and full-wave systems,  going beyond sophisticatedly designed tight-binding models.

\clearpage
\section{A two-band toy model of fully-linked EC pair}
\begin{figure*}[b!]
\includegraphics[width=0.87\textwidth]{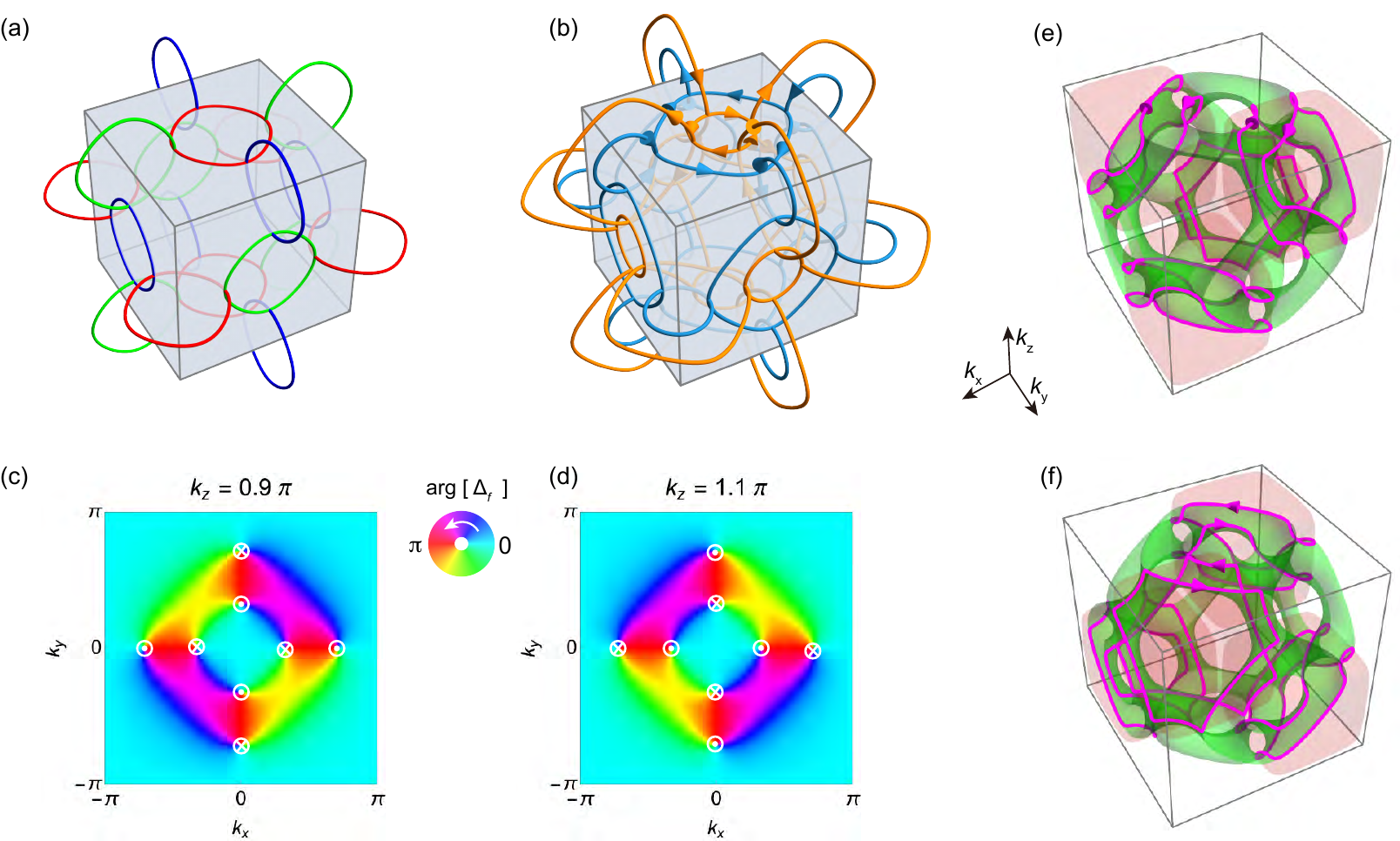}
\caption{\label{fig-toymodel} 
(a)The nodal chain of the 2-band toy model \eqref{Eq-H0M3PT} with $v_x=v_y=v_z=1$ and the gray box denotes the first BZ. (b) The fully-linked EC pair of the 2-band toy model \eqref{Eq-3MD} with $\gamma_x=\gamma_y=0.4$. The directions of the ELs near the plant $k_z =\pi$ are marked by arrows and other ELs' directions can be determined by the \mdagger[x,y,z] symmetries. The phase of the discriminant on the plane (c) $k_z = 0.9\pi$ and (d) $k_z =1.1 \pi$. The fully-linked EC pairs in (b) are untied when breaking the \mdagger[x,y,z] by introducing perturbation (e) $m_x=0.01$ and (f) $m_x=-0.01$. The light green and pink surfaces represent $\mathrm{Re}[\Delta_f(\mathbf{k})]=0 $ and $\mathrm{Im}[\Delta_f(\mathbf{k})]=0 $, respectively. 
}
\end{figure*} 

In this section, with the help of a simple two-band toy model, we will introduce an interesting EC configuration simultaneously protected by three mirror-adjoint symmetries \mdagger[x,y,z], which has been observed in the PCs (see Fig. 5(h) of the main text).
We first consider a Hermitian model respecting three $M$ symmetries \eqref{MzH-definition},
\begin{equation}\label{Eq-H0M3PT}
\mathcal{H}_0(\mathbf{k})=\left[v_z\left(\cos k_x+\cos k_y+\cos k_z\right)\right] \sigma_z+\left(v_z \sin k_x \sin k_y \sin k_z\right) \sigma_x,
\end{equation}
where $v_{x}$ and $v_z$ are coefficients. 
Note that besides the three mirror symmetries $\hat{M}_{x,y,z}=\sigma_z$, the model \eqref{Eq-H0M3PT} also respects the $\mathcal{PT}$ symmetry. And if we introduce $\mathcal{PT}$-symmetric non-Hermiticity into~\eqref{Eq-H0M3PT}, the EP will form robust surfaces in the 3D momentum space~\cite{zhou2019Exceptional}.
Therefore, to obtain robust ELs, we need to break the $\mathcal{PT}$ symmetry first. We modified \eqref{Eq-H0M3PT} into a model as
\begin{equation}\label{Eq-H0M3PT}
\mathcal{H}'_0(\mathbf{k})=\mathcal{H}_0(\mathbf{k}) + \left(v_y \sin k_x \sin k_y \sin k_z\right) \sigma_y,
\end{equation}
which is no more $\mathcal{PT}$-symmetric. But the three mirror symmetries $\hat{M}_{x,y,z}=\sigma_z$ remain intact and can protect nodal lines on the mirror planes as shown in Fig.~\ref{fig-toymodel} (a). Then, by introducing non-Hermiticity into the above model, the three mirror symmetries are modified into \mdagger[x,y,z] symmetries,
\begin{equation}\label{Eq-3MD}
\mathcal{H}_1(\mathbf{k})=\mathcal{H}'_0(\mathbf{k})+ i \gamma_x \sigma_x + i\gamma_y \sigma_y.
\end{equation}
As shown in Fig.~\ref{fig-toymodel} (b), the nodal ring on each mirror plane is split into two nested ERs. Each inner (outer) ring connects with the other four outer (inner) rings on its orthogonal mirror planes, forming two sets (orange and cyan) of ECs which are fully linked in the BZ. In Fig.~\ref{fig-toymodel} (c) and (d), we plot the phase of the discriminant on the planes $k_z =0.9 \pi$ and $k_z =1.1 \pi$, respectively. The phase singularities denote the position where the ELs pierce through the plane and the signs of the phase winding numbers indicate the direction of the ELs, which have been marked by arrows in Fig.~\ref{fig-toymodel} (b). The directions of ELs on the plane $k_z = \pi$ can also be determined similarly. By introducing a perturbation that breaks the \mdagger[x,y,z] symmetries, we obtain
\begin{equation}\label{Eq-3MD-brokrn}
\mathcal{H}_2(\mathbf{k})=\mathcal{H}_1+ m_x \sigma_x,
\end{equation}
and all the chain points are opened as shown in Figs.~\ref{fig-toymodel} (e) and (f), and the links between the two sets of chains are also untied.


\section{Example of multi-band exceptional chains}

\begin{figure*}[t!]
\includegraphics[width=0.95\textwidth]{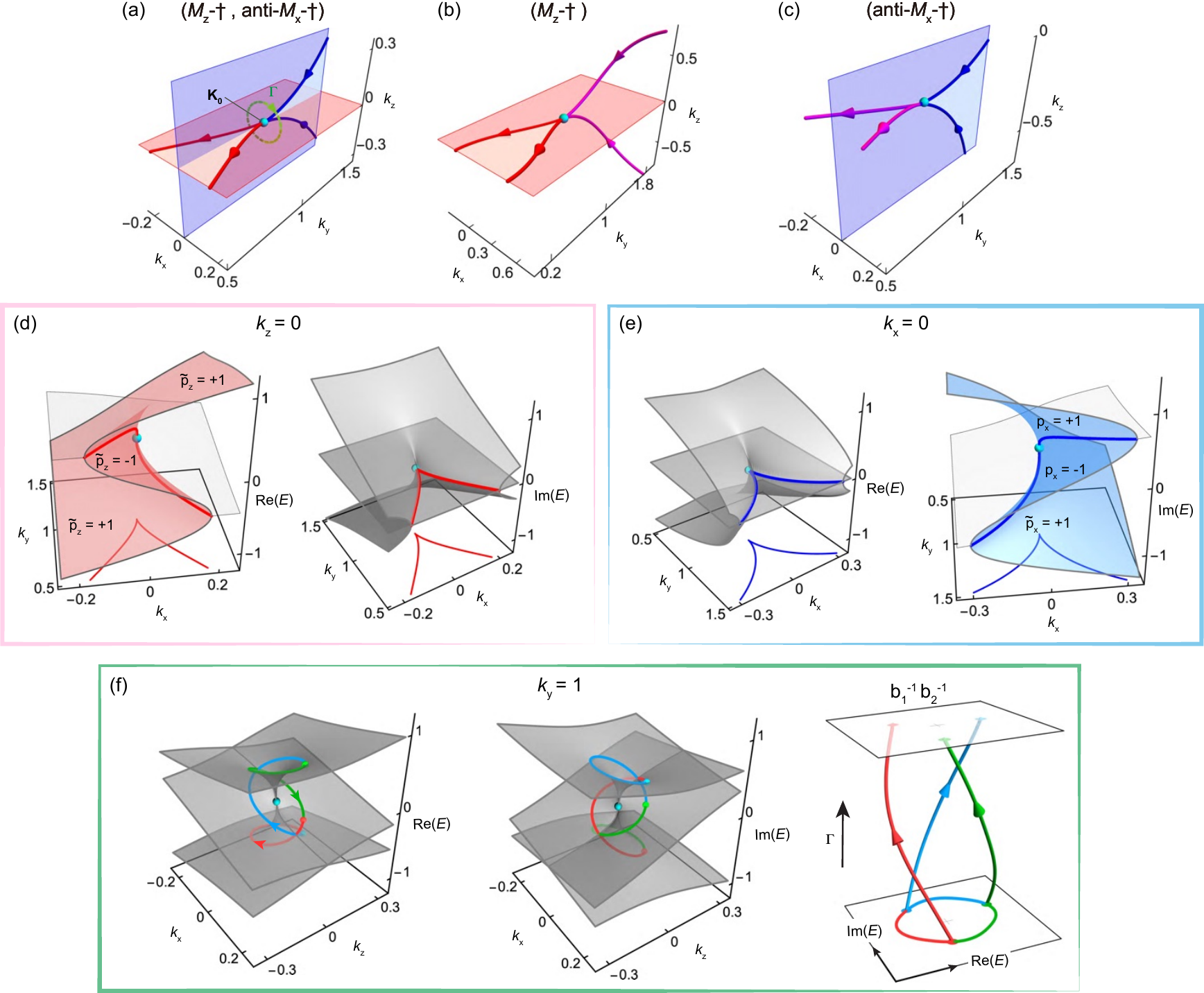}
\caption{\label{fig-3Dbands-cusp} 
The triple-band ECs protected by (a) both \mdagger[z] and anti-\mdagger[x], (b) \mdagger[z] and (c) anti-\mdagger[x], where the light red (blue) shadows represent the region where all three bands are in the exact phase, the cyan dots denote the EP3 chain points. The band structure of the plane (d) $k_z=0$ and (e) $k_x=0$, where the red (blue) folded surface represents the bands in the exact phase. (f) The band structure on the plane $k_y =1$ and the energy braiding of the three bands along the dashed green loop $\Gamma$ in (a).}
\end{figure*}

To demonstrate that our theory of ECs is not limited to two-band models but applicable to the cases of any number of bands, we offer a three-band model to exemplify the formation of symmetry-protected multi-band ECs. The model is given by the following $\mathbf{k}$-space non-Hermitian Hamiltonian:
\begin{equation}\label{3band hamiltonian}
    \mathcal{H}(\mathbf{k})=\begin{pmatrix}
    k_xk_z +i k_y & -1 & 1+ik_z \\
    -1 & k_x & -1\\
   -1+ik_z & -1 & -k_xk_z-ik_y
    \end{pmatrix},
\end{equation}
which has two important symmetries, i.e., the \mdagger[z]\ symmetry
\begin{equation}
    \hat{M}_z\mathcal{H}(\hat{m}_z\mathbf{k})^\dagger\hat{M}_z=\mathcal{H}(\mathbf{k}),\qquad\text{with}\quad \hat{M}_z=\begin{pmatrix}0 & 0 &1\\ 0 & 1 & 0\\ 1 & 0 & 0  \end{pmatrix},
\end{equation}
and the anti -\mdagger[x]\ symmetry (which can be viewed as the combined symmetry of sublattice and \mdagger[x])
\begin{equation}
    \hat{M}_x\mathcal{H}(\hat{m}_x\mathbf{k})^\dagger\hat{M}_x=-\mathcal{H}(\mathbf{k}),\qquad\text{with}\quad \hat{M}_x=\begin{pmatrix}1 & 0 &0\\ 0 & -1 & 0\\ 0 & 0 & 1  \end{pmatrix}.
\end{equation}
From the three eigenvalues of $\hat{M}_z$, i.e. $+1,+1,-1$, we know that in the exact phases in the mirror plane $k_z=0$, two bands have a positive \mdagger[z]-parity $\tilde{p}_z=+1$ while the left band has a negative \mdagger[z]-parity $\tilde{p}_z=-1$ as shown in Fig. \ref{fig-3Dbands-cusp} (d). Thus, according to \textbf{Theorem}~\ref{theorem-mdagger}, order-2 ELs will appear in $k_z=0$ formed by two bands with opposite \mdagger[z]-parities. In particular, since the eigenenergies in $k_z=0$ are either real or complex-conjugated paired, \textbf{the characteristic polynomial of $\mathcal{H}(\mathbf{k}_{m_z})$ in $k_z=0$ must be a real polynomial}:
\begin{equation*}
\begin{split}    
    f(E,\mathbf{k}_{m_z})&=\det\qty[E-\mathcal{H}(\mathbf{k}_{m_z})]=E^3+c_2(\mathbf{k}_{m_z}) E^2+c_1(\mathbf{k}_{m_z}) E+c_0(\mathbf{k}_{m_z}),
\end{split}
\end{equation*}
where $c_2=-\Tr[\mathcal{H}(\mathbf{k}_{m_z})]=-\sum_{i=1}^3E_i$, $c_1=-\frac{1}{2}\qty(\Tr[\mathcal{H}(\mathbf{k}_{m_z})^2]-\Tr[\mathcal{H}(\mathbf{k}_{m_z})]^2)$, and $c_0=-\det[\mathcal{H}(\mathbf{k}_{m_z})]=-\prod_{i=1}^3E_i$ are all real numbers ($E_i$ ($i=1,2,3$) denote the three eigenenergies).
Then, the \textbf{catastrophe theory for real polynomials} tells us that \textbf{a real polynomial of degree 3 can support a stable triple degeneracy (i.e., an EP of order 3 (EP3)) at the cusp of two order-2 ELs in a 2D parameter space}, where the two ELs are formed by two different pairs of bands, i.e., the \nth{1} and \nth{2} bands and the \nth{2},\nth{3} bands, respectively, sorted by the real parts of the eigenenergies.  This cusp degeneracy is indeed a class of elementary catastrophe singularities, labeled as $A_3$ as per the ADE classification.  It can thus be concluded that \textbf{the \mdagger[z]\ symmetry can fix EP3 cusps as the junction of two ELs lying inside the $z$-mirror plane}, as evidenced by the band structure of the Hamiltonian~\eqref{3band hamiltonian} plotted in Fig.\ref{fig-3Dbands-cusp}(d).

The anti-\mdagger[x]\ symmetry can induce almost the same consequences in the $k_x=0$ mirror plane as those induced by \mdagger[z]\ symmetry in the $k_z=0$ plane. In fact, after reexpressing the Hamiltonian $\mathcal{H}(\mathbf{k})'=i\mathcal{H}(\mathbf{k})$, the anti-\mdagger[x]\ symmetry alters to a \mdagger[x]\ symmetry for  $\mathcal{H}(\mathbf{k})'$, so the eigenstates in the $k_x=0$ plane either have purely imaginary eigenenergies (exact phase) or appear pairwise with ``anti-complex-conjugate'' eigenenergies $E_1=-E_2^*$ (broken phase). Moreover, \textbf{anti-\mdagger[x]\ symmetry can also protect a cusp of two ELs stably fixed in the $k_x=0$ plane}, where the two ELs are formed by two different pairs of bands with opposite \mdagger[x]-parities in the nearby exact phase, as shown in Fig. \ref{fig-3Dbands-cusp}(e).

As shown in Fig.~\ref{fig-3Dbands-cusp}(a), along the crossing line of $k_x=0$ and $k_z=0$, the presence of both \mdagger[z] and anti-\mdagger[x]\ symmetries demand one of the eigenenergies always equals zero, $E_1(0,k_y,0)\equiv 0$, while the other two eigenenergies take either real opposite values $E_2(0,k_y,0)=-E_2(0,k_y,0)\in\mathbb{R}$ or imaginary opposite values $E_2(0,k_y,0)=-E_2(0,k_y,0)\in i\mathbb{R}$, corresponding to the (\mdagger[z]-exact,anti-\mdagger[x]-broken) phase ($k_y < 1$) and the (\mdagger[z]-broken,anti-\mdagger[x]-exact) phase ($k_y > 1$), respectively. \textbf{As a result, an EP3 ($\mathbf{K}_0$ in Fig.~\ref{fig-3Dbands-cusp}(a)) can be fixed on this crossing line at the transition point of the two phases, forming the chain point of the two EL cusps on the respective mirror planes.} 
In particular, the \mdagger[z] and anti-\mdagger[x] symmetries require the two ELs connecting the cusp on each plane symmetrically lie on the two sides of the central line $k_x=k_z=0$, despite forming by different pairs of bands. And both symmetries impose the constraint on the DN: $\mathcal{D}(\Gamma)=-\mathcal{D}(\hat{m}_i\Gamma)$ ($i\in\qty{x,z}$), so \textbf{the two ELs on each plane must be directed either both inwards to or both outwards from the EP3}. Consequently, the source-free principle implies once the two ELs in the $k_x=0$ plane both inflow towards the EP3, the two ELs in the $k_z=0$ plane must both outflow from the cusp, forming a triple-band EC. The identical directions of the two ELs on each plane are also verified by the eigenenergies braiding along the loop $\Gamma$ around the EP3 chain point $\mathbf{K}_0$. As displayed in Fig.~\ref{fig-3Dbands-cusp}(f), the braiding along $\Gamma$ is $b(\Gamma)=b_1^{-1}b_2^{-1}$. Therefore, the DN along the loop is $\mathcal{D}(\Gamma)=-2$ according to \textbf{Theorem}~\ref{theorem-algebraic length}, indicating the ELs inflow to the chain point along the negative direction of the $k_y$ axis.

More intriguingly, the triple-band EC cannot be destroyed by perturbatively breaking either one of the two symmetries. This is because the preserved symmetry stays protecting the cusp of two ELs on the corresponding mirror plane. And the orientations of the two in-plane ELs cannot be changed abruptly, hence two out-of-plane ELs have to germinate from the EP3 cusp to balance the two in-plane ELs so that the source-free principle of ELs at the  EP3 is satisfied. In Figs.~\ref{fig-3Dbands-cusp} (b) and (c), we present two examples of the multi-band ECs protected solely by the \mdagger[z] symmetry and the anti-\mdagger[x]\ symmetry, respectively, which are generated by the following perturbed Hamiltonians
\begin{align}
    \text{\mdagger[z]-symmetric:}&\quad \mathcal{H}_1(\mathbf{k})=\begin{pmatrix}
    k_xk_z +i k_y & -1 & 1+ik_z \\
    -1 & k_x-\red{0.25} & -1\\
   -1+ik_z & -1 & -k_xk_z-ik_y
    \end{pmatrix},\\
    \text{anti-\mdagger[x]-symmetric:}&\quad \mathcal{H}_2(\mathbf{k})=\begin{pmatrix}
    k_xk_z +i k_y & -1 & 1+ik_z \\
    -1 & k_x+\red{0.25i} & -1\\
   -1+ik_z & -1 & -k_xk_z-ik_y
    \end{pmatrix}.
\end{align}

\section{Photonic crystal realizations of exceptional chains}

\subsection{Hermitian nodal chains }
\begin{figure*}[h!]
\includegraphics[width=0.7\textwidth]{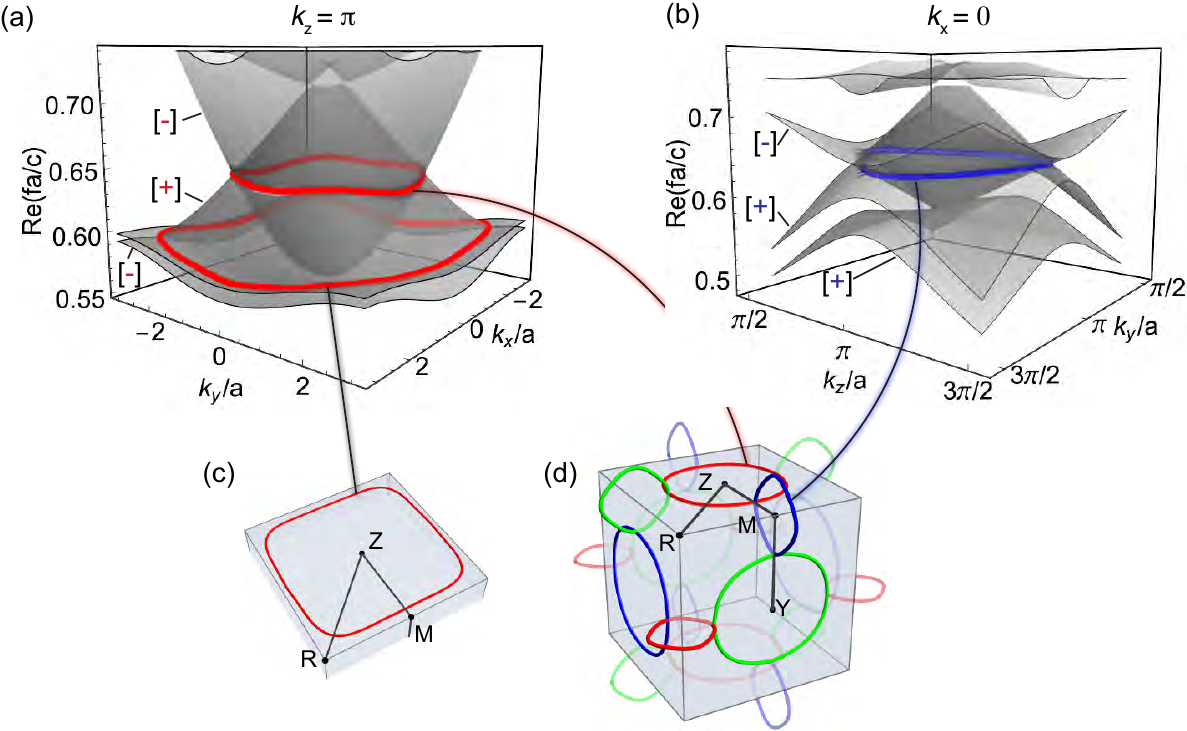}
\caption{\label{fig-2Dbands-nodalchain} 
The 2D band diagrams on the mirror plane (a) $k_z=\pi$ and (b) $k_x=0$ for the PCs in Fig. 5(a) of the main text. (c) The retrieved nodal line formed by the first and second band is plotted in the BZ. (d) The retrieved nodal line formed by the second and third band is plotted in the BZ. }
\end{figure*} 

In this subsection, we will show how to retrieve the nodal lines in Fig. 5(g) of the main text by numerically sweeping the 2D band structures on the mirror planes using COMSOL. We plot the 2D band structure on the mirror planes $k_z=\pi$ in Fig.~\ref{fig-2Dbands-nodalchain} (a), where the $M_z$ mirror parities $p=\pm$ of the bands are labeled. We see that the lowest two bands with opposite mirror parities intersect near the frequency $fa/c=0.6$, forming the nodal ring on $k_z=\pi$ in the right panel of Fig. 5(g) of the main text. The second and third bands with opposite mirror parities intersect near the frequency $fa/c=0.65$, forming a smaller nodal ring, which is plotted in the BZ as shown in the left panel of Fig. 5(g) of the main text. Similarly, as shown in Fig.~\ref{fig-2Dbands-nodalchain} (b), the nodal ring forming by the second and third bands on the mirror plane $k_x=0$ is connected with the nodal ring on the plane $k_z=\pi$, forming nodal chains, as shown in Fig.~\ref{fig-2Dbands-nodalchain}(d). Note that the lower two bands in Fig.~\ref{fig-2Dbands-nodalchain} (b) have the same $M_x$ parities, forbidding the formation of nodal rings. Therefore, the nodal ring on the plane $k_z=\pi$ formed by the lower two bands is an isolated ring as shown in Fig.~\ref{fig-2Dbands-nodalchain}(c).
Other nodal rings on mirror planes $k_{x,y,z}=0,\pi$ can also be retrieved by sweeping the 2D band structures in a similar way. Accordingly, we obtained the globally connected nodal chains in the left panel of Fig. 5(g) of the main text. 

\subsection{Orthogonal exceptional chains}

\begin{figure*}[h!]
\includegraphics[width=0.9\textwidth]{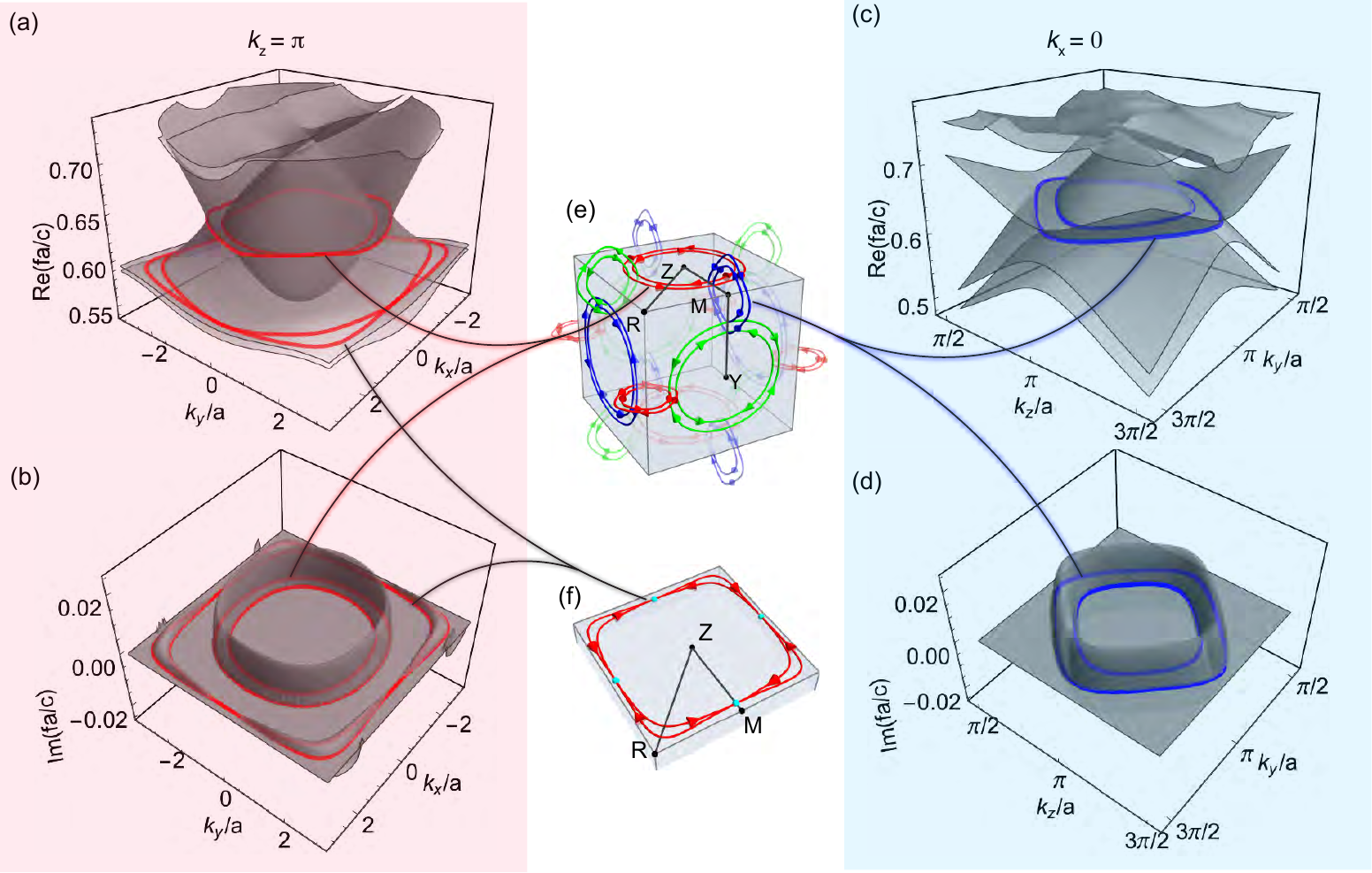}
\caption{\label{fig-2Dbands-MD+MD} 
The real parts of 2D band diagrams on the mirror plane (a) $k_z=\pi$ and (c) $k_x=0$ for the PCs in Fig. 5(b) of the main text. (b) shows the imaginary parts of the bands 1-3 in (a). (d) shows the imaginary parts of the bands 2-3 in (c). The retrieved orthogonal ECs and planar ECs in the BZ are plotted in (e) and (f), respectively. }
\end{figure*} 
By introducing non-Hermiticity (loss and gain) into the PCs in the above subsection, the nodal lines will split into ELs. Figure~\ref{fig-2Dbands-MD+MD} plots the real and imaginary parts of the 2D band diagrams on the plane (a, b) $k_z=\pi$ and (c, d) $k_x=0$. We see that the nodal rings in Fig.~\ref{fig-2Dbands-nodalchain} formed by the second and third bands (ordered by the real parts of the bands) have split into a pair of ERs. And the annulus region between the two ERs on mirror planes $k_z=\pi$ and $k_x=0$ is the broken phase, where the two bands have identical real parts and opposite imaginary parts.  
Plotting the ELs on all the mirror planes formed by the second and third bands in the BZ, we obtained the orthogonal ECs as shown in Fig. ~\ref{fig-2Dbands-MD+MD}(e). 

\begin{figure*}[h!]
\includegraphics[width=0.95\textwidth]{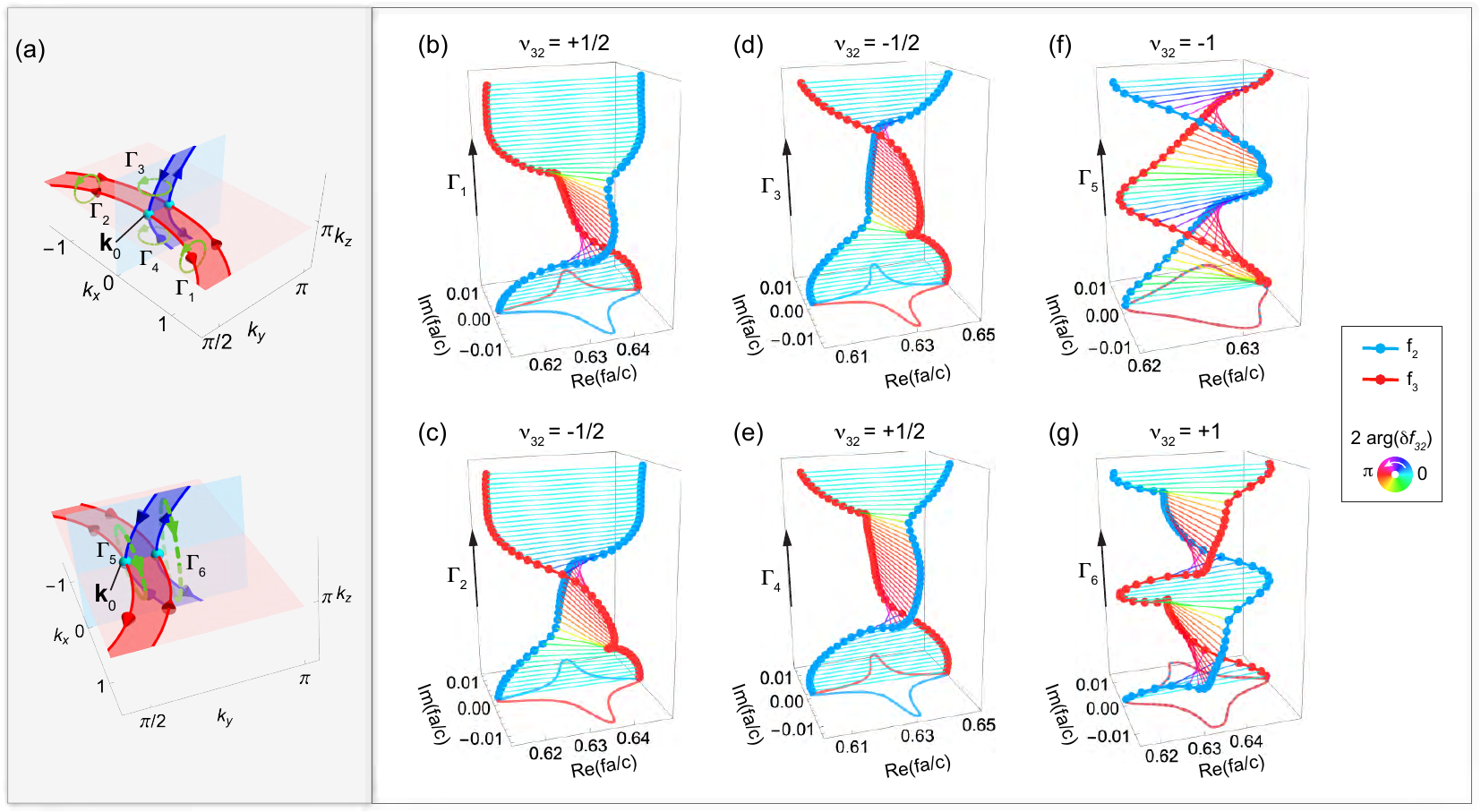}
\caption{\label{fig-orientation-MD+MD-orthogonal} 
(a) The zoom-in of the orthogonal EC configuration near the $M$ point in Fig.~\ref{fig-2Dbands-MD+MD}(e), and the green loops with arrows denote the 1D winding loops encircling ELs. (b-g) The eigenfrequency braiding along loops in (a), and the energy vorticities are labeled at the top. }
\end{figure*} 
To determine the orientations of the ELs and verify the source-free principles of the orientable ELs in the PCs, we 
choose 6 1D winding loops encircling the ELs near the chain points as shown in Fig.~\ref{fig-orientation-MD+MD-orthogonal} (a). We plot the eigenfrequency braiding of the second and third bands that forms the ELs along different 1D loops in Figs.~\ref{fig-orientation-MD+MD-orthogonal} (b-g). 
For example, Fig. ~\ref{fig-orientation-MD+MD-orthogonal} (a) plots the eigenfrequencies in the complex plane 
along the loop $\Gamma_1$, where the verticle axis represents the loop parameters. 
We see that after one cycle, the left strand $f_2$ (cyan) crosses from above to reach the initial position of the right strand $f_3$ (red), which can be denoted as braid $b(\Gamma_1) = b_2^{n_2}$ with algebraic length $n_2 = 1$. The two strands are connected by bars whose colors represent twice of the relative eigenfrequency's phase $\mathrm{arg}(\delta f) = \mathrm{arg}(f_3-f_2)$.
Therefore, according to Eq. ~\eqref{Eq-relation-nvD}, the algebraic length of the braid, the discriminant number, and the energy vorticity  along the loop $\Gamma_1$ are related by $n_2 = \mathcal{D} = 2\nu_{32} = 1$. The positive sign of $\nu_{32}$ 
means the encircled EL has the same direction in compliance with the right-hand rule of the loop. Similarly, along the loop $\Gamma_2$, the left strand $f_2$ (cyan) crosses from below to reach the initial position of the right strand $f_3$ (red), and therefore, the encircled EL has the reverse direction in compliance with the right-hand rule of the loop. The reversal of the direction when the EL go through the plane $k_x=0$ is enforced by the \mdagger[x] symmetry. 
The orientations of other ELs near the chain point $\mathbf{k}_0$ are also marked in Fig.~\ref{fig-orientation-MD+MD-orthogonal} (a). We can see that as two blue ELs on the plane $k_x = 0$ flow into the chain point $\mathbf{k}_0$, two red ELs on the plane $k_z=\pi$ must flow out from the chain point, which demonstrates the source-free principles for the ELs in PCs.  

\subsection{Planar exceptional chains}
\vspace{-10pt}
\begin{figure*}[h!]
\includegraphics[width=0.99\textwidth]{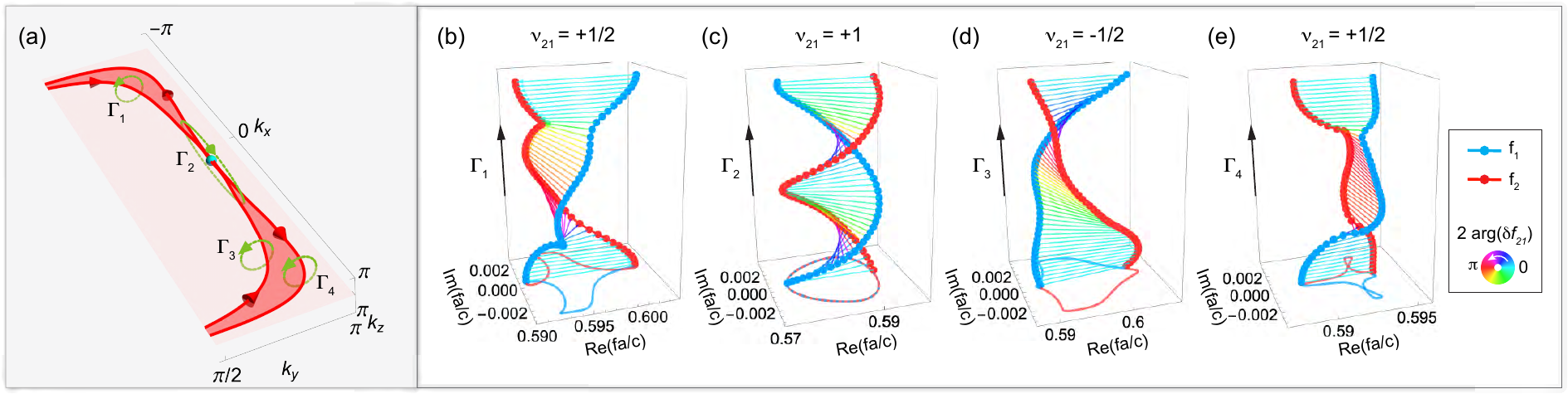}
\caption{\label{fig-orientation-MD+MD-planar} 
(a) The zoom-in of the planar EC configuration near the $M$ point in Fig.~\ref{fig-2Dbands-MD+MD}(f), and the green loops with arrows denote the 1D winding loops encircling ELs. (b-g) The eigenfrequency braiding along loops in (a), and the energy vorticities are labeled at the top.}
\end{figure*} 

Now we turn to the planar EC formed by the first and second bands. As shown in Figs. \ref{fig-2Dbands-nodalchain} and \ref{fig-2Dbands-MD+MD}, after introducing non-Hermiticity, the nodal line is split, and
since the two bands on the mirror plane $k_x=0$ has the same parity, the degenerate points on the line $ZM$ must be non-defective (see Corollary \ref{corollary-mdagger}). We plot the eigenfrequency braiding along 4 different winding loops in Fig.~\ref{fig-orientation-MD+MD-planar}, thereby determining the orientation of the encircled ELs.  

\vspace{-20pt}
\subsection{Double-earring exceptional chains}

\begin{figure*}[h!]
\includegraphics[width=0.9\textwidth]{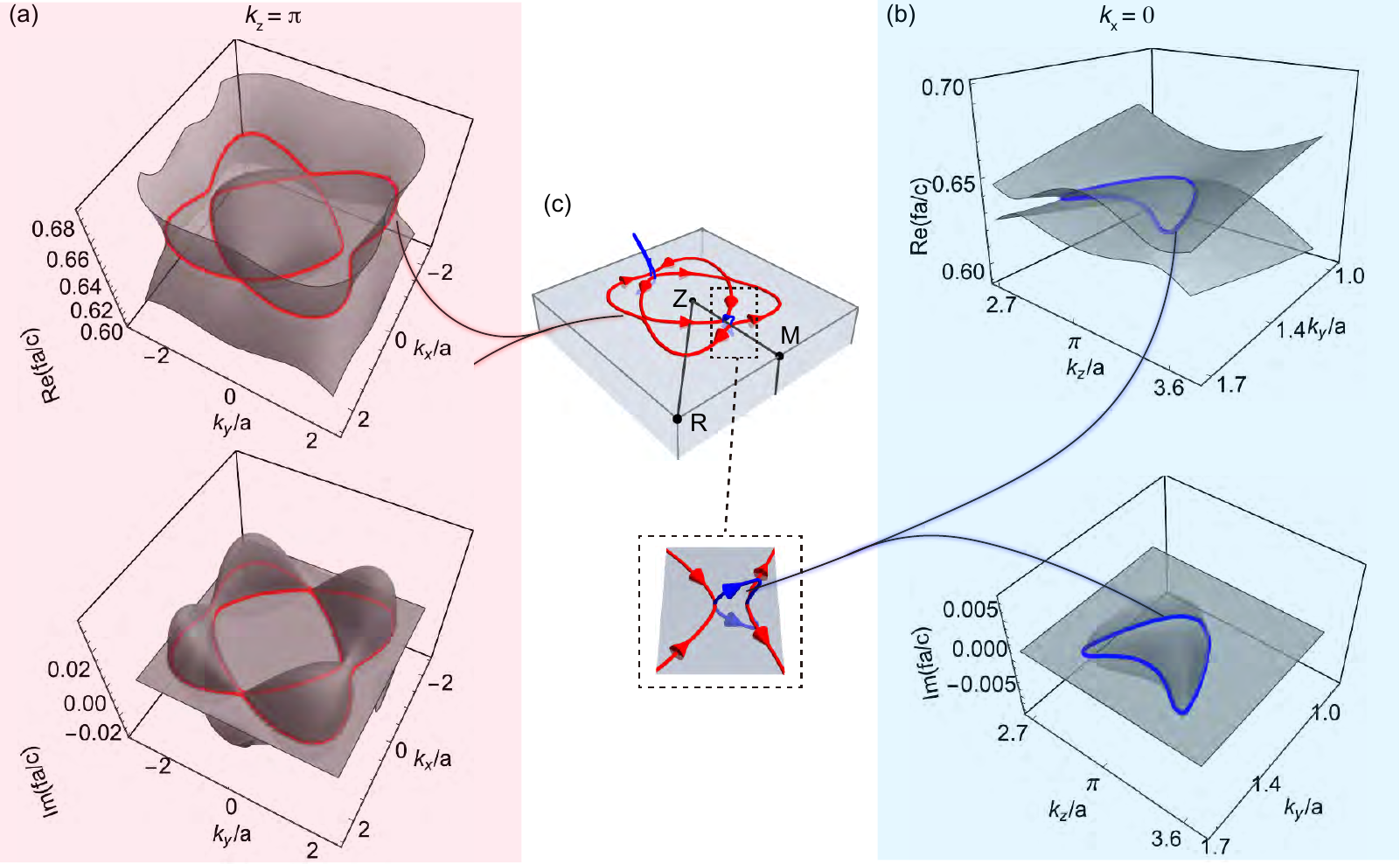}
\caption{\label{fig-2Dbands-MD+C2T} 
The diagrams of the two bands forming ELs on the mirror plane (a) $k_z=\pi$ and (b) $k_x=0$ for the PCs in Fig. 5(c) of the main text. The retrieved double-earring ECs in the BZ are plotted in (c). }
\end{figure*} 

\begin{figure*}[h!]
\includegraphics[width=0.99\textwidth]{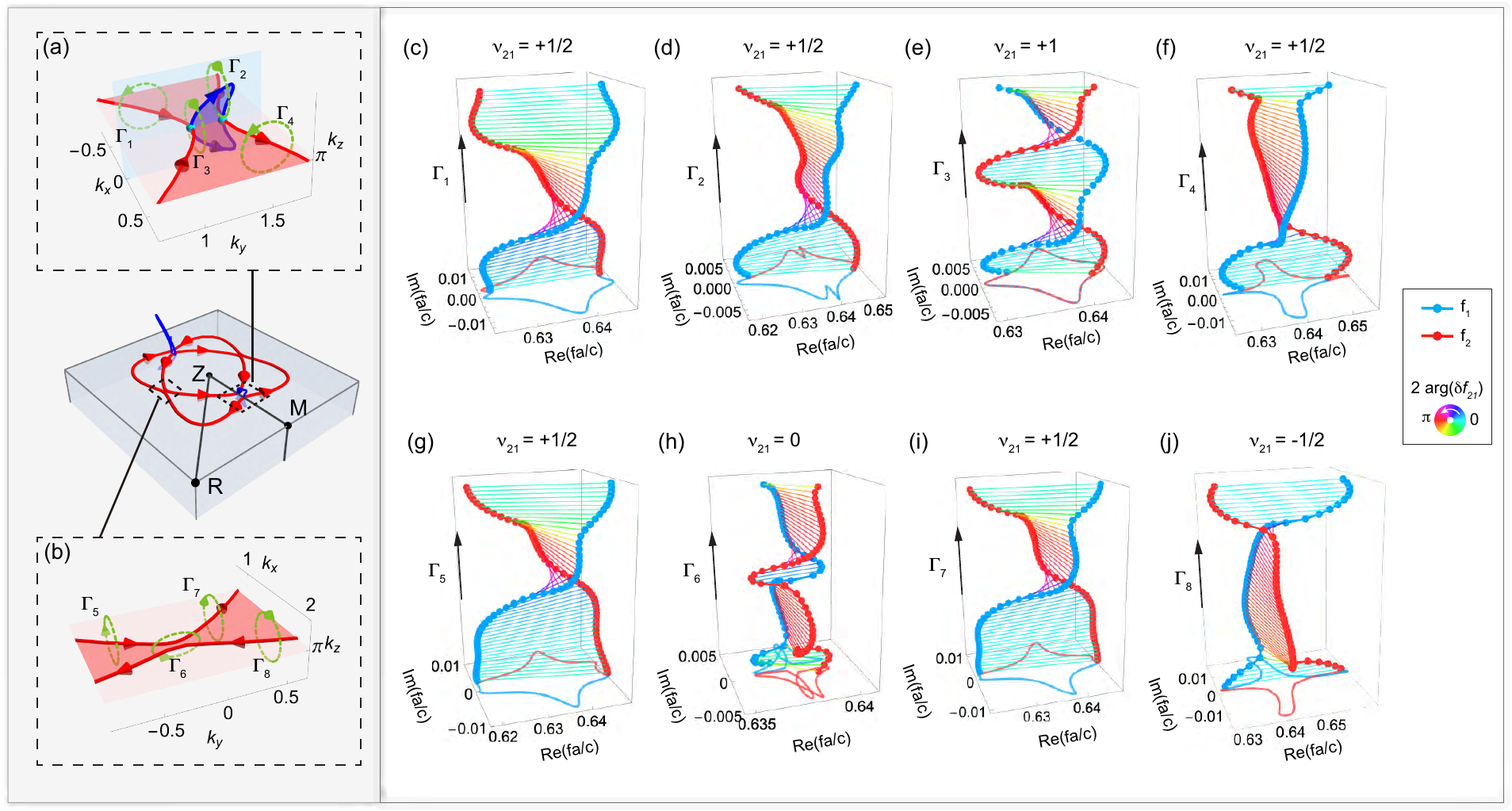}
\caption{\label{fig-orientation-MD+C2T} 
(a,b) The zoom-in of the EC configuration, and the green loops with arrows denote the 1D winding loops encircling ELs. (c-j) The eigenfrequency braiding along loops in (a), and the energy vorticities are labeled at the top. }
\end{figure*} 

In Fig.~\ref{fig-2Dbands-MD+C2T}, we plot the 2D band diagrams on the mirror planes $k_z=\pi$ and $k_x=0$ for the PCs in Fig. 5(c) of the main text. Note that we only show the two bands forming ELs and label the eigenfrequencies of the two bands as $f_1$ and $f_2$ based on the value of their real parts. By sweeping the 2D band diagrams on the mirror planes, we retrieve the degenerate lines and plot them in the BZ as shown in Fig.~ \ref{fig-2Dbands-MD+C2T} (c), which form double-earring ECs. 

Then, we choose four winding loops near the chain points as shown in Fig.~\ref{fig-orientation-MD+C2T} (a). By plotting the eigenfrquency braiding in Fig.~\ref{fig-orientation-MD+C2T} (c-f), we determine the orientation of the ELs, and find that ELs reverse their direction as passing through the mirror planes, which are enforced by the \mdagger[z] and $C_{2x}\mathcal{T}$ symmetries. 
In addition, we notice that the two ELs are very close around $k_y=0$, so to check whether there is a planar chain or not, we zoom in on this part in Fig.~\ref{fig-orientation-MD+C2T} (b). As shown in Fig.~\ref{fig-orientation-MD+C2T} (h), the braiding along loop $\Gamma_6$ is trivial, indicating that the ELs are not chained together.

\newpage

\bibliography{references.bib}